\documentclass[sigconf, nonacm]{acmart}

\usepackage{graphicx}
\usepackage[list=true, font=small, labelfont=bf, 
labelformat=brace, position=top]{subcaption}

\usepackage{tabularx}
\usepackage{makecell}
\usepackage{multirow}

\usepackage{float}
\usepackage{enumitem}
\usepackage{footmisc}

\usepackage{algorithm} 
\usepackage[noend]{algpseudocode} 
\usepackage{amsmath,amsfonts}
\usepackage{bm} 

\setlist[description]{leftmargin=\parindent,}

\newcommand{\Break}{\State \textbf{break}}

\DeclareMathOperator{\argmin}{argmin}
\DeclareMathOperator{\argmax}{argmax} 

\DeclareMathOperator{\gain}{gain}
\DeclareMathOperator{\ANNS}{ANNS}
\DeclareMathOperator{\KNN}{KNN}
\DeclareMathOperator{\GQ}{GQ}

\DeclareMathOperator{\AND}{\overline{\delta_N}}
\DeclareMathOperator{\DEG}{DEG}

\DeclareMathOperator{\MRNG}{MRNG}
\DeclareMathOperator{\lune}{lune}

\DeclareMathOperator{\va}{\large \text{v}_A}
\DeclareMathOperator{\vb}{\large \text{v}_B}
\DeclareMathOperator{\vc}{\large \text{v}_C}
\DeclareMathOperator{\vd}{\large \text{v}_D}
\DeclareMathOperator{\ve}{\large \text{v}_E}
\DeclareMathOperator{\vf}{\large \text{v}_F}

\DeclareMathOperator{\opt}{opt}

\DeclareMathOperator{\extend}{ext}

\newcommand\recallAtVar[1]{\operatorname{recall@}\hspace{-0.2em}{#1}}

\newcommand\vldbdoi{XX.XX/XXX.XX}
\newcommand\vldbpages{XXX-XXX}
\newcommand\vldbvolume{17}
\newcommand\vldbissue{1}
\newcommand\vldbyear{2024}
\newcommand\vldbauthors{\authors}
\newcommand\vldbtitle{\shorttitle} 
\newcommand\vldbavailabilityurl{URL_TO_YOUR_ARTIFACTS}
\newcommand\vldbpagestyle{plain}

\begin{document}
\title{Fast Approximate Nearest Neighbor Search with a Dynamic Exploration Graph using Continuous Refinement}
%


\author{Nico Hezel}
\orcid{0000-0002-3957-4672}
\affiliation{%
  \institution{HTW Berlin, Germany}
}
\email{hezel@htw-berlin.de}

\author{Kai Uwe Barthel}
\orcid{0000-0001-6309-572X}
\affiliation{%
  \institution{HTW Berlin, Germany}
}
\email{barthel@htw-berlin.de}

\author{Konstantin Schall}
\orcid{0000-0003-3548-0537}
\affiliation{%
  \institution{HTW Berlin, Germany}
}
\email{konstantin.schall@htw-berlin.de}

\author{Klaus Jung}
\orcid{0000-0002-3600-6848}
\affiliation{%
  \institution{HTW Berlin, Germany}
}
\email{klaus.jung@htw-berlin.de}

%
%
%

\begin{abstract}
For approximate nearest neighbor search, graph-based algorithms have shown to offer the best trade-off between recall and search time. We propose the \textit{Dynamic Exploration Graph} (DEG), which is superior to existing algorithms in terms of search and exploration efficiency by combining two new ideas: First, a single undirected even-regular graph is incrementally built by partially replacing existing edges to integrate new vertices and to update old neighborhoods at the same time. Secondly, an edge optimization algorithm is used to continuously improve the quality of the graph. 
Combining this ongoing refinement with the graph construction process leads to a well-organized graph structure at all times, resulting in: (1) increased search efficiency, (2) predictable index size, (3) guaranteed connectivity and therefore reachability of all vertices, and (4) a dynamic graph structure. 
In addition we investigate how well existing graph-based search systems can handle indexed queries where the seed vertex of a search is the query itself. 
Such exploration tasks, despite their good starting point, are not necessarily easy. High efficiency in approximate nearest neighbor search (ANNS) does not automatically imply good performance in \textit{exploratory search}.
Extensive experiments show that our new Dynamic Exploration Graph significantly outperforms existing algorithms for indexed and unindexed queries.
\end{abstract}

\maketitle              

\pagestyle{\vldbpagestyle}
\begingroup\small\noindent\raggedright\textbf{PVLDB Reference Format:}\\
\vldbauthors. \vldbtitle. PVLDB, \vldbvolume(\vldbissue): \vldbpages, \vldbyear.\\
\href{https://doi.org/\vldbdoi}{doi:\vldbdoi}
\endgroup
\begingroup
\renewcommand\thefootnote{}\footnote{\noindent
This work is licensed under the Creative Commons BY-NC-ND 4.0 International License. Visit \url{https://creativecommons.org/licenses/by-nc-nd/4.0/} to view a copy of this license. For any use beyond those covered by this license, obtain permission by emailing \href{mailto:info@vldb.org}{info@vldb.org}. Copyright is held by the owner/author(s). Publication rights licensed to the VLDB Endowment. \\
\raggedright Proceedings of the VLDB Endowment, Vol. \vldbvolume, No. \vldbissue\ %
ISSN 2150-8097. \\
\href{https://doi.org/\vldbdoi}{doi:\vldbdoi} \\
}\addtocounter{footnote}{-1}\endgroup

\ifdefempty{\vldbavailabilityurl}{}{
\vspace{.3cm}
\begingroup\small\noindent\raggedright\textbf{PVLDB Artifact Availability:}\\
The source code, data, and/or other artifacts have been made available at \url{https://github.com/Visual-Computing/DynamicExplorationGraph}.
\endgroup
}

\section{Introduction}
Nearest neighbour search (NNS) seeks to determine the closest data points to a query within a particular set. 
The attribute of closeness can be interpreted by a distance function and is evaluated based on various selected features of the points being compared. 
The data points are therefore represented by feature vectors in a high-dimensional space, where less similar points are further apart.
NNS is applied to many problems, including pattern recognition, statistical classification, computer vision, content-based image retrieval, spell checking, and data compression. 
When the feature vector space is discrete and sparse (e.g., for text documents), special data structures such as an inverted index \cite{Knuth1997} can be used. 
However, image, video, and audio data are often represented by dense continuous feature vectors, usually obtained from deep learning models \cite{Simonyan2015}. 
The metric and feature vectors are often predefined, which makes any linear search very costly if the number of data points or the dimension of their features is very high \cite{Li2020}. 
To overcome this problem, many applications \cite{Zhao2019,Johnson2019,Sugawara2016} employ an approximated nearest neighbor search (ANNS) \cite{Wei2020,Wang2021}. 
Such a search involves either compressing the data or leveraging additional data structures to skip a considerable portion of the data points. While this approach introduces inherent inaccuracies, it substantially accelerates the search process.
A slight variation of ANNS is used in product recommender systems \cite{Park2015} and visual image browsing systems \cite{Barthel2019} where similar items to a selected dataset item need to be retrieved.
Interactive image search and browsing systems with many concurrent users would therefore benefit from efficient algorithms for approximate nearest neighbor search and exploration.

\subsection{Background}

\textbf{Exact Search vs Approximate Search.} Due to the curse of dimensionality exact nearest neighbor search is very inefficient for modern datasets with high dimensional data points \cite{Indyk1998}.
Algorithms for approximate nearest neighbor search speed up the process by building an index based on the dataset and precompute as much information as possible.
\textit{Quantization-based approaches} \cite{Jegou2011,Weber1998,Andre2015} transform the input data into a lower dimensional feature space to do a coarse search first. 
\textit{Hash-based searches} \cite{Indyk1998,Gong2020,Huang2017} partition the space with hyper-planes and check only data points located on the same side of each hyperplane as the query. 
\textit{Tree-based algorithms} \cite{Arora2018,Fu2000,Naidan2015} also partition the search space, where each subspace can be re-partitioned recursively, resulting in a hierarchical data structure. 
\textit{Graph-based approaches} \cite{Hajebi2011,Dong2011,Malkov2014,DPG2020,Cong2016,Malkov2020,Iwasaki2018,Cong2019,Cong2021} represent all data points with vertices in a graph and connect similar vertices with edges.
In recent surveys \cite{Aumuller2020,Shimomura2021,DPG2020}, graphs tend to have a better trade-off between accuracy and efficiency than any other type of algorithm. 
Therefore, we will focus only on graph-based approaches in the remainder of the paper.
\\
\\
\textbf{Problems of Graph-based Search Methods.} Most of the existing graph approaches can be categorizes into three groups \cite{Wang2021Survey}. Graphs like kGraph \cite{Dong2011} and EFANNA \cite{Cong2016} approximate a \textit{k-nearest neighbor graph} (KNNG) by connecting each vertex to the closest $k$ other vertices, making navigation to other graph regions rather difficult. 
Another group of graphs (e.g. ONNG \cite{Iwasaki2018}, DPG \cite{DPG2020}, NSG \cite{Cong2019}, NSSG \cite{Cong2021}) prunes the edges of existing KNNGs by imposing constraints on the distribution of the neighborhood. This selection process reduces the size of the graph and makes navigation efficient again. 
However, the construction time is cumulative (first building a KNNG and then pruning its edges) and adding new data points is impossible without repeating the pruning procedure.
The last group combines the gathering of good neighbor candidates and the pruning process in one step. 
HNSW \cite{Malkov2020} is the fastest graph in this group, but introduces a hierarchical data structure. On its lower layers there is no guarantee of \textit{strong connectivity}, which might trap the search process in local minima \cite{Cong2021}. 
\\
\\
\textbf{Evaluation of Exploration Queries.} General exploration, where the search starts at a specific indexed data point and similar vertices need to be retrieved, is rarely considered in literature.
This form of \textit{exploratory search} is used in various browsing applications, including interactive visual image navigation systems \cite{Navigu2023} and image/video search systems \cite{Vibro2023}. 
While some graphs \cite{Dong2011,Cong2019,Malkov2020} are inherently optimized for this type of search query, no experiments have been conducted to determine the performance of such queries for different graph algorithms. 
The efficiency of indexed queries cannot be deduced from the quality of unindexed queries, as shown in Section \ref{sec:explorationExperiements}.
\\
\\
\textbf{Requirements for ANNS Graphs.} In summary an ideal graph-based search system should be able to work with any distance function and generic feature space regardless of the dimensionality. 
For dynamic datasets where data points are continuously removed or added, efficient manipulation of the existing graph is crucial. 
The time between the insertion of a new vertex and the ability to find it, should be kept low. 
Furthermore, the index (graph and auxiliary data) should be as compact as possible and scalable. 
In order to handle unindexed and indexed queries, no changes to the hyperparameters or the construction algorithm should be necessary. 
The efficiency of retrieving many similar data points for a query should be as high as possible.

\subsection{Contribution}

In this paper, we propose a new type of neighborhood graph called \textit{Dynamic Exploration Graph} (DEG), which consists of a single graph component with no additional data structures. 
The DEG is an even-regular, undirected, weighted graph where the edge weights represent the distances between the feature vectors of adjacent vertices.  
New vertices are added incrementally by partially replacing edges of existing vertices to update their neighborhood and ensure regularity.
Unlike other approaches an \textit{edge optimization scheme} improves existing connections after data points have been added.
This dynamic update process continuously lowers the average neighbor distance of the graph by connecting vertices to more similar neighbors.
Neither the algorithm for adding vertices nor the one for updating edges affects the regularity or connectivity of the graph.
\\
\\
Our main contributions are as follows:
\begin{itemize}

\item In Section \ref{sec:requirementsOfGraphAlgorithms}, we define and analyze which graph properties are necessary to use existing graphs in situations like exploration or changing datasets.

\item Section \ref{sec:graphConstructionAndOptimization} introduces a new metric for assessing the quality of small graph changes and the fundamentals of the DEG.

\item An incremental graph construction strategy for even-regular, undirected, weighted graphs is presented in Section \ref{sec:incrementalConstruction}, which guarantees graph connectivity and regularity. 

\item To improve the search and exploration efficiency a dynamic edge optimization algorithm reconnects existing vertices to lower the average neighbor distance. A detailed description can be found in Section \ref{sec:dynamicEdgeOptimizations}. 

\item Comprehensive experiments in Section \ref{sec:searchExperiments} show that the DEG is up to 35\% more efficient than the current state of the art in approximate nearest neighbor search tasks.

\item A protocol for testing the exploration quality of graphs is established and used in Section \ref{sec:explorationExperiements}. The DEG again is up to 50\% more efficient for various recall ranges.


\end{itemize}

\section{Preliminaries}
\label{sec:preliminaries}

\subsection{Notation}

Unless otherwise specified the following notations apply to all equations in the paper. Let $P$ denote a finite dataset of points representing points in an $m$-dimensional vector space $\mathbb{R}^m$. Let $\delta(\cdot, \cdot)$ denote the distance between two points. During the search phase one of the two points is the query while the other is a data point of $P$. The query can be an element of $P$ or an arbitrary element of $\mathbb{R}^m$. The $k$-nearest neighbors in $P$ for a given query $q$ are denoted by $\KNN(P, q)$ with $|\KNN(P, q)| = k$ and
\begin{equation} \label{eq:knn}
\begin{gathered}
\forall x \in \KNN(P, q) \quad
\forall x^\prime \in P \setminus \KNN(P, q) \quad
\delta(x, q) \leq \delta(x^\prime, q) 
\end{gathered}
\end{equation}

To define an approximate $k$-nearest neighbor search $\ANNS(P, q)$, the distance comparison of Equation \ref{eq:knn} can be modified to $\delta(x, q) \leq (1 + \lambda)\delta(x^\prime, q)$ for $\lambda > 0$. A common way to measure the quality of an ANNS search algorithm is to compute the average recall rate of all queries in a test set $Q$.
\begin{equation}
\recallAtVar{k} = \frac{1}{|Q|} \sum_{q \in Q} \frac{\lvert \ANNS(P, q) \cap \KNN(P, q) \rvert}{k}
\label{eq:recallAtK}
\end{equation}
The recall rate will be $1$ if the approximate search delivers the same $k$ elements as the exact search. $0$ indicates that all of the retrieved elements are different from the true $k$-nearest neighbors. When comparing ANNS algorithms, the recall rate must be considered in relation to the search time (queries per second).
\\
\\
\textbf{Proximity Graph.} Let $G = G(V, E)$ be a directed graph, where $V$ is the set of vertices and $E$ the set of edges defining the relationship among the vertices. For every data point in $P$ exists a vertex in $V$. Furthermore let $(v,u) \in E$ denote an edge connecting $v \in V$ with $u \in V$. 
Edges are considered short or long if the distance to the adjacent vertex $\delta(u,v)$ is small or large, respectively. 
Let the neighbors $N(G, v)$ refer to the set of adjacent vertices of $v$ in $G$.
The edges connecting $v$ to the vertices in $N(G, v)$ are defined as outgoing edges, their number is called the \textit{out-degree} of vertex $v$.
All edges coming from other vertices to $v$ are called incoming edges. Their number is called the \textit{in-degree} of $v$.
\\
\\
\textbf{Approximated Nearest Neighbor Search.} There are two commonly used graph search algorithms, called greedy-search and range-search \cite{Wang2021Survey}. The later is used by the DEG and depicted in Algorithm \ref{alg:rangeSearch}. While the greedy-search is similar, it does not use the range-search factor $\varepsilon$ and instead increases $k$ internally to explore more vertices. The final search result is stored in $R$.

Both algorithms start at one or many seed vertices $S \subset V$ which are either predefined, randomly selected, or chosen in some other way. 
In each iteration, also called \textit{hop}, the element in $S$ closest to the query is removed and its neighbors are analyzed. Depending on their closeness to the query (see Algorithm \ref{alg:rangeSearch}) they are added to $S$ and  $R$. An additional set $C$ helps to not check any vertex twice.

The goal is to maximize $\recallAtVar{k}$ while keeping the number of checked vertices $\lvert C \rvert$ as small as possible. 
The efficiency of navigating the graph from one vertex to another, commonly referred to as "navigation speed" is the ratio of hops and checked vertices $\lvert C \rvert$.
\\
\begin{algorithm}\small
    \caption{RangeSearch($G, S, q, k, \varepsilon$)} \label{alg:rangeSearch}
	\begin{algorithmic}[1]
	\Require graph $G$, set of seed vertices $S \subset V$, query $q \in \mathbb{R}^m$ 
	, number of search results $k \in \mathbb{N}$, search range factor $\varepsilon \in \mathbb{R}^+$
    \Ensure $R$ is a list of vertices approximately closest to $q$
    \State $C \gets \emptyset$ \Comment{set of checked vertices}
    \State $R \gets S$ \Comment{result list}
    \State $r \gets \infty$ \Comment{search radius}
    \While {$S \neq \emptyset$}
        \State $s \gets \argmin_{x \in S} \delta(x, q)$ \Comment{find closest vertex to $q$ in $S$}
        \State $S \gets S \setminus \{s\}$
        \If{$\delta(s, q) > r(1 + \varepsilon)$} \Comment{vertex to check is too far way}
            \State \Return $R$
        \EndIf
        
        \State $N \gets N(G,s)$ \Comment{neighbors of $s$}
        \State $N \gets N \setminus C$ \Comment{unchecked neighbors of $s$}
        \ForAll {$n \in N$}
            \If{$\delta(n, q) \leq r(1 + \varepsilon)$} \Comment{check close neighbors} 
                \State $S \gets S \cup \{n\}$
                \Comment{analyze their neighborhood later} 
                \If{$\delta(n, q) \leq r$} \Comment{is neighbor close enough?} 
                    \State $R \gets R \cup \{n\}$
                    \Comment{add to result list} 
                    \If{$|R| > k$} \Comment{limit the result list size}
                        \State $R \gets R \setminus \{ \argmax_{x \in R} \delta(x, q) \}$
                        \State $r \gets \max_{x \in R} \delta(x, q)$
                    \EndIf
                \EndIf
            \EndIf
        \EndFor
        \State $C \gets C \cup N$ \Comment{add checked neighbors to check list}
    \EndWhile
    \State \Return $R$
	\end{algorithmic} 
\end{algorithm}
\\
\textbf{Graph Quality.} A commonly mentioned graph metric is the \textit{graph quality} \cite{Dong2011}, which compares the similarity of the neighborhood of a vertex to its actual nearest vertices. The quality can be calculated as follows:
\begin{equation}
\label{eq:graph_quality}
\GQ(G) = \frac{1}{|V|} \sum_{v \in V} \frac{|N(G, v) \cap \KNN(V, v)|}{|N(G, v)|}
\end{equation}
\noindent where $|N(G, v)| = |\KNN(V, v)|$ for every $v$.
\subsection{Scope}
To balance a focused and a comprehensive comparison, we apply some key constraints.
\vspace{3mm} 
\\
\textbf{Graph-based ANNS.} Only graph based algorithms are considered. Although some effective other algorithms in terms of index size and construction time have been proposed, their search performance is far inferior to graph based approaches \cite{Aumuller2020,Shimomura2021}.
\vspace{3mm} 
\\
\textbf{Dynamic datasets and algorithms.} 
Experiments involving changing datasets and fully dynamic graphs will be covered in a future paper,  as it requires new testing protocols and datasets.
\vspace{3mm} 
\\
\textbf{Hardware.} The focus is on in-memory algorithms running on a single CPU thread, although some hardware-specific \cite{Groh2019,Zhao2020,Subramanya2019,Chen2021} and distributed \cite{Deng2019} approaches are applicable to our method.

\begin{table*}[ht!]
\begin{tabular*}{\textwidth}{@{\extracolsep{\stretch{1}}}*{9}{c}}
\hline
    \textbf{Property} 
    & \textbf{kGraph} 
    & \textbf{DPG}
    & \textbf{ONNG}
    & \textbf{EFANNA}   
    & \textbf{NSG} 
    & \textbf{NSSG}
    & \textbf{HNSW}
    & \textbf{DEG} \\
\hline
    Incremental Graph
    & No
    & No
    & No
    & No
    & No
    & No
    & \textbf{Yes}
    & \textbf{Yes} \\
    Fully Dynamic Graph
    & No
    & No
    & No
    & No
    & No
    & No
    & No
    & \textbf{Yes} \\
    Search Reachability
    & No
    & No
    & \textbf{Yes}
    & No
    & \textbf{Yes}
    & \textbf{Yes}
    & \textbf{Yes}
    & \textbf{Yes} \\
    Graph Connectivity
    & No
    & No
    & No
    & No
    & No
    & No
    & No
    & \textbf{Yes} \\
\hline
\end{tabular*}
\caption{Properties of graphs optimized for ANN search. The Dynamic Exploration Graph (DEG) is proposed in Section \ref{sec:graphConstructionAndOptimization}.}
\vspace{-2.0em}
\label{tab:graph_props}
\end{table*}

\section{Related Work}

Graph-based ANN methods have gained significant attention in recent years. The most promising techniques involve approximating one or more fundamental graphs, such as the \textit{k-Nearest Neighbor Graph} (KNNG) \cite{Paredes2005}, \textit{Relative Neighborhood Graph} (RNG) \cite{Toussaint1980}, or \textit{Delaunay Graph} (DG) \cite{Delaunay1933}. However, constructing these graphs efficiently requires prior knowledge of the data distribution and feature space \cite{Navarro2002}.

The KNNG is a directed graph where all vertices are adjacent to their best possible neighbors using Equation \ref{eq:knn}. 
However, applying this equation for all vertices of the dataset would result in a complexity of $O(n^2)$.
The authors of \textit{kGraph} \cite{Dong2011} proposed to acquire potential neighbors according to the simple idea "a neighbor of a neighbor is probably also a neighbor".
Starting from a \textit{random graph}, each vertex selects the most similar vertices from its adjacent vertices and their neighborhoods.
This process is called \textit{NN-Descent} and is repeated several times, successively increasing the \textit{graph quality}. 
Another graph approximating KNNG is the \textit{Extremely Fast Approximate Nearest Neighbor Search Algorithm} (EFANNA) \cite{Cong2016}. 
Instead of starting with a random graph, EFANNA first creates multiple KD trees \cite{Bentley1975} and uses ANNS to find initial neighbors for all vertices. 
Subsequently, the graph is optimized with a few iterations of \textit{NN-expansion}, a variant of \textit{NN-descent}. 
According to \cite{Peng2019}, the search performance of a KNNG is rather inefficient because the neighbors of a vertex are very similar, which makes reaching other regions of the graph rather slow. 

To circumvent this problem, a \textit{Delaunay Graph} (DG) can be used. A DG guarantees that the result of a \textit{greedy or range-search} is always the nearest neighbor of the query. The disadvantage of a DG is the high number of edges, which for high-dimensional data almost results in a complete graph \cite{Harwood2016}.
In \cite{Malkov2014} a \textit{Navigable Small World} (NSW) graph is proposed to approximate the \textit{Delaunay Graph}. New vertices are incrementally added and connected with undirected edges to good neighbor candidates discovered by a \textit{greedy-search}. As the number of vertices increases, shorter edges are created, improving search accuracy. The initial longer edges become shortcuts to different parts of the graph. Since no edges are deleted or replaced, this process generates hub vertices with high edge numbers, leading to a poly-logarithmic search complexity as the graph size grows \cite{Ponomarenko2014}.

The same authors proposed the improved \textit{Hierarchical Navigable Small World} (HNSW) \cite{Malkov2020} graph by approximating a \textit{Delaunay Graph} and a \textit{Relative Neighborhood Graph} (RNG). 
A RNG is similar to the DG but considers the neighbors distributions and applies restrictions to the connected neighbors, thereby reducing the number of distance calculations during ANNS. 
HNSW produces an approximation of RNG by building a hierarchical graph and adding longer edges to higher layers for faster navigation, in addition to imposing an upper bound of connections per vertex at the bottom layer. Its search complexity scales logarithmically, but its hierarchical advantages decrease with increasing intrinsic dimensionality \cite{Peng2019}.

Instead of building an entirely new graph, the \textit{Diversified Proximity Graph} (DPG) \cite{DPG2020} uses an existing \textit{kGraph} and converts all the edges into undirected edges. 
In a second step, edges to adjacent neighbors of a vertex are removed if their angular similarity to each other is too high.
This approximation of a RNG \cite{Wang2021Survey} evenly distributes the neighboring vertices in all directions and therefore increases the \textit{navigation speed} within the graph. 
No extra data structure is necessary, but depending on the size of the initial \textit{kGraph} there is no guarantee  sufficient edges will be removed to create a manageable graph index.
The concept of pruning the edges of an existing graph is used in other graphs as well.
The \textit{Optimized Nearest Neighbors Graph} (ONNG) \cite{Iwasaki2018} applies several in- and out-degree adjustments on an \textit{Approximate k-Nearest Neighbor Graph} (ANNG) \cite{Iwasaki2010} to approximate a \textit{Delaunay Graph} and \textit{Relative Neighborhood Graph}.
The ANNG is an undirected graph following the same construction procedure as the NSW, except that it uses \textit{range search} instead of \textit{greedy search}.
After replacing all undirected edges in ANNG with bidirectional edges, ONNG retains only the shortest edges per vertex through a degree adjustment process and searches for redundant paths to further reduce the number of neighbors. 
Fast navigation during ANNS is only possible by using an additional \textit{VP-tree} \cite{Fu2000} to find a good starting vertex in the graph.

The \textit{Navigating Spreading-out Graph} (NSG) \cite{Cong2019} further reduces the number of edges by approximating a \textit{Monotonic Relative Neighborhood Graph} (MRNG). 
A MRNG guarantees a monotonic path for every pair of vertices ${p,q}$ such that the path $v_1, v_2, ..., v_k$ with $v_1=p$ and $v_k=q$ comes closer to $q$ with every step $\delta(v_i,q) > \delta(v_{i+1},q)$ for $i = 1,...,k - 1$. 
MRNG never converges to a fully connected graph like DG, but has a high indexing complexity of $O(n^2 log(n) + n^2c)$, where c is the average out-degree.
The Navigating Spreading-out Graph on the other hand is an approximation of MRNG and uses EFANNA as an initial graph to find good neighbor candidates with the help of ANNS and various selection strategies.
The same authors improved NSG in the \textit{Navigating Satellite System Graph} (NSSG) \cite{Cong2021} where they speed up the construction phase by replacing the neighbor candidate acquisition process with \textit{NN-descent}. 
In both cases, a \textit{VP-tree} is used to combine potential graph components into a single component during the construction phase.

\section{Requirements of graph algorithms} \label{sec:requirementsOfGraphAlgorithms}

When evaluating graph-based ANNS systems, it is important to distinguish between measurable performance criteria, special graph properties such as the ability to handle dynamic (changing) dataset, and the applicability of the graph to other use cases (e.g. user guided exploration, recommendation, label propagation).

The achievable performance depends partially on the distance metric, the dimensionality and the number as well as the distribution of the data points in the feature space \cite{DPG2020}. 
Another significant influence is given by the architecture of the graph.
The performance of a graph can be measured by the \textbf{search speed} and its \textbf{recall rate} as well as its required \textbf{construction time} and \textbf{memory consumption}.
Various measurements for different graphs and datasets are documented in Section \ref{sec:searchExperiments} and Section \ref{sec:constructionTimeAndMemoryConsumption}. 

Every architecture also implies certain limitations in terms of construction and extension as well as navigation, reachability and exploration possibilities. 
In addition to pure search tasks, there are other use cases that can be addressed with graph-based search systems.
Here, a \textit{exploratory search} should be mentioned, which does not start at the general seed vertex, but at a result vertex of the previous search and does not allow some vertices to be included in the result list, as they have already been shown to the user of the application. 
This scenario requires a connected graph and the ability to start the search at any vertex. 

\subsection{Graph properties}
Many general graph properties have been investigated in literature \cite{Wang2021Survey}. 
We will therefore focus on the properties required for dynamic datasets in interactive search and exploration systems.

\textbf{Incremental Graphs} can be extended at any time with new vertices and do not need knowledge about the entire dataset. 
For large, continuously growing datasets an incremental updating scheme will save a lot of resources. 

\textbf{Fully Dynamic Graphs} are necessary when there is a need to continuously delete and add data points. It is not sufficient to flag deleted vertices and ignored them in search results, since they still consume memory and must be visited during ANNS to maintain a high accuracy while decreasing the search speed.

\textbf{Search Reachability} of all vertices during ANNS is generally ensured by starting at a pre-selected seed vertex, from which there is a path to every other vertex.

\textbf{Graph Connectivity}. A \textit{Strongly Connected Graph} or an undirected graph with a \textit{single connected component} is necessary, if a search must be able to start at any vertex. Inherently, all connected graphs provide full search reachability.
\\
\\
Table \ref{tab:graph_props} summarizes the graph properties for the graphs described in the Related Work section. 
In Appendix \ref{appendix:evaluatingGraphProperties} a more detailed explanation and justification of the table entries is given. 

\section{Dynamic Exploration Graph} \label{sec:graphConstructionAndOptimization}

To meet the specifications outlined in Section \ref{sec:requirementsOfGraphAlgorithms}, we propose the Dynamic Exploration Graph (DEG), which is designed to be both compact and efficient for navigation and exploration, while ensuring full reachability from every vertex. The required connectivity property is obtained by incrementally building an undirected graph \cite{Malkov2014}. The main challenge with undirected graphs lies in attaining a well-balanced distribution of neighboring vertices, without the formation of hubs (vertices with high degrees) \cite{Wang2021Survey}. Such distribution is accomplished by reducing the average neighborhood distance and the even regularity property of the graph. Furthermore, the DEG approximates a Monotonic Relative Neighborhood Graph \cite{Cong2021}, which reinforces the desired distribution of vertices and helps the construction process.
The following sections:
\begin{itemize}
\item define the basic properties of the DEG and proposes a metric to evaluate small graph changes
\item describe how the DEG approximates the Monotonic Relative Neighborhood Graph 
\item present two algorithms for adding new vertices to the graph and for optimizing existing edges
\end{itemize}

\subsection{DEG Foundations}
In this section, the main principles of the DEG are presented. 
First, the limitation of the number of edges per vertex and its advantages are explained, followed by a description of how the quality of a neighborhood distribution can be measured. 
Finally, we prove connectivity can be guaranteed if certain graph properties are not violated during graph manipulation.
\\
\\
\textbf{Even Regularity}. By design the DEG is a even-regular graph with undirected edges and no loops. Its regularity $d$ is denoted by $\DEG_d$. Due to the edge and degree constraints, the smallest possible $\DEG_d$ is a complete graph $K_{d+1}$ with $d+1$ vertices. 
The number of edges in any undirected regular graph is $|E| = (|V| \cdot d)/2$, which can be derived from the handshaking lemma \cite{Euler1736}. Adding a new vertex to the graph therefore increases the edge count by $d/2$. 
An odd regularity is not suited for the DEG, since it would require an even amount of vertices at any time to stay regular. 
The degree of each vertex in the DEG should be at least 4.
A degree of 2 or 0 would form either circles or no edges altogether.

Without the regularity constraint the performance of the DEG would be similar to NSW, as the number of hops during the search would increase poly-logarithmically with the size of the graph \cite{Malkov2020}. 
\\
\\
\textbf{Edge Quality}. When a new vertex is added to the graph, the regularity constraint forces $d/2$ existing edges to be replaced by $d$ new edges, changing the neighborhoods of the vertices involved.  
Section \ref{sec:incrementalConstruction} describes this procedure in more detail.
In order to guide the graph construction algorithm, it is crucial to determine whether the quality of the edges improves for a specific action.
The following section will review the existing \textit{graph quality} metric and highlight its limitations. Subsequently, a new metric is proposed which is used by the DEG. 

\begin{figure}[h!]
\centering
\includegraphics[trim=3cm 6cm 6cm 6.5cm,clip=true,width=\columnwidth]{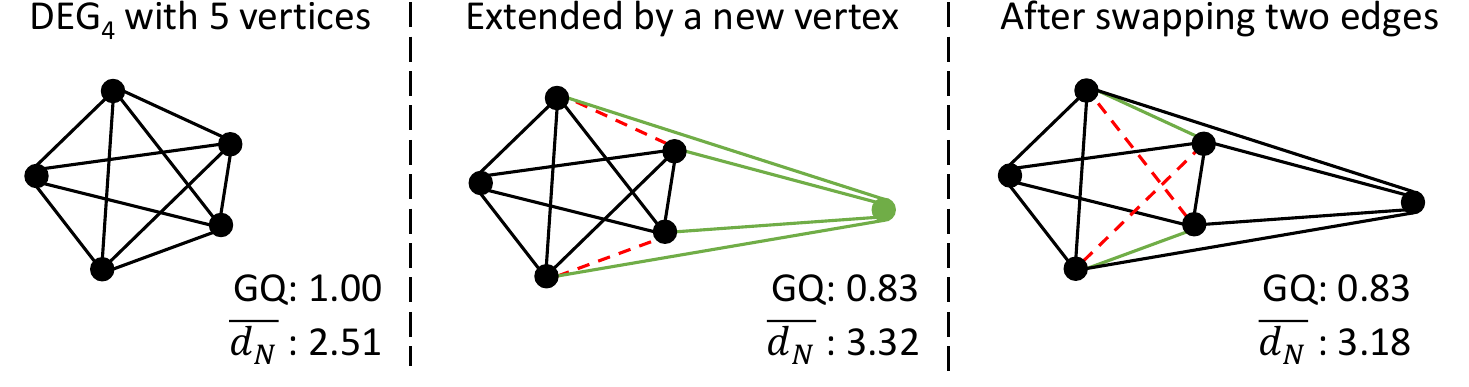} 
\vspace{-0.5cm}
\captionsetup{font=small}
\caption{A 2D toy example of a $DEG_{\bm{4}}$. Left: the smallest possible $DEG_{\bm{4}}$ ($\bm{K_5}$). Center: a new vertex is added to the graph. Right: two edges have been swapped to improve the graph but the graph quality (GQ) stays the same. 
} \label{fig:deg_basics}
\end{figure}

Figure \ref{fig:deg_basics} on the left shows a $\DEG_4$ with 5 vertices which is equal to the complete graph $K_5$. 
A complete graph always has a perfect \textit{graph quality} (GQ) of 1, since all vertices are connected to each other.
In the center of Figure \ref{fig:deg_basics}, a new vertex (shown in green) is added to the graph.
To maintain regularity, two edges (red) were replaced by four new ones (green), which decreases the \textit{graph quality} in this case. 
On the right of Figure \ref{fig:deg_basics}, two existing edges have been swapped. 
Although the four involved vertices have been connected to closer vertices, the \textit{graph quality} remained the same.
Such insensitivity is often observed during minor changes and making clear instructions for action impossible.
\\
We propose the \textbf{Average Neighbor Distance} ($\AND$) as a better metric to measure the quality of an undirected graph. 

\begin{definition}
Let $G(V,E)$ be a $d$-regular undirected graph and $N(G, u) \subset V$ the set of vertices adjacent to $u \in V$. The \textit{average neighbor distance} of a set of vertices $U \subset V$ is:
\end{definition}
\begin{equation} \label{eq:averageNeighborDistance}
\AND(U) = \frac{1}{|U|} \sum_{u \in U} \frac{1}{d} \hspace{-0.1cm}\sum_{v \in N(G, u)} \delta(u, v)
\end{equation}
\\
For $U = V$ the average neighbor distance of the entire graph is calculated. 
A low distance indicates a high similarity between connected vertices. 
To avoid constantly recalculating the distances of adjacent vertices, they can be saved as the weights of the edges: $w_{u,v} = \delta(u, v)$. 
Changing an undirected edge $(u,v)$ in a DEG affects the average neighbor distance of both vertices $u$ and $v$.
When swapping the endpoints of two edges $(u1,v1)$ and $(u2,v2)$, it is sufficient to compare the sum of the edge weights before ($w_{u1,v1} + w_{u2,v2}$) and after ($w_{u1, v2} + w_{u2, v1}$) the change to know if the average neighbor distance of the graph is reduced by this operation. 
This efficient calculation is extensively used in our dynamic edge optimization scheme and when connecting new vertices to the graph.
Figure \ref{fig:deg_basics} illustrates how the metrics $\GQ$ and $\AND$ indicate a graph deterioration after adding a new vertex, but only $\AND$ was able to detect an improvement after two edges were swapped.
\\
\\
\textbf{Connectivity}. 
Another property of the DEG is its guaranteed connectivity, which is a prerequisite for exploration tasks with varying starting points. 
In addition to its even regularity and undirected edges, the DEG is also an Eulerian graph, satisfying Euler's theorem. According to the theorem, a connected graph has an Euler cycle if and only if every vertex has an even degree.
The closed trail or circuit representing the Euler cycle includes all edges of the graph and can be formed from any vertex. 
As a result, each vertex has at least two paths to reach all other vertices. 
The DEG therefore does not have bridges and guarantees 2-edge-connectivity, allowing one edge to be removed without disconnecting the graph. 
Even when removing more edges the probability of a disconnection is very low (see Appendix \ref{appendix:connectivityGuarantees}).
The graph manipulation methods for \textit{adding vertices} and \textit{swapping edges} presented in Section \ref{sec:incrementalConstruction} and \ref{sec:dynamicEdgeOptimizations} are designed to always preserve connectivity.
\\
\\
\textbf{Approximation of DG and MRNG}. 
The properties of the DEG and the construction process described in section \ref{sec:incrementalConstruction} make the DEG an approximation of a \textit{Delaunay Graph} (DG) and a \textit{Relative Neighborhood Graph}. Moreover, the resulting edge distribution is very similar to the current state-of-the-art algorithm HNSW. The exact reasoning for this can be found in Appendix \ref{appendix:approximatingDGandRNG}.

In addition Algorithm \ref{alg:checkMRNG} is used during the construction phase to identify potential neighbors which are part of a \textit{Monotonic Relative Neighborhood Graph} (MRNG). Approximating a MRNG helps the range search to approach the query with less hops. The derivation of this algorithm is provided in Appendix \ref{appendix:approximatingDGandRNG}.

\begin{algorithm}\small
    \caption{checkMRNG($G, v1, v2$)} 
    \label{alg:checkMRNG}
    \begin{algorithmic}[1]
    \Require graph $G$, vertex $v1 \in V$ of $G$, vertex $v2 \in V$ of $G$
    \Ensure an edge between $v1$ and $v2$ is MRNG conform
    \ForAll {$u \in N(G, v1) \cap N(G, v2)$}
        \If{$\delta(v1, v2) > max(w(v1, u), w(v2, u))$)} 
            \State \Return false
        \EndIf
    \EndFor
    \State \Return true
    \end{algorithmic}     
\end{algorithm}
\vspace{-0.5cm}

\subsection{Incremental Construction} 
\label{sec:incrementalConstruction}

\begin{figure}[h!]
\centering
\includegraphics[trim=4cm 3.5cm 5.5cm 3cm,clip=true,width=\columnwidth]{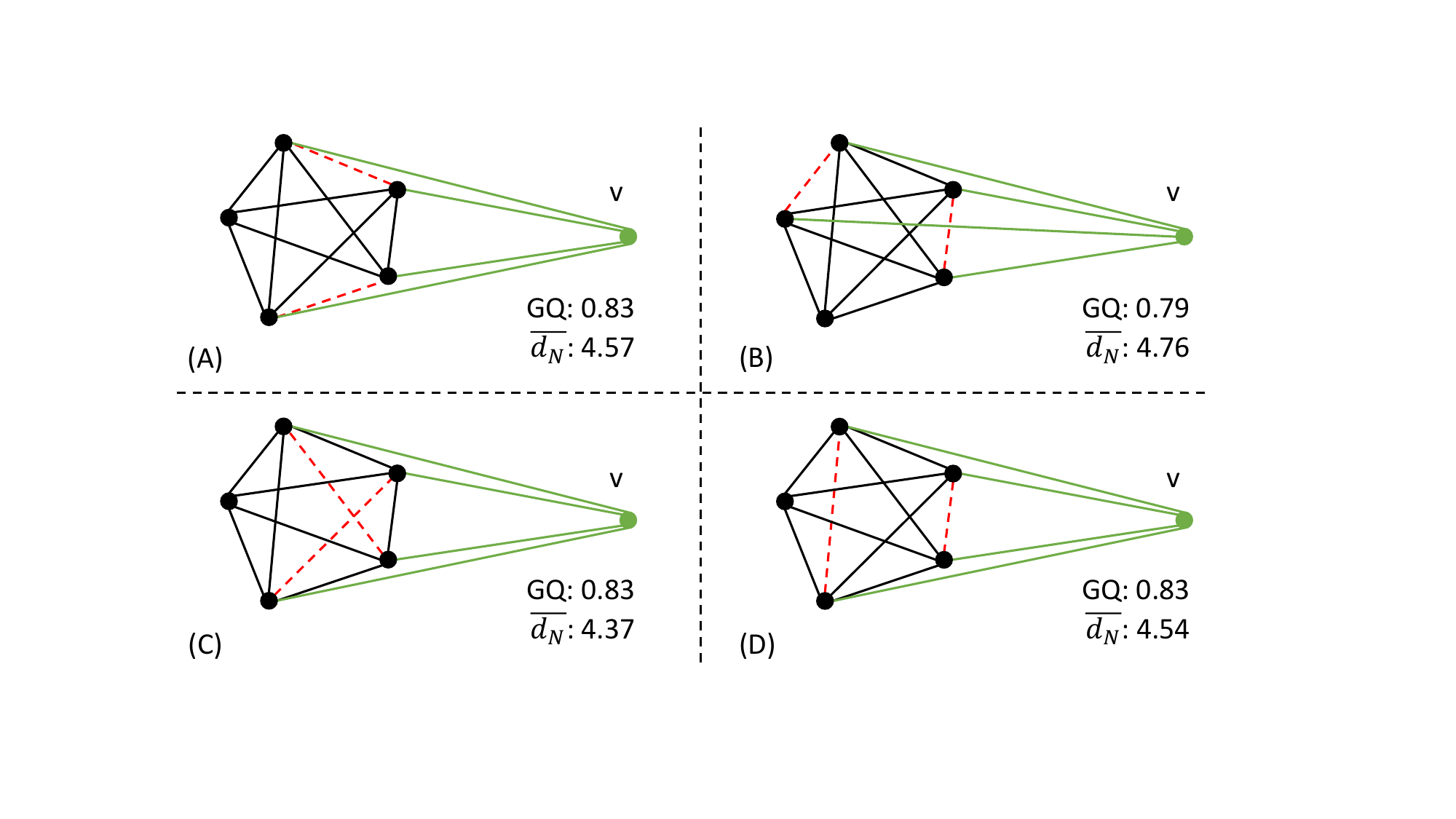}
\captionsetup{font=small}
\caption{A 2D example for extending a $\DEG_4$, showing the final graphs for various selection schemes for choosing the neighbor $n$ in step (4): (A) the neighbor closest to $v$. (B) the neighbor with the shortest edge. (C) the neighbor with the longest edge. (D) the neighbor where the $\AND(V)$ decreases the most.} \label{fig:add_vertex}
\end{figure}

The following section describes how new vertices are added to the Dynamic Exploration Graph. 
Similar to other KNNG algorithms, the new vertex $v$ of data point $p$ tries to connect itself to the best results of a search.
To add a new vertex to the graph, $d$ new edges have to be added and $d/2$ edges have to be removed to ensure the graph properties are not violated:

\begin{enumerate}
\item Starting from the smallest $\DEG_d$ with $d+1$ vertices an arbitrary vertex of the graph is selected as start seed $S$. 

\item A $RangeSearch(\DEG_d, S, v, k_{\extend}, \varepsilon_{\extend})$ is performed for the new vertex $v$. 
The parameter $\varepsilon_{\extend}$ restricts the search range and $k_{\extend}$ is the size of the search result.

\item The best vertex $b$ of the search result not yet connected to the new vertex $v$ is chosen.

\item Based on a selection criterion depicted in Figure \ref{fig:add_vertex} and described later in this section, the edge $(b,n)$ is removed from $b$ where $n$ is not adjacent to $v$. 

\item The two vertices $b$ and $n$ now have $d-1$ neighbors and will be connected to the new vertex $v$, allowing $b$ to reach $n$ via a detour over $v$. Therefore the graph restores its connectivity and regularity.

\item The steps 3-6 are repeated until $v$ has enough edges. 
\end{enumerate}

\begin{algorithm}[h!]\small
    \caption{ExtendGraph($G, d, v, S, k_{\extend}, \varepsilon_{\extend}$)}
    \label{alg:extendGraph}
    \begin{algorithmic}[1]
    \Require current graph $G$, even degree $d \in \mathbb{N}$ and $d \geq 4$, new vertex $v$, set of seed vertices $S$, number of search results $k_{\extend} \in \mathbb{N}$ with $k_{\extend} \geq d$, search range factor $\varepsilon_{\extend} \in \mathbb{R}^+$
    \Ensure find neighbor for $v$ in $G$
	\State $U \gets \emptyset$ \Comment{new neighbors of $v$}
	\State $S \gets$ RangeSearch($G, S, v, k_{\extend}, \varepsilon_{\extend}$)
	\Comment{$k$ vertices similar to $p$}
    \State \textit{skipRNG} $\gets$ false \Comment{two phases: with and w/o RNG checks}
    \While {$|U| < d$}
        \State $B \gets S \setminus U$ \Comment{copy of $S$ minus $U$}
        \While {$|U| < d$ AND $B \neq \emptyset$}        
            \State $b \gets \argmin_{x \in B} \delta(x, v)$ \Comment{closest vertex to $v$ in $B$}
            \State $B \gets B \setminus \{b\}$
            
            \If{\textit{skipRNG} $=$ true OR checkMRNG($G, v, b$)} 
                \State $N \gets N(G,b) \setminus U$ \Comment{neighbor candidates of $b$}
                \State $n \gets$ vertex in $N$ with longest edge to $b$
                \State $U \gets U \cup \{n\} \cup \{b\}$
                \State remove edge $(n,b)$
            \EndIf
        \EndWhile
        \State \textit{skipRNG} $\gets$ true 
    \EndWhile
    \ForAll {$e \in U$}
        \State add edge $(v,e)$
    \EndFor    
    
    \State optimizeEdge(G,v,u,...) for $u \in U \setminus S$\Comment{optional}
    \end{algorithmic}    
\end{algorithm}
Figure \ref{fig:add_vertex} illustrates four ways to select vertex $n$ in step (4) to reduce the average neighbor distance. 
Variant (A) tries to connect the new vertex with the most similar other vertices. 
In (B) the shortest edge of the best vertex is removed, following the idea the incident vertex is likely to be similar and a good candidate for the new vertex. 
In (C), the longest edge of the best vertex is replaced and the two incident vertices are connected to the new vertex. While half of the new neighbors may not be ideal, the quality of the existing neighborhoods is only slightly affected. 
Scheme (D) replaces the edge of the best vertex for which the average neighbor distance of the graph is the lowest.

The best $\AND$ values were obtained by approach (C) and (D) in Figure \ref{fig:add_vertex}.
Although in scheme (D), vertex $n$ is chosen at each step to improve the average neighbor distance, its selection impacts the subsequent vertex $b$. Scheme (D) therefore does not guarantee to always achieve the best results once the new vertex is connected to the graph.
In our experiments with larger graphs (Appendix \ref{appendix:neighborChoices})), we found for datasets with a high local intrinsic dimensionality scheme (C) is preferable and otherwise (D) gives the best results.
It should be noted the graph quality (GQ) for scheme (A), (C) and (D) is the same, making it not suitable as a metric for small graph changes.

The complete graph extension procedure for the neighbor selection scheme (C), is described in Algorithm \ref{alg:extendGraph}.
Since it is possible for all selected vertices $b$ and $n$ to be also elements of the search result, the minimum set size $k_{\extend}$ should be at least $d$.
As mentioned in the last section, additional MRNG compliance tests accelerate the convergence to a graph with good search results. 
Depending on $k_{\extend}$ and the distribution of the data points, there may not be enough neighbor candidates satisfying the MRNG requirements. If $|U| < d$ after the first pass in Algorithm \ref{alg:extendGraph}, the MRNG tests are disabled and the retrieval of good $b$ vertices is repeated.

The new neighbors of $v$ which are not in $S$ might not be the closest possible neighbors. The optional part in Algorithm \ref{alg:extendGraph} therefore replace these neighbors, using the dynamic edge optimization process described in the next section.

\subsection{Dynamic Edge Optimization} \label{sec:dynamicEdgeOptimizations}
During graph extension, the neighbor choice often favors the new vertex, causing existing vertices to lose good short edges in some cases. To address this problem, we developed a second algorithm that tries to continuously improve the edges of the graph. The following section explains this functionality, identifies potential pitfalls, and discusses how to maintain regularity and connectivity.

The goal of the \textit{dynamic edge optimization} process is to minimize the \textit{average neighbor distance}. 
The optimization tries to improve the graph by swapping edges, where a \textit{gain} is determined by the difference of the neighborhood distances before and after the swap. If this \textit{gain} is positive, the average neighbor distance will decrease with the swap. 
The shorter the edges, the more likely the vertices are connected to their best nearest neighbors. 
The number of edges in the graph remains the same, since only the incident vertices of the edges are swapped. 

\begin{figure}[t!]
\centering
\includegraphics[trim=0.5cm 1.3cm 0cm 1.5cm,clip=true,width=\linewidth]{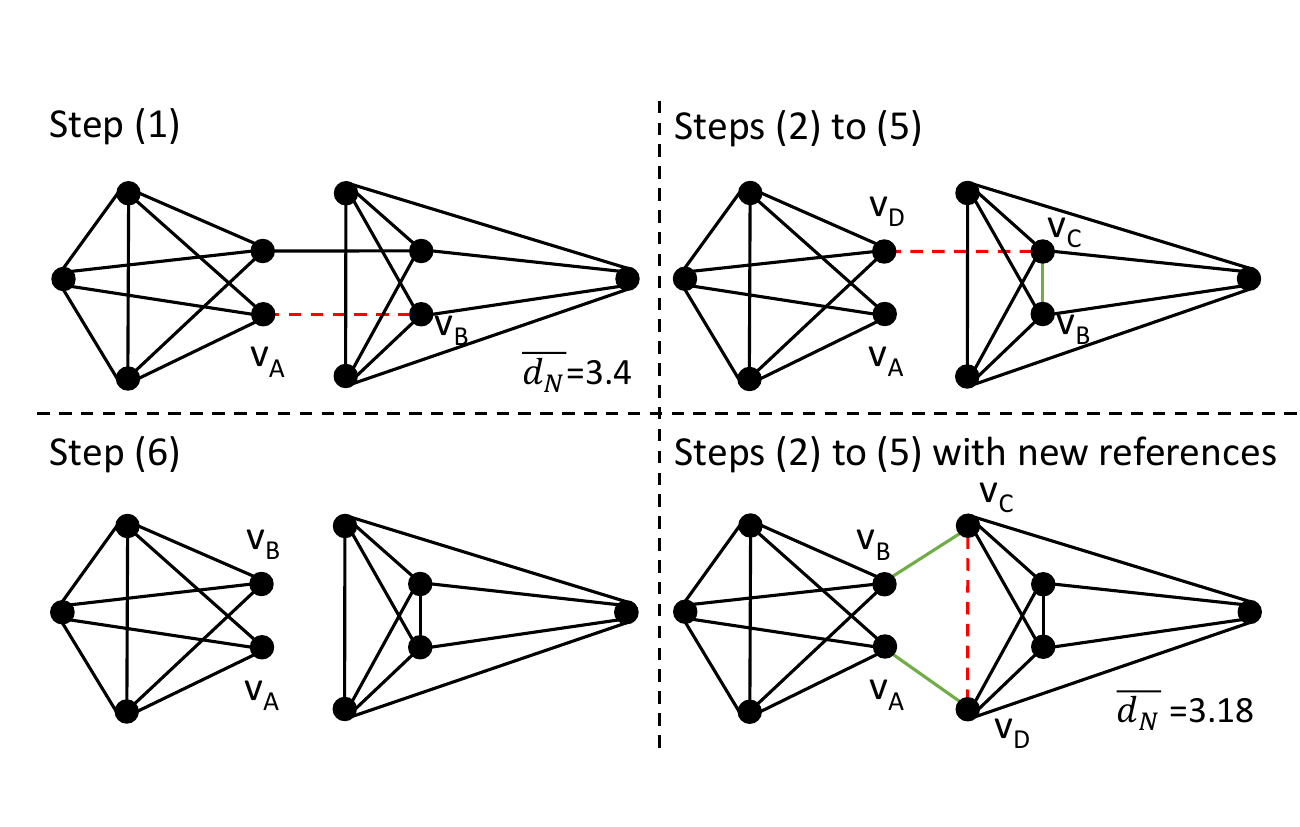}
\captionsetup{font=small}
\vspace{-0.6cm}
\caption{The dynamic edge optimization process is illustrated by an 2D example. Starting from a valid $\DEG_4$ in Step(1) the algorithms tries to replace the edge $(\va, \vb)$ with a better edge constellation.} \label{fig:dynamic_edge_optimization}

\end{figure}

The process to optimize an edge $(\va, \vb)$ can be outlined as follows and is demonstrated in Figure \ref{fig:dynamic_edge_optimization}:

\begin{enumerate}
    \item First the edge $(\va, \vb)$ is removed. This deletion may result in the loss of the 2-edge connectivity, but there will be at least one remaining path between $\va$ and $\vb$.   
    \item A RangeSearch with $S = \{\va\}$ is performed to find a good neighbor for $\vb$. From the result set, vertex $\vc$ and its neighbor $\vd$ are selected such that $\vc \neq \va$, $\vc \neq \vb$, and $\vd \neq \vb$, $N(G, \vb) \cap {\vc} = \emptyset$, where the $\gain = \delta(\va, \vb) - \delta(\vb, \vc) + \delta(\vc, \vd)$ is maximized.
    \item The edge $(\vc, \vd)$ is replaced by $(\vb, \vc)$. For $\va$ the vertices $\{\vb, \vc\}$ may become unreachable. $\vd$ might not be able to reach $\{\vb, \vc\}$. 
    \item Attempt to restore the regularity of vertex $\va$ and $\vd$.
    
    \begin{description}  
         \item[Case a:] If $\va = \vd$, the vertex is missing two edges.  
         A RangeSearch with $S = \{\vb, \vc\}$ for query $\va$ is performed, $\ve$ is selected from the result set and $\vf$ from its neighborhood such that $\va \neq \ve$, $\va \neq \vf$, $N(G, \va) \cap \{\ve, \vf\} = \emptyset$ and $(\gain +\ \delta(\ve, \vf) - \delta(\va, \ve) - \delta(\va, \vf))$ is maximized. If the final gain is positive, the edge $(\ve, \vf)$ is replaced with the two edges $(\va, \ve)$ and $(\va, \vf)$.
         \item[Case b:] If $\va \neq \vd$, the vertices $\va$ and $\vd$ can be connected if: $N(G, \va) \cap {\vd} = \emptyset$, $\gain -\ \delta(\va, \vd) > 0$ and there is a path from $\va$ or $\vd$ to $\vb$ or $\vc$.
    \end{description}
       
    \item If $\va$ and $\vd$ could not be connected, steps (2) to (4) are repeated recursively by referencing $\vb$ to the vertex previously denoted by $\vd$ and starting the neighborhood search of step (2) at the two previous vertices $\vb$ and $\vc$. 

     
    \item If no solution is found after a few iterations (typically 5 are enough), all previous changes are reverted.
\end{enumerate}

Step (1) may cause the loss of 2-edge connectivity, and step (3) may disconnect the graph into two components. Therefore, steps (4a) and (4b) are intended to reconnect the components and restore the 2-edge connectivity. 
If neither (4a) nor (4b) produces a valid solution, step (5) is executed to repeat steps (2) to (4). In this repetition, the vertex labels and search seeds are swapped to ensure that step (3) never results in more than two graph components. It will always reconnect the two potential graph component of the last iteration, but might also create a new isolated component.
There are three ways the process can terminate: 1) (4a) finds a valid vertex/edge combination; 2) (4b) discovers a suitable edge in an already connected graph; or 3) after several iterations, step (6) stops the process and reverts all changes. Regardless of the outcome, the final graph retains its original properties.

The entire edge optimization process is also implemented in Algorithm \ref{alg:optimizeEdge}. In combination with Algorithm \ref{alg:dynamicEdgeOptimization}, which iteratively identifies suboptimal edges and improves them using Algorithm \ref{alg:optimizeEdge}, the average neighbor distance can be reduced.
This improvement is demonstrated in section \ref{sec:qualityOfEdges}, where it can be seen that the search efficiency continues to increase over time.

\begin{algorithm}\small
    \caption{optimizeEdge($G, v1, v2, i_{\opt}, k_{\opt}, \varepsilon_{\opt}$)} \label{alg:optimizeEdge}
    \begin{algorithmic}[1]
    
    \Require current graph $G$, edge between  $v1$ and $v2$, max number of changes $i_{\opt}$, search parameters $k_{\opt}$ and $\varepsilon_{\opt}$
    \Ensure try to improve the edge $(v1, v2)$ 
    \State $M \gets \emptyset$ \Comment{history of modifications}
    \State $\gain \gets w_{v1, v2}$ \Comment{positive change of the graph edges}
    \State Remove edge $(v1, v2)$ from $G$ and add to $M$
    \State $v3 \gets v1, v4 \gets v1$ \Comment{needed for multiple iterations}
    \While {$|M| < i_{\opt}$}

        \State $b \gets \gain$ \Comment{best current gain}
        \State $S \gets$ RangeSearch($G, \{v3, v4\}, v2, k_{\opt}, \varepsilon_{\opt}$)
        \ForAll {$s \in S$}
            \If {$s \neq v1$ and $s \neq v2$ and $N(G, v2) \cap {s} = \varnothing$}
                \ForAll {$n \in N(G, s)$}
                    \If {$n \neq v2$ and $b < \gain - \delta(s, v2) + w_{s, n}$}
                        \State $b \gets \gain - \delta(s, v2) + w_{s, n}$
                        \State $v3 \gets s$, $v4 \gets n$
                    \EndIf
                \EndFor
            \EndIf
        \EndFor
        \If {$b == \gain$}
            \Break
        \EndIf        
        \State $\gain \gets b$
        \State Add edge $(v2, v3)$ to $G$ and add to $M$
        \State Remove edge $(v3, v4)$ from $G$ and add to $M$            
        
        \If {$v1 == v4$}         
            \State $b \gets 0$ \Comment{best current gain}
            \State $S \gets$ RangeSearch($G, \{v2, v3\}, v1, k_{\opt}, \varepsilon_{\opt}$)
            \ForAll {$s \in S$}
                \If {$s \neq v1$ and $N(G, v1) \cap {s} = \varnothing$}
                    \ForAll {$n \in N(G, s) \setminus {v1}$}
                        \If {$b < \gain + w_{s, n} - \delta(s, v1) - \delta(n, v1)$}
                            \State $b \gets \gain + w_{s, n} - \delta(s, v1) - \delta(n, v1)$
                            \State $v5 \gets s$, $v6 \gets n$
                        \EndIf
                    \EndFor
                \EndIf
            \EndFor  
            \If {$b > 0$}
                \State $\gain \gets b$
                \State Add edge $(v1, v5)$ and $(v1, v3)$ to $G$
                \State Remove edge $(v5, v6)$ from $G$
                \State \Return 
            \EndIf   

        \ElsIf {$N(G, v1) \cap {v4} = {v4}$ and $0 < (gain - \delta(v1, v4))$}
            \State $S1 \gets$ RangeSearch($G, \{v2, v3\}, v1, k_{\opt}, \varepsilon_{\opt}$)
            \State $S4 \gets$ RangeSearch($G, \{v2, v3\}, v4, k_{\opt}, \varepsilon_{\opt}$)
            \If {$S1 \cap {v1} = {v1}$ or $S1 \cap {v4} = {v4}$}    \Comment{found a path}
                \State Add edge $(v1, v4)$ to $G$ 
                \State \Return 
            \EndIf
        \EndIf

        \State $v \gets v4$, $v4 \gets v3$, $v3 \gets v2$, $v2 \gets v$ \Comment{prepare next iteration}
    \EndWhile
    \State Revert changes in $M$
    \end{algorithmic} 
\end{algorithm}

\begin{algorithm}\small
    \caption{dynamicEdgeOptimization($G, i_{\opt}, k_{\opt}, \varepsilon_{\opt}$)} \label{alg:dynamicEdgeOptimization}
    \begin{algorithmic}[1]
    	\Require current graph $G$, max number of changes $i_{\opt}$, search parameters $k_{\opt}$ and $\varepsilon_{\opt}$
        \Ensure the worst edge of a random vertex in $G$ was swapped if the average neighbor distance is improved
        
        \State $v1 \gets rnd(G)$ \Comment{random vertex in $G$}
        \State $N \gets N(G, v1)$ \Comment{neighbors of $v1$}

        \ForAll {$v2 \in N$}
            \If {$checkMRNG(G, v1, v2) == False$}
                \State $optimizeEdge(G, v1, v2, i_{\opt}, k_{\opt}, \varepsilon_{\opt})$  
            \EndIf   
        \EndFor        
        \State $v2 \gets$ vertex in $N$ with longest edge to $v1$
        \State $optimizeEdge(G, v1, v2, i_{\opt}, k_{\opt}, \varepsilon_{\opt})$    
    \end{algorithmic} 
\end{algorithm}

\subsection{Implementation}
\label{sec:implementation}

The implementation of the DEG along with all the graphs files constructed in the experiments, are available on github\footnote{https://github.com/Visual-Computing/DynamicExplorationGraph}. 
The code of the DEG is based on HNSW and NSG using the same SIMD instructions \cite{SIMD} and sequential memory allocation. Since the number of edges per vertex is fixed, the required memory to index a static dataset can be pre-allocated. Furthermore is it possible to jump to a vertex memory region using the vertex index information stored in each neighbor list, making the algorithms very memory efficient. 

When loading the graph for search purposes the weights are omitted. 
The seed vertex for the range-search is determined by computing the median vertex of the graph. This vertex is only used during test time and not in the construction phase.

\subsection{Complexity}
First, we investigate the space complexity of DEG's data structure. Second, the search time complexity is approximated by the properties of the graph and applied to estimate the indexing complexity.
\\
\\
\textbf{Space Complexity}
The amount of memory allocated by the DEG depends on the use case. If new vertices are to be added, the edge weights are required, resulting in a memory consumption of $O(|V| \cdot (m + 2 \cdot d))$, where $m$ is the dimensions of the feature space and $d$ is the regularity of the graph. If the DEG is no longer manipulated, then the weights can be disposed to reduce the memory requirement to $O(|V| \cdot (m + d))$. 
\\
\\
\textbf{Search Time Complexity}
\label{sec:analyticalSearchTimeComplexity} The greedy and range search algorithms consist of two phases. In the approach phase, the algorithms navigate from the starting seed vertex to the vertex closest to the query. Upon reaching the vertex, the exploration phase begins to search for other vertices similar to the query within the nearby neighborhood.

Approximately $O(h \cdot o)$ vertices are examined in each phase, where $h$ is the average number of hops and $o$ is the average number of evaluated neighbors per hop. 
For the DEG with $o \leq d$ and during the approach phase, $h$ scales almost logarithmically to the number $n$ of vertices in the graph. This property is derived from DEG's approximation of MRNG \cite{Cong2016}, which calculates the expected length of a monotonic path as $h = (n^\frac{1}{m} \cdot log(n^\frac{1}{m}))$. For high-dimensional datasets with $m > 10$ and $n < 10^9$, the expected number of hops is less than the logarithm of $n$, resulting in a search complexity of $O(log(n) \cdot d)$ for the approach phase. This phase corresponds to a 1-NN (k=1 nearest neighbor) search without backtracking.

For all graphs studied in this paper, the exploration phase has a theoretical worst-case complexity of $O(n \cdot d)$. However, in most prior empirical studies \cite{Cong2016,Malkov2020}, the overall $k$-NN search performance can be approximately generalized by the 1-NN performance (approach phase). 
This aligns with our empirical results in Section \ref{sec:scalabilityAndComplexity} where the general search complexity of the DEG is $O(n^\frac{1}{m} \cdot log(n^\frac{1}{m})) = O(h)$ with $m$ being close to the intrinsic dimensionality of the tested dataset.
\\
\\
\textbf{Indexing Time Complexity}
The complexity of indexing the DEG depends on the search complexity. Algorithm \ref{alg:extendGraph} adds a vertex to the graph using an approximate nearest neighbor search, while Algorithm \ref{alg:optimizeEdge} optimizes some of the new edges.

The edge optimization algorithm performs $t$ iterations to find a good swap combination. Each iteration involves up to three searches: one to reconnect potential graph components and either two more searches to extend the graph or two path searches to check connectivity. A path search is a ANNS but can terminate early upon finding a path. Thus, the worst-case complexity of Algorithm \ref{alg:optimizeEdge} is $O(t \cdot p \cdot h)$, with $t$ swap iterations, $p=3$ potential search tasks, and an anticipated path length of an average search denoted as $h$.

Overall, the total indexing complexity for adding a data point is $O(h + d \cdot t \cdot p \cdot h)$, combining the search for neighbor candidates of the new vertex and the subsequent optimization of its $d$ edges.

\section{Evaluation}
In this section, we present an in-depth evaluation of our new graph by conducting extensive experiments with public and frequently cited datasets.

\subsection{Datasets}
\label{sec:datasets}
As not all current state-of-the-art algorithms can scale to datasets with a billion vectors, we focus our evaluation on the million vector datasets SIFT1M \cite{Jegou2011} and GloVe \cite{Pennington2014} in addition to two smaller datasets, Enron \cite{Enron} and Audio \cite{Audio}. All of them are widely used in the related literature \cite{Wang2021Survey, Cong2021, DPG2020} and cover different media types (image, audio, and text). The local intrinsic dimension (LID) \cite{Costa2005} is computed for each dataset to better reflect its difficulty, since the data may be on a low-dimensional manifold. This allows an evaluation of how well the algorithms generalize across different data distributions. Further details can be found in Table \ref{tab:datasets}. The datasets are organized as base and query data points. The assessed systems index all base data points and conduct an approximate neighborhood search using the query data.

\begin{table}[h]
    \centering
    \begin{tabularx}{\linewidth} { 
   >{\arraybackslash}l 
   >{\centering\arraybackslash}X  
   >{\centering\arraybackslash}X 
   >{\centering\arraybackslash}X 
   >{\centering\arraybackslash}X
   >{\arraybackslash}r  }
        \hline
        \textbf{Dataset} & $\bm{m}$ & \mbox{\textbf{\# Base}} & \mbox{\textbf{\# Query}} & \mbox{\textbf{TopK}} & \textbf{LID}\\ 
        \hline
        Audio \cite{Audio} & 192 & 53,387 & 200 & 20 & 5.6\\
        Enron \cite{Enron} & 1,369 & 94,987 & 200 & 20 & 11.7\\ 
        SIFT1M \cite{Jegou2011} & 128 & 1,000,000 & 10,000 & 100 & 9.3\\
        GloVe \cite{Pennington2014} & 100 & 1,183,514 & 10,000 & 100 & 20.0\\
        \hline
    \end{tabularx} 
    \vspace{0.5em}
    \caption{Details of the used datasets: dimensionality of the feature vectors (m), number of data points (\# Base) and queries (\# Query), number of provided true nearest neighbors per query (TopK), local intrinsic dimension of the dataset (LID).}
    \label{tab:datasets}
    \vspace{-2.0em}
\end{table}

\subsection{Evaluation metrics} 
\label{sec:evaluationMetrics}
As discussed in Section \ref{sec:requirementsOfGraphAlgorithms} the most important metrics of ANNS systems are: the \textbf{index build time}, the \textbf{memory consumption} and the \textbf{relation between search speed and recall rate}.
The measured total time to build the index includes constructing kd-trees, base graphs for edge pruning, and any neighborhood optimization steps. 
For the memory consumption a distinction is made between the maximal required memory during the construction phase and the test phase. 
During the test phase the $k$ nearest neighbors (e.g. 100-NN) of all database queries will be retrieved. 
The average $\recallAtVar{k}$ according to Equation \ref{eq:recallAtK} and the queries per second (QPS) are plotted in a diagram. 

\subsection{Experimental Setup} \label{sec:setup}
In order to ensure a comprehensive comparison, we evaluated the DEG with seven state-of-the-art graph-based algorithms (\textbf{kGraph\footnote{https://github.com/aaalgo/kgraph}}, \textbf{EFANNA\footnote{https://github.com/ZJULearning/efanna}}, \textbf{DPG\footnote{https://github.com/DBAIWangGroup/nns\_benchmark}}, \textbf{ONNG\footnote{https://github.com/yahoojapan/NGT}}, \textbf{HNSW\footnote{https://github.com/nmslib/hnswlib}}, \textbf{NSG\footnote{https://github.com/ZJULearning/nsg}}, \textbf{NSSG\footnote{https://github.com/ZJULearning/SSG}}),  using the serial scan from FAISS \label{footnote:faiss} \cite{Johnson2019} as baseline. 
\\
\\
All experiments were conducted on a machine with a Ryzen 2700x CPU, operating at a constant core clock speed of 4GHz, and 64GB of DDR4 memory running at 2133MHz. The source code of the different methods were compiled and tested on the same machine with AVX2 instructions enabled. 
To avoid potential run-time discrepancies arising from variations in multi-threading implementations, a single CPU thread was used for all experiments. 
This approach was also necessary because not all algorithms support multi-threaded indexing and querying. 
Additionally, the graph index was created entirely in memory to prevent potential IO bottlenecks.

\subsection{DEG Parameters} 
\label{sec:parameters}
In this section, only the indexing parameters of the DEG are explained, all settings of the other algorithms can be found in Appendix \ref{appendix:parameters}. The search and indexing performance of the DEG is influenced by several parameters, including the graph's degree $d$, the values for $k$ and $\varepsilon$ used in the range-search (Algorithm \ref{alg:rangeSearch}), and the maximum number of changes $i_{\opt}$ allowed during an edge optimization attempt. Additionally, the values of $k$ and $\varepsilon$ can differ between Algorithm \ref{alg:extendGraph} and Algorithm \ref{alg:optimizeEdge}, referred to as $k_{\extend}$ and $\varepsilon_{\extend}$ and $k_{\opt}$ and $\varepsilon_{\opt}$, respectively.

In our experiments, we focused on achieving high-accuracy search results above 95\% and modify the parameters accordingly. Our aim is to balance between construction time, memory consumption, and search quality with the settings from Table \ref{tab:deg_parameters}. It can be observed for datasets with a small LID, lesser neighbors per vertex are sufficient and while higher graph degrees ($d > 30$) help in the case of GloVe (see Section \ref{sec:searchLimit}) it also increases memory consumption and the construction time. 

\begin{table}[h!]
    \centering
    \begin{tabularx}{\linewidth} { 
   >{\arraybackslash}l 
   >{\centering\arraybackslash}X  
   >{\centering\arraybackslash}X  
   >{\centering\arraybackslash}X  
   >{\centering\arraybackslash}X  
   >{\centering\arraybackslash}X  
   >{\centering\arraybackslash}X  }
        \hline
        \textbf{Dataset} & 
        $\bm{d}$  & 
        $\bm{k_{\extend}}$ & $\bm{\varepsilon_{\extend}}$ & $\bm{k_{\opt}}$ &  $\bm{\varepsilon_{\opt}}$ &
        $\bm{i_{\opt}}$ \\ 
        \hline
        Audio \cite{Audio} & 20 & 40 & 0.3 & 20 & 0.001 & 5\\
        Enron \cite{Enron} & 30 & 60 & 0.3 & 30 & 0.001 & 5\\ 
        SIFT1M \cite{Jegou2011} & 30 & 60 & 0.2 & 30 & 0.001 & 5\\
        GloVe \cite{Pennington2014} & 30 & 30 & 0.2 & 30 & 0.001 & 5\\
        \hline
    \end{tabularx}
    \vspace{0.5em}
    \caption{Parameters of the DEG for various datasets. $\bm{d}$ is the degree of the graph and $\bm{k}$ the size of the search result list during the vertex extension ($\extend$) or edge optimization ($\opt$) of the graph. $\bm{\varepsilon}$ is the search radius factor of the range-search.}
    \label{tab:deg_parameters}
    \vspace{-2.0em}
\end{table}

\subsection{Comparing the search performance} \label{sec:searchExperiments}

\begin{figure*}[ht!]
    \begin{subfigure}{0.24\textwidth}
        \includegraphics[trim=2.0cm 4.0cm 2.0cm 6.5cm,clip=true,width=\textwidth]{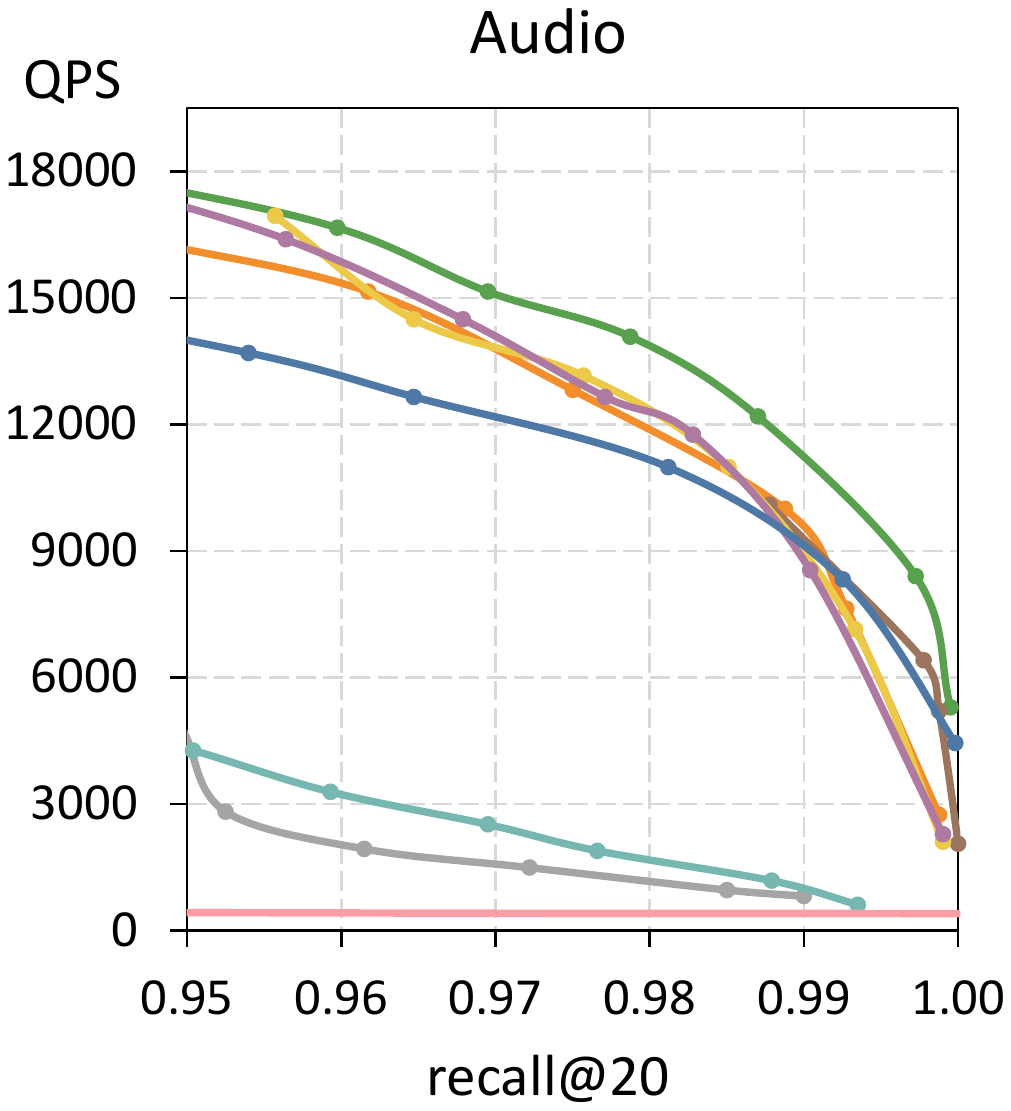}
    \end{subfigure}
    \hfill
    \begin{subfigure}{0.24\textwidth}
        \includegraphics[trim=2.0cm 4.0cm 2.0cm 6.5cm,clip=true,width=\textwidth]{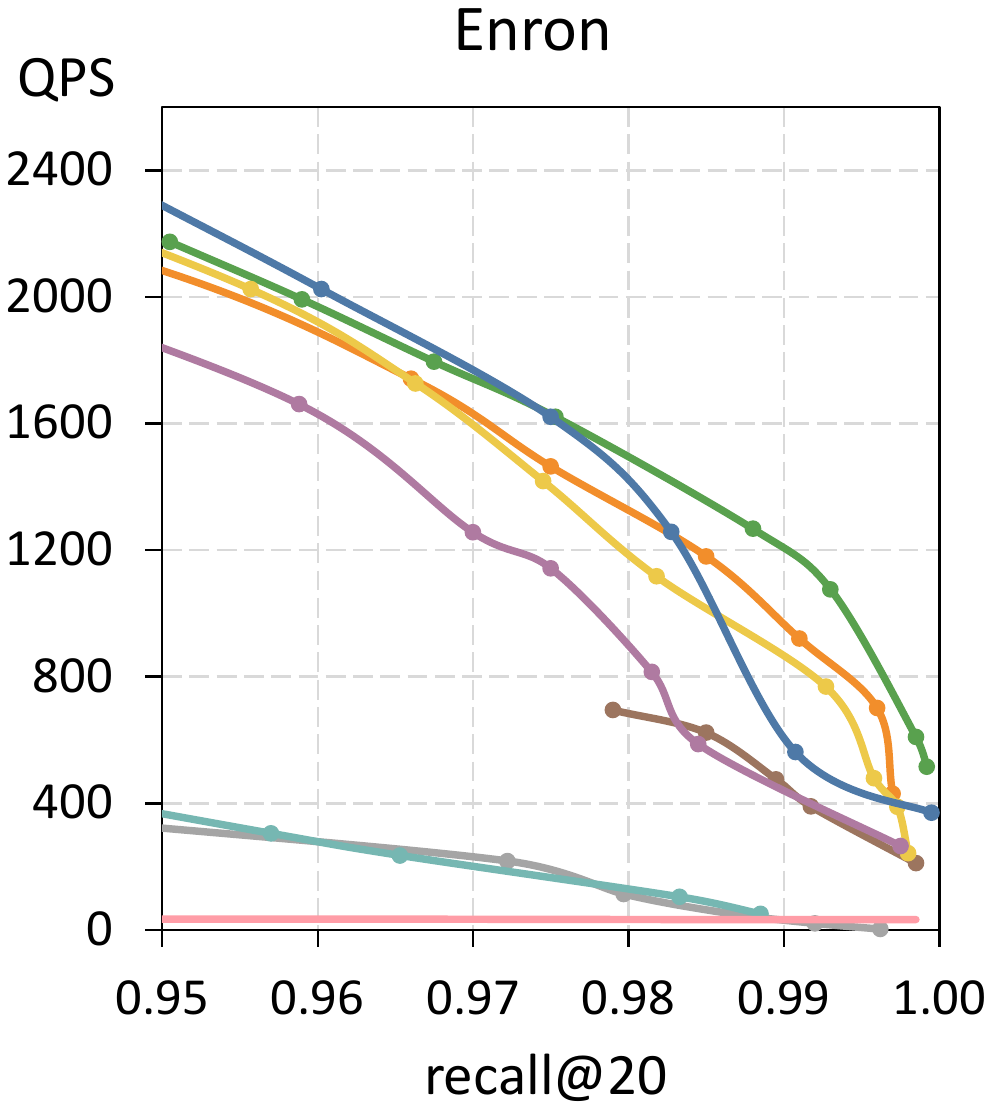}
    \end{subfigure}
    \hfill
    \begin{subfigure}{0.24\textwidth}
        \includegraphics[trim=2.0cm 4.0cm 2.0cm 6.5cm,clip=true,width=\textwidth]{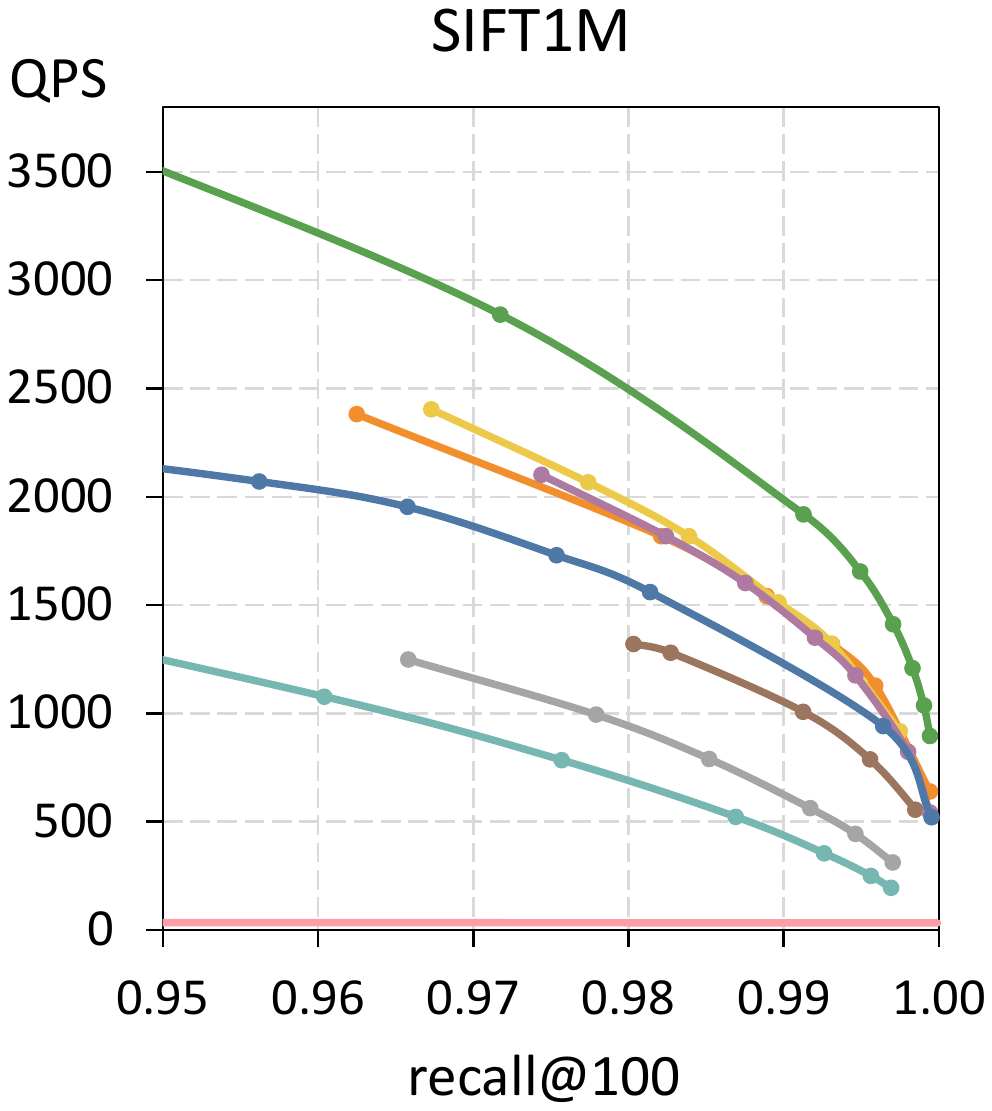}
    \end{subfigure}
    \hfill
    \begin{subfigure}{0.24\textwidth}
        \includegraphics[trim=2.0cm 4.0cm 2.0cm 6.5cm,clip=true,width=\textwidth]{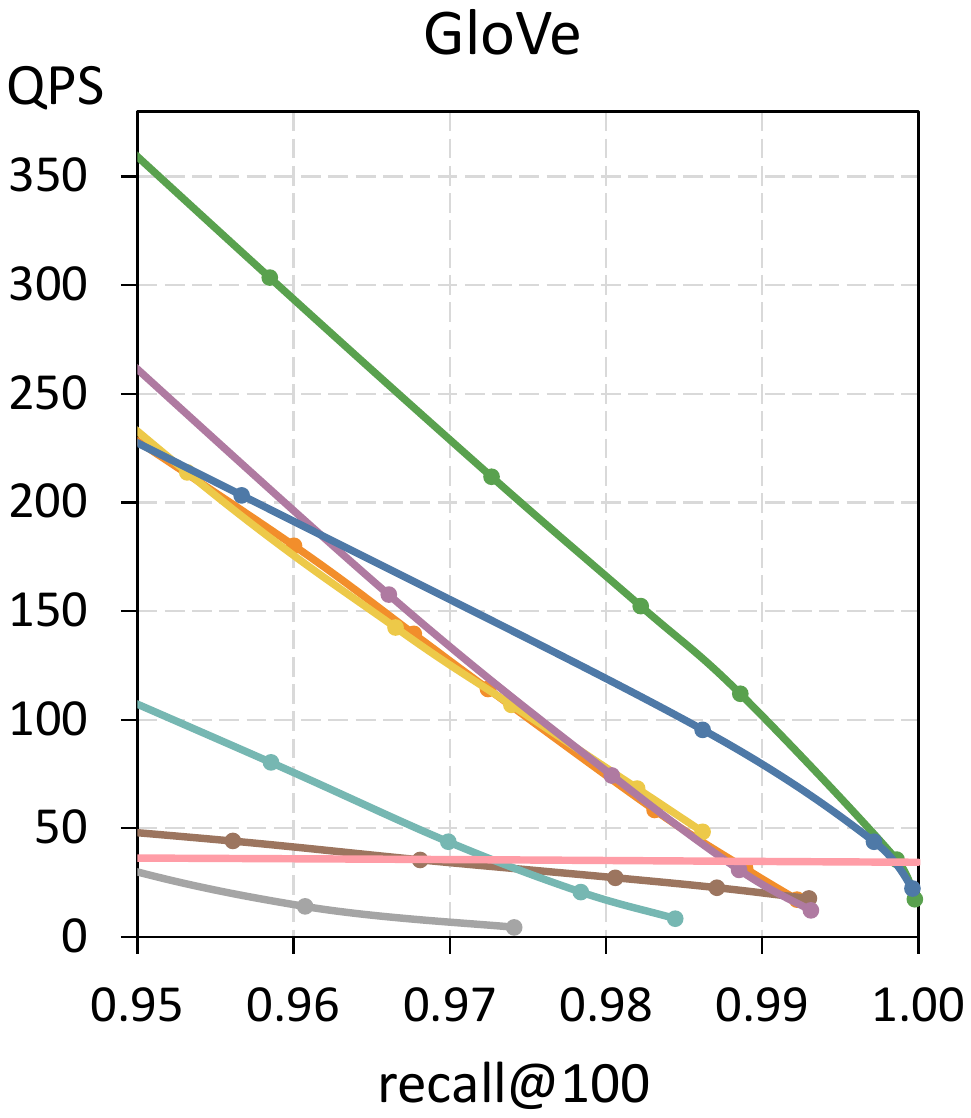}
    \end{subfigure}
    \begin{subfigure}{0.8\textwidth}
        \includegraphics[trim=1.5cm 13.85cm 2cm 15.5cm,clip=true,width=\textwidth]{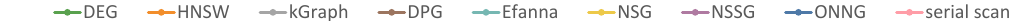}
    \end{subfigure}
    \vspace{-0.2cm}
    \caption{The number of queries per second (QPS) in relation to the recall@K for approximate nearest neighbor search was tested for different graphs and datasets (higher is better). The proposed Dynamic Exploration Graph (DEG) is more efficient than the current state of the art. }
    \label{fig:search_performance}
\end{figure*}

\begin{table*}[ht!]
    \centering
    \begin{tabularx}{\textwidth} { |
   >{\centering\arraybackslash}l  |
   >{\centering\arraybackslash}X  
   >{\centering\arraybackslash}X   
   >{\centering\arraybackslash}X  
   >{\centering\arraybackslash}X  |
   >{\centering\arraybackslash}X  
   >{\centering\arraybackslash}X   
   >{\centering\arraybackslash}X  
   >{\centering\arraybackslash}X  |
   >{\centering\arraybackslash}X  
   >{\centering\arraybackslash}X   
   >{\centering\arraybackslash}X  
   >{\centering\arraybackslash}X  |
   >{\centering\arraybackslash}X  
   >{\centering\arraybackslash}X   
   >{\centering\arraybackslash}X  
   >{\centering\arraybackslash}X  |}
        \hline
        \multirow{2}{*}{\textbf{Algo.}} &
        \multicolumn{4}{c|}{\textbf{Audio (41 MB)}} &
        \multicolumn{4}{c|}{\textbf{Enron (520 MB)}} &
        \multicolumn{4}{c|}{\textbf{SIFT1M (516 MB)}} &
        \multicolumn{4}{c|}{\textbf{GloVe (478 MB)}} \\ 
        \cline{2-17}
        &
        $\bm{IT}$ \textbf{(min)} & 
        $\bm{PM_c}$ \textbf{(MB)} & 
        $\bm{PM_s}$ \textbf{(MB)} & 
        $\bm{FS}$ \textbf{(MB)} &
        $\bm{IT}$ \textbf{(min)} & 
        $\bm{PM_c}$ \textbf{(MB)} & 
        $\bm{PM_s}$ \textbf{(MB)} & 
        $\bm{FS}$ \textbf{(MB)} &
        $\bm{IT}$ \textbf{(min)} & 
        $\bm{PM_c}$ \textbf{(MB)} & 
        $\bm{PM_s}$ \textbf{(MB)} & 
        $\bm{FS}$ \textbf{(MB)} &
        $\bm{IT}$ \textbf{(min)} & 
        $\bm{PM_c}$ \textbf{(MB)} & 
        $\bm{PM_s}$ \textbf{(MB)} & 
        $\bm{FS}$ \textbf{(MB)} \\ 
        \hline
        DEG & 0.4 & \textbf{52} & \textbf{48} & 50 & 14.2 & \textbf{561} & 538 & 543 & 28.2 & \textbf{780} & \textbf{665} & 756 & 84.4 & \textbf{821} & 678 & 762 \\
        HNSW & 0.7 & 67 & 67 & 53 & 6.4 & 576 & 576 & 552 & 35.8 & 892 & 892 & 660 & 54.2 & 985 & 985 & 781 \\   
        kGraph & \textbf{0.2} & 120 & 86 & 79 & \textbf{2.2} & 721 & 622 & 611 & 15 & 3984 & 1656 & 1568 & 41.8 & 6111 & 2674 & 2139 \\
        DPG & 0.7 & 203 & 69 & 48 & 5.8 & 794 & 651 & 558 & 17.2 & 4085 & 1116 & 707 & \textbf{26.5} & 4753 & 1886 & 926 \\
        EFANNA & 0.6 & 240 & 55 & 50 & 12.4 & 880 & 545 & 535 & \textbf{10.6} & 2356 & 748 & 720 & 83.7 & 8410 & 2447 & 2376 \\
        NSG & 1.3 & 227 & 50 & \textbf{45} & 18.1 & 949 & \textbf{536} & \textbf{526} & 30.6 & 3797 & 671 & \textbf{635} & 112.1 & 8410 & \textbf{603} & \textbf{544} \\
        NSSG & 1.9 & 380 & 50 & \textbf{45} & 10.9 & 820 & 537 & 527 & 22.4 & 4017 & 710 & 686 & 90.5 & 8410 & 642 & 583 \\
        ONNG & 4.2 & 297 & 116 & 60 & 24.5 & 1540 & 693 & 574 & 292.4 & 7297 & 1112 & 947 & 3644.5 & 6588 & 1612 & 1282 \\
        \hline
    \end{tabularx}
    \vspace{0.1cm}
    \caption{Single threaded indexing time ($\bm{IT}$) and peak memory consumption including the feature vector data during construction ($\bm{PM_c}$) and search ($\bm{PM_s}$) of graph-based approaches together with their resulting file size ($\bm{FS}$)}
    \label{tab:indexing_speed_and_memory}
    \vspace{-0.5cm}
\end{table*}

All evaluated search systems index the feature vectors of the base data and subsequently retrieve the $k$ most similar feature vectors from the index for each query of the query data. The search results are compared to the ground truth data to calculate the average recall@K rate. Depending on the parameters of the search algorithm, the recall and search speed changes. The FAISS (serial scan) curve was generated using a reduced base dataset in order to get correct timings for lower recall rates. 

Our analysis reveals a close alignment between the results depicted in Figure \ref{fig:search_performance} and the findings reported in literature \cite{Cong2021, Wang2021Survey}. 
The new Dynamic Exploration Graph is significantly faster than the current state of the art. With a recall of 99\%, the approximate improvements are as follows: Audio 15\%, Enron 25\%, SIFT1M 30\% and for GloVe 35\%.
\\
\\
\textbf{Observations:}

1) In terms of search speed, k-NN graphs like kGraph and EFANNA are among the slowest, while approximations of a Relative Neighborhood Graph achieve the best results.

2) As the local intrinsic dimension of datasets increases, the "curse of dimensionality" poses challenges in achieving favorable recall rates while maintaining acceptable search speeds. With the rise in local intrinsic dimension, the performance gap between the DEG algorithm and other methods also widens.

3) The Audio and Enron dataset only provide the top 20 most similar data points per query, leading to noticeable fluctuations in the recall-curves for small changes in the search settings.

4) 
Regardless of time or memory constraints, further adjustments to the graphs construction parameters does not result in higher search efficiency. 
The DEG is the only exception, where a higher vertex degree can boost the search speed by up to 30\% (see Section \ref{sec:searchLimit} for more details).

\subsection{Run-time and memory consumption} \label{sec:constructionTimeAndMemoryConsumption}

Besides achieving high search quality and speed, it is desirable that the graph and any additional data structures can be created as quickly as possible with minimal memory overhead.

Table \ref{tab:indexing_speed_and_memory} documents the single-threaded indexing time (IT) and peak memory (PM) requirements of all tested graph, including the peak memory during the construction phase ($PM_C$) and the subsequent search phase ($PM_S$). These numbers include the feature vectors and give a general idea of the memory requirements of the different approaches.

After creating and storing the graph to disk, the file size (FS) is measured, which again includes the feature vectors from the base dataset. While aiming to use minimal disk space, it is important to consider the size of the dataset in relation to the file size of the graph. For instance, if the feature vectors have hundreds or thousands of dimensions, a graph with 2-3 dozen edges per vertex account for only a small fraction of the total memory requirement.
\\
\\
\textbf{Observations:}

\begin{figure*}[ht!]
    \begin{subfigure}{0.24\textwidth}
        \includegraphics[trim=2.0cm 4.0cm 2.0cm 6.5cm,clip=true,width=\textwidth]{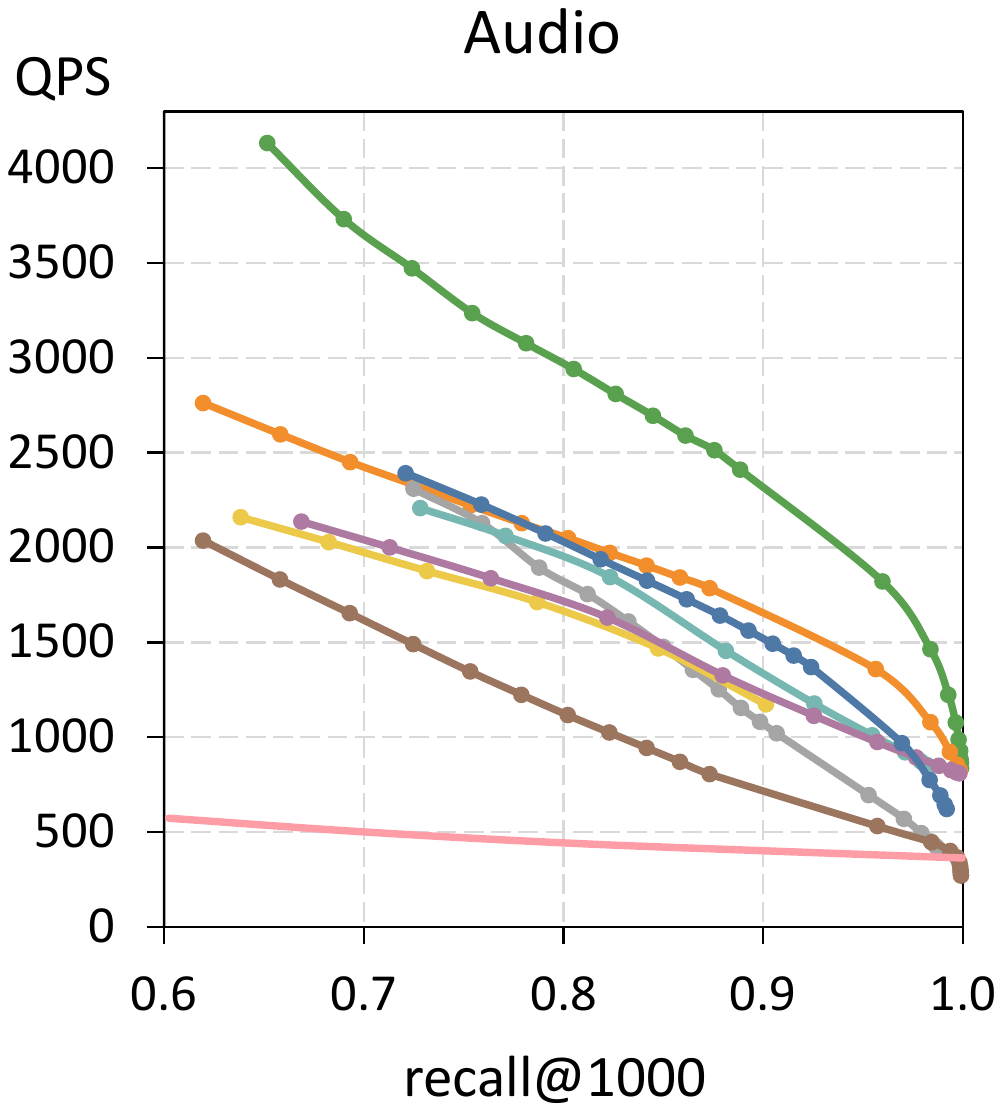}
    \end{subfigure}
    \hfill
    \begin{subfigure}{0.24\textwidth}
        \includegraphics[trim=2.0cm 4.0cm 2.0cm 6.5cm,clip=true,width=\textwidth]{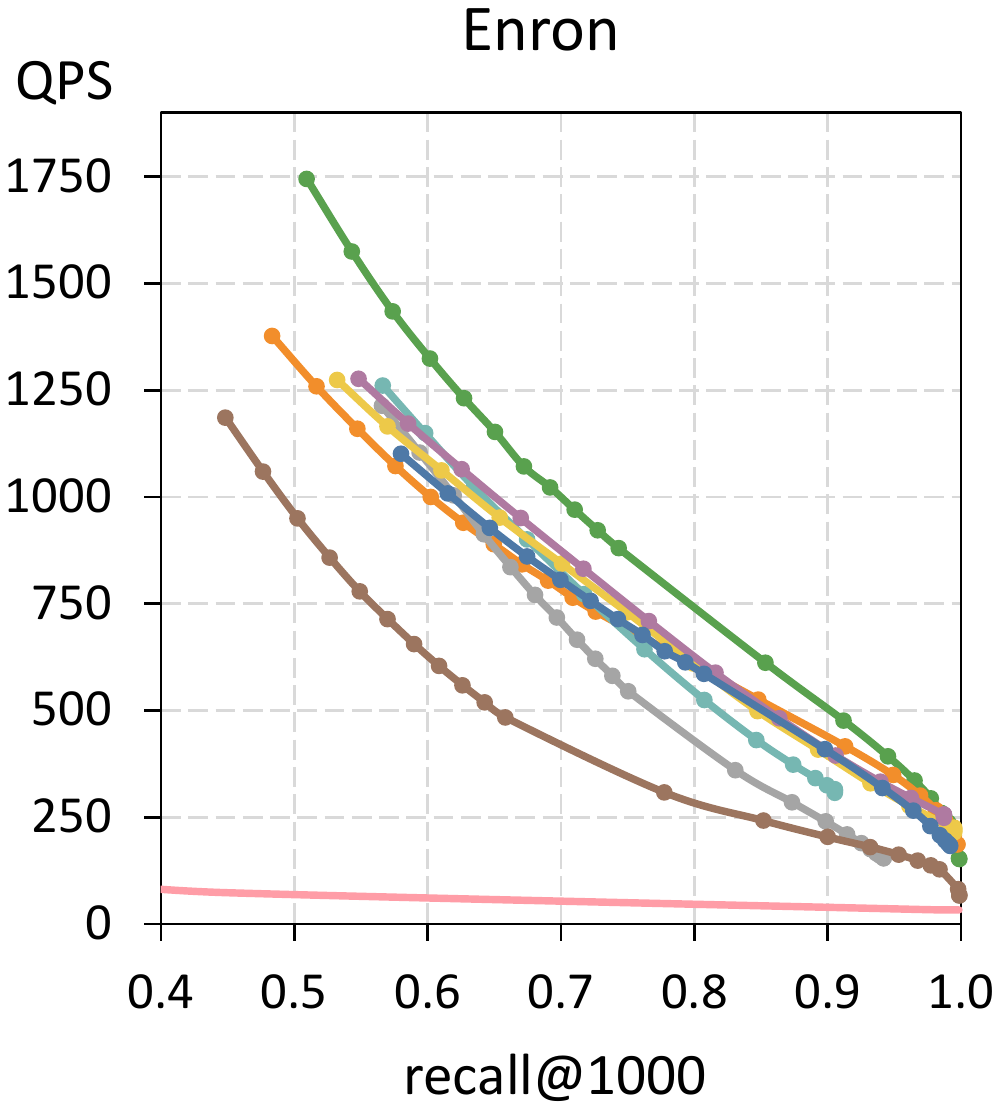}
    \end{subfigure}
    \hfill
    \begin{subfigure}{0.24\textwidth}
        \includegraphics[trim=2.0cm 4.0cm 2.0cm 6.5cm,clip=true,width=\textwidth]{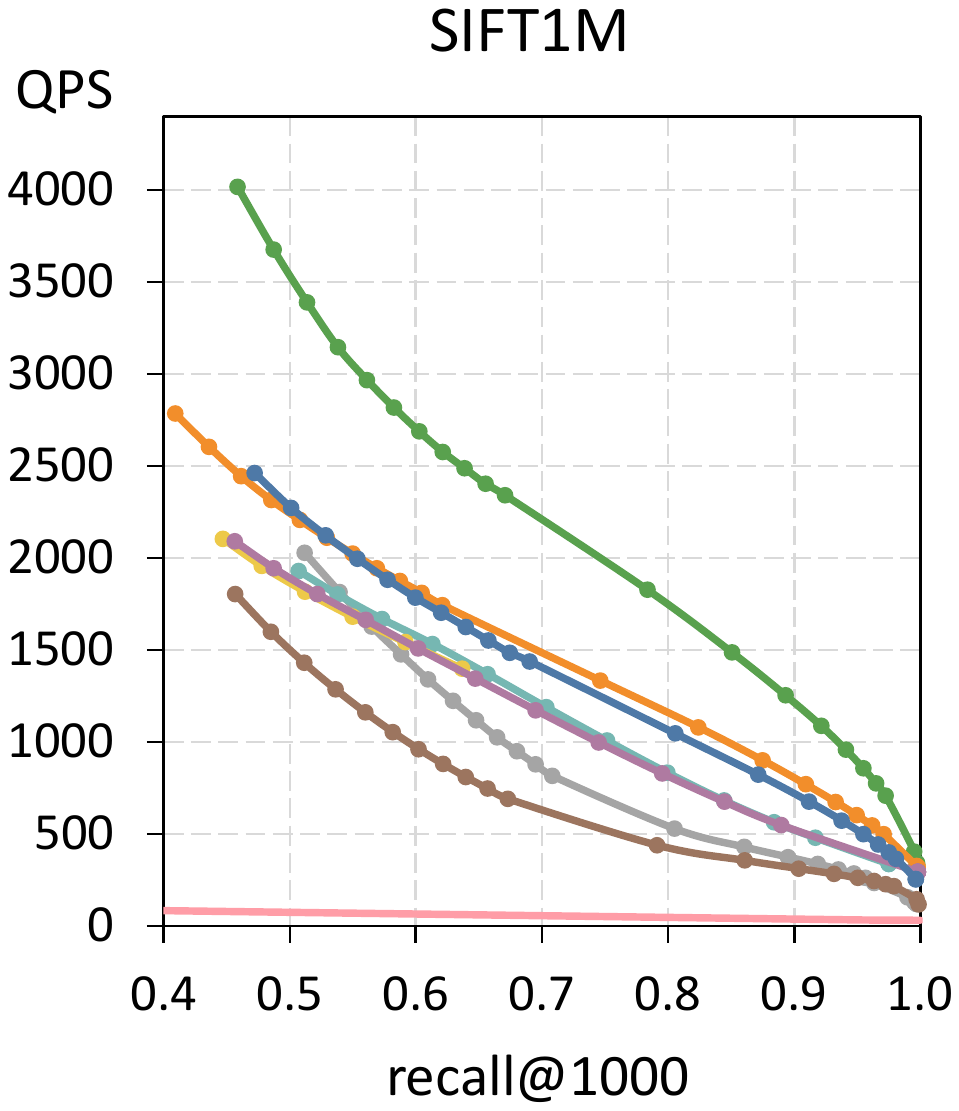}
    \end{subfigure}
    \hfill
    \begin{subfigure}{0.24\textwidth}
        \includegraphics[trim=2.0cm 4.0cm 2.0cm 6.5cm,clip=true,width=\textwidth]{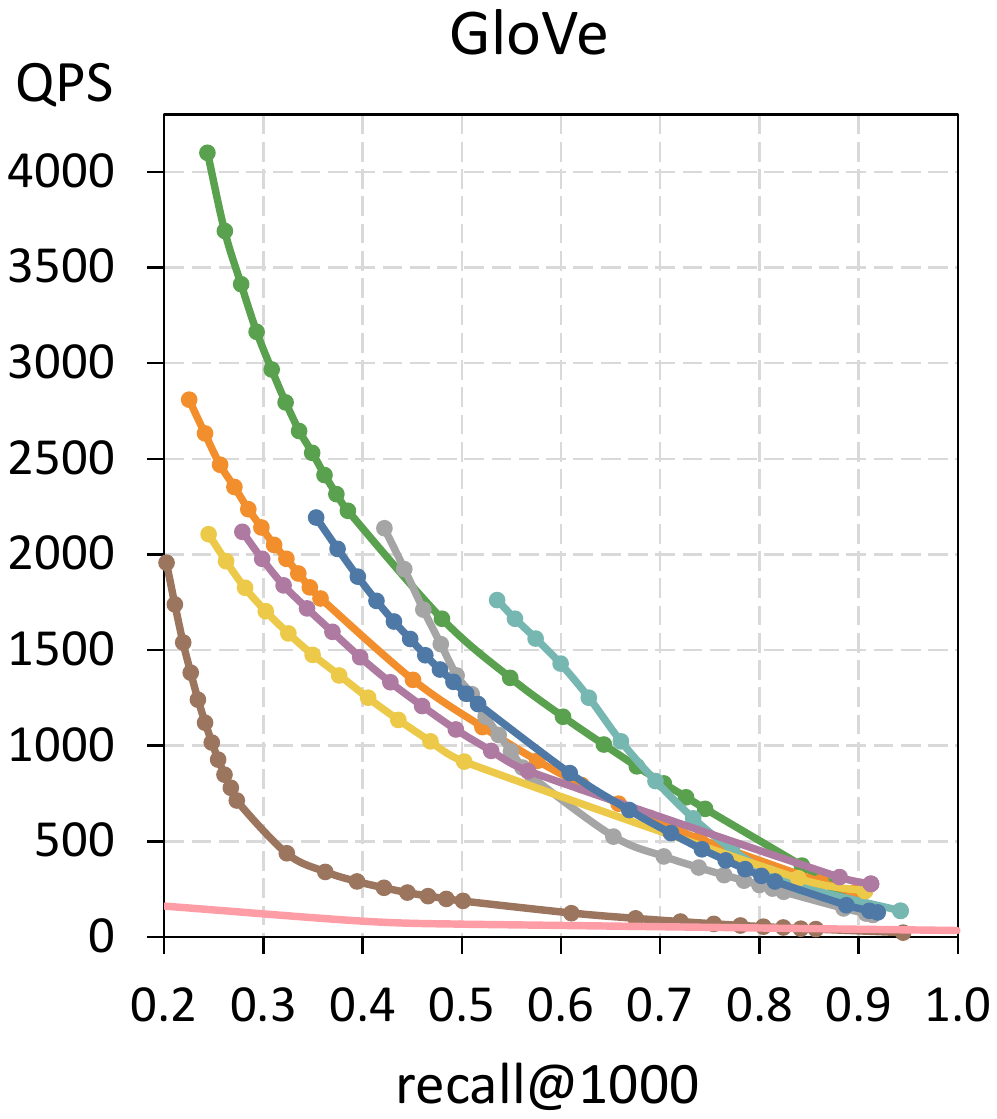}
    \end{subfigure}  
    \begin{subfigure}{0.8\textwidth}
        \includegraphics[trim=1.5cm 13.85cm 2cm 15.5cm,clip=true,width=\textwidth]{figures/recall_vs_qps_legend.pdf}
    \end{subfigure}
    \vspace{-0.2cm}
    \caption{Exploration performance measured by recall@1000 in relation to the number of queries per second (QPS). Various graphs are compared on four datasets. Top right is better.}
    \label{fig:explore_performance}
    
\end{figure*}

1) During the construction phase, memory usage tends to be higher for techniques like DPG, NSG, NSSG, and ONNG. These methods involve creating an initial graph and then pruning its edges, which requires storing both the original and modified graphs in memory simultaneously.

2) Typically, the memory requirement during the search phase slightly exceeds the size of the index file, suggesting only minimal additional data structures are generated. However, DPG and ONNG are exceptions as they demand significantly more memory. Conversely, the memory requirements of the DEG are lower since it stores edge weights in the file but omits them during search.

3) k-NN graphs such as kGraph and EFANNA typically have a high number of edges and therefore have a large index file. On the other hand, NSG and NSSG prune most of the edges and have the smallest files, followed by HNSW and DEG.

4) Incremental graphs (e.g. HNSW and DEG) and k-NN graphs (e.g. like kGraph and EFANNA) generally exhibit lower indexing times (IT) than graphs which prune edges. Although the pruning process is often quite fast, it necessitates an initial graph with numerous edges, which may take a significant amount of time to construct.

5) It should be noted, that while graphs like kGraph, EFANNA and DPG are among the fastest to build their indices, they have also the slowest search performance. ONNG, in turn, demonstrates favorable search times for certain datasets and recall ranges, yet requires a lot of time to build its index.

\subsection{Exploration Quality} \label{sec:explorationExperiements}

The experiments conducted in Section \ref{sec:searchExperiments} adhere to the established test protocols for approximate k-NN search, as documented in the existing literature \cite{Aumuller2020}. The search aims to find data points within the graph that are similar to a non-indexed query. Typically, the process begins by selecting one or multiple seed vertices, and then approaches the closest vertices to the query using Algorithm \ref{alg:rangeSearch}. 

For the following experiment all queries are part of the index and the seed vertex is set to the corresponding query. This kind of search is required in recommendation systems, where numerous similar data points must be identified for a given data point, or in user-guided browsing applications, where search queries are continually refined in a feedback loop. In both cases, it is crucial to avoid presenting users with data they have already seen. As a result, extensive search result lists are frequently required, and recall rates becomes less critical as the last entries become increasingly dissimilar. We refer to such searches as \textit{exploratory search}.

The graphs and search algorithms tested in this section are identical to those in Section \ref{sec:searchExperiments}. Only the queries and the seed vertices have been changed. For each dataset 10000 vertices from the base dataset where randomly chosen to form the respective test set of the exploration tasks. The 1000 most similar neighbors are searched per query and the entire recall range is considered.

As depicted in Figure \ref{fig:explore_performance}, it is evident that the DEG outperforms the other graphs consistently across various recall rates. Depending on the desired recall, the DEG can achieve up to a 50\% higher number of queries per second compared to the second-best graph.
\\
\\
\textbf{Observations:}

1) The nature of k-NN graphs like EFANNA and k-graph to strive for a high \textit{graph quality}, allowing them to surpass some sparser graphs in the low-recall range. However, the advantage of k-NN graphs diminishes in the high-recall range due to the lack of effective navigation edges and the existence of numerous hub vertices (see Appendix \ref{appendix:graphStatistics} for more statistics).

2) The ranking of the best performing graphs varies between exploration tasks and normal search tasks. This suggests that the effectiveness of the graph in approximated nearest neighbor search does not necessarily translate to its exploration qualities.

3) EFANNA and kGraph exhibit several source vertices with zero incoming edges. Those vertices limit the general reachability between vertices and can decrease the exploration quality.

\section{Ablation Study} \label{sec:ablation}


\subsection{Empirical Scalability} \label{sec:scalabilityAndComplexity}

\begin{figure}[h!]
    \begin{subfigure}{0.48\columnwidth}
        \includegraphics[trim=6.5cm 2cm 6.5cm 2.2cm,clip=true,width=\columnwidth]{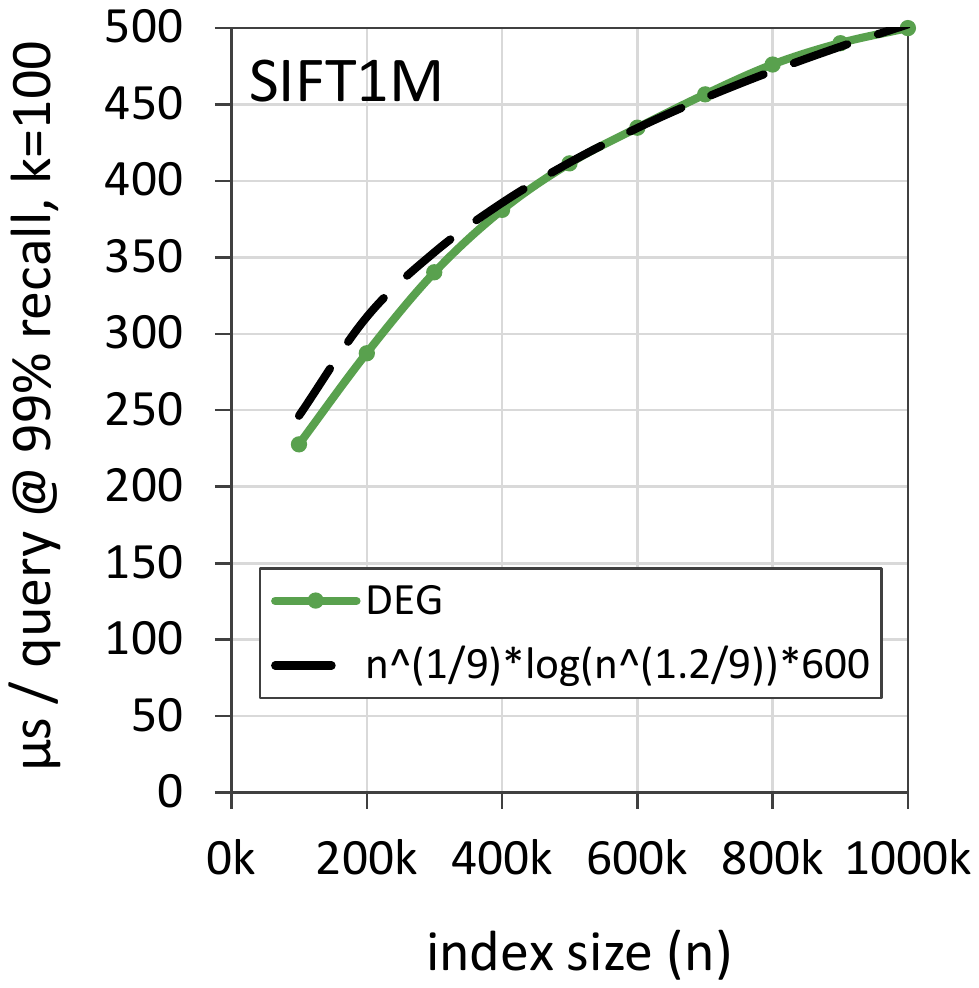}
    \end{subfigure}
    \hfill
    \begin{subfigure}{0.48\columnwidth}
        \includegraphics[trim=6.5cm 2cm 6.5cm 2.2cm,clip=true,width=\columnwidth]{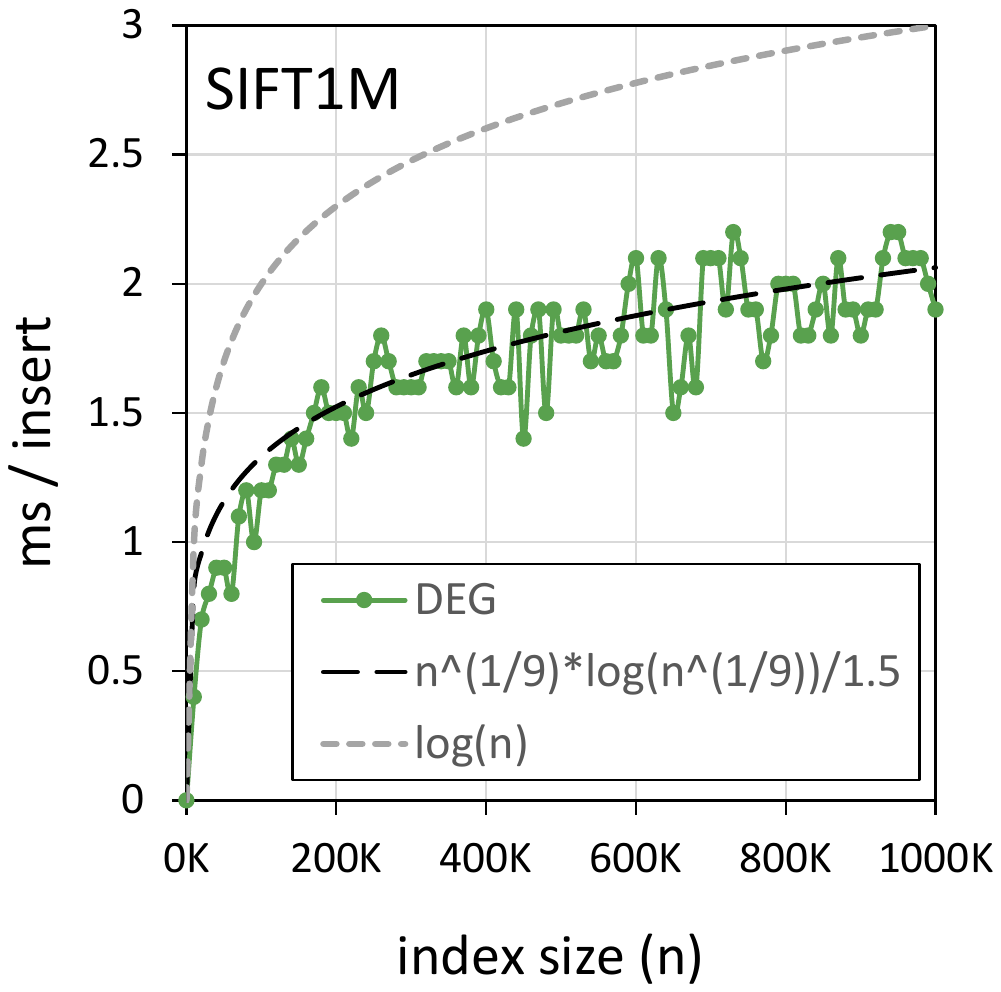}
    \end{subfigure}
    \vspace{-1.0em}
    \caption{Influence of index size on indexing speed (right) and search speed (left) to attain a 99\% recall rate at k=100.}    \label{fig:build_and_search_time_scaling}
\end{figure}

The DEG construction Algorithm \ref{alg:extendGraph} has six parameters: $d$ the degree of the graph; $k_{\extend}, \varepsilon_{\extend}$ for adding new vertices and $k_{\opt}, \varepsilon_{\opt}, i_{\opt}$ to swap existing edges. In our experiments, we found the optimal parameters will not change with the amount of data, as long as the data distribution stays the same. Therefore we randomly sub-sample sets of diﬀerent size and construct a DEG for each of them using the parameters from Table \ref{tab:deg_parameters}. 
The indexing time and search time on the test set to reach a recall of 99\% and $k = 100$ for all these subsets are recorded. 
The curves of Figure \ref{fig:build_and_search_time_scaling} show how the time increases with the number of vertices $n$ in the index for the dataset SIFT1M. 
In order to extrapolate the data and make prediction about the required time in larger datasets, various functions to fit the curves have been developed. 
Notably, the search time complexity concerning the number of vertices can be expressed as $O(n^{1/9} \cdot log(n^{1/9}))$. This agrees with our theoretical analysis in Section \ref{sec:analyticalSearchTimeComplexity} and is similar to HNSW, NSG and NSSG in \cite{Wang2021Survey}. 
The time complexity for adding a new vertex to an index is very similar to the search time complexity, as depicted in Figure \ref{fig:build_and_search_time_scaling} on the right. This is because the graph extension and optimization algorithms execute only a few search requests during the process.
Even if the SIFT data had 1 billion data points, the time to add another vertex would be just around 6.6 ms on the tested hardware and still be faster than $log(n)$. 
In case of a data distribution similar to GloVe the time required to add the billionth vertex would be approximately 53 milliseconds, and the scalability can be described by the equation $O(n^{5.3/20} \cdot log(n^{5.3/20}))$.

%

\subsection{Quality of edges} 
\label{sec:qualityOfEdges}

The following section investigates the effectiveness of edge optimization in producing fast search graphs. For this purpose, an even-regular undirected graph with random edges was generated for the SIFT1M dataset and then optimized using Algorithm \ref{alg:dynamicEdgeOptimization}. The used parameters can be found in Table \ref{tab:deg_parameters}.

\begin{figure}[t!]
    \begin{subfigure}{0.49\columnwidth}
        \includegraphics[trim=2cm 5.5cm 2cm 7cm,clip=true,width=\columnwidth]{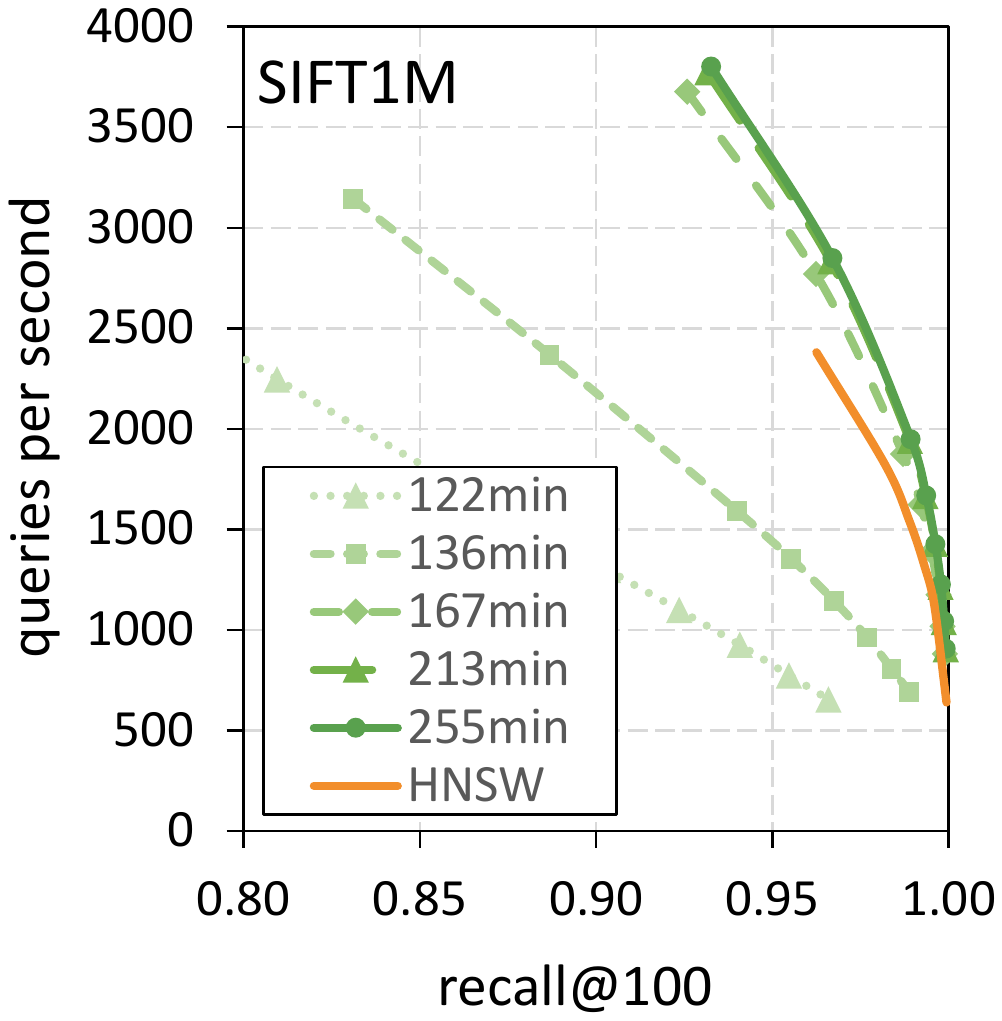}        
    \end{subfigure}
    \hfill
    \begin{subfigure}{0.49\columnwidth}
        \includegraphics[trim=2cm 5.5cm 2cm 7cm,clip=true,width=\columnwidth]{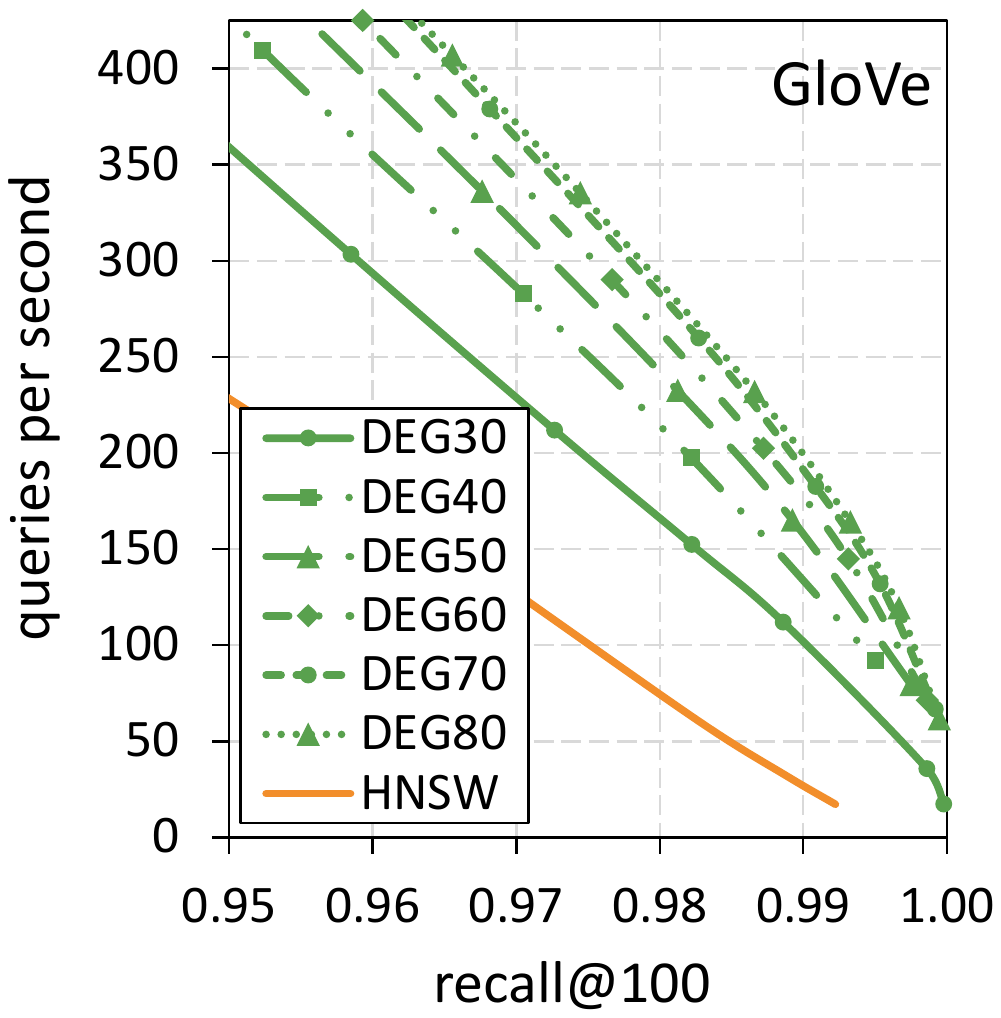}    
    \end{subfigure}  
    \vspace{-0.5em}
    \caption{Left: Edge optimization transforms a random even-regular undirected graph into an efficient DEG. Right: Increasing the number of edges on specific datasets can be beneficial. HNSW (in orange) was added for reference.}  
    \label{fig:swap_scaling_sift}
    \label{fig:edge_count_scaling_glove}
\end{figure}

As the algorithm improves the edges of random vertices, the graph's search quality increases with each iteration. 
The process was run in a single thread and the search quality was evaluated after different numbers of iterations.
The results are shown in Figure \ref{fig:swap_scaling_sift}, where each curve represents a Dynamic Exploration Graph optimized for a given duration. 
It took over two hours to get useful connections and competitive results. Then, after another half hour, the graph returned better search results than the state of the art. Although further edge optimization is possible, the return is diminishing.



\subsection{No edge count limit} 
\label{sec:searchLimit}

The indexing parameters in Table \ref{tab:deg_parameters} were selected to quickly construct an efficient search graph with low memory requirements. Additional edge optimization can generate superior graphs at the expense of increased computation time, as outlined in Section \ref{sec:qualityOfEdges}. However, for complex datasets like GloVe, it is possible to obtain even better results if there are no limitations on time and memory.

This section explores the impact of the number of edges on search quality. As the graph degrees for the Audio, Enron, and SIFT1M datasets in the previous sections are already close to optimal, their experiments are not presented due to space limitations. On the other hand, constructing the DEG for the GloVe dataset, which has a high Local Intrinsic Dimensionality, still has room for improvement. For further investigation, a series of tests were conducted using the parameters from Table \ref{tab:deg_parameters}, but with a higher degree ($d$).

The data depicted in Figure \ref{fig:edge_count_scaling_glove} reveals that the DEG with 80 edges per vertex and at a recall rate of 99\% can search nearly four times faster than HNSW. If the number of edges increases even further, the search speed gradually declines. Although it is feasible to construct such massive graphs, it takes a lot of time and memory. A good balance for most datasets is found when between 20 and 30 edges per vertex.

\section{Conclusion}

This paper introduces the Dynamic Exploration Graph (DEG), an even-regular undirected graph constructed by two algorithms: graph extension and edge optimization. The former allows for incremental expansion of the graph with new vertices, whereas the latter regularly updates the neighborhood of existing vertices to reduce the \textit{average neighbor distance} and create a balance between short and long edges. This balance enables quick navigation in other graph regions and connects similar vertices. The DEG approximates MRNG, providing logarithmic search and exploration complexity for most practical datasets with $n \leq 10^{9}$. Extensive experiments have demonstrated the memory and search efficiency and highlighted the significance of new exploration tests for recommender and interactive image navigation systems. Compared to existing methods, the DEG performs 15-50\% faster in the high-recall range and requires the least amount of memory during the indexing phase. The fundamental idea of the DEG is continuous self-optimization to further improve its performance. Future work will focus on how to handle dynamic datasets.

%
%
%
\bibliographystyle{ACM-Reference-Format}
\bibliography{references}


\begin{thebibliography}{60}


\ifx \showCODEN    \undefined \def \showCODEN     #1{\unskip}     \fi
\ifx \showDOI      \undefined \def \showDOI       #1{#1}\fi
\ifx \showISBNx    \undefined \def \showISBNx     #1{\unskip}     \fi
\ifx \showISBNxiii \undefined \def \showISBNxiii  #1{\unskip}     \fi
\ifx \showISSN     \undefined \def \showISSN      #1{\unskip}     \fi
\ifx \showLCCN     \undefined \def \showLCCN      #1{\unskip}     \fi
\ifx \shownote     \undefined \def \shownote      #1{#1}          \fi
\ifx \showarticletitle \undefined \def \showarticletitle #1{#1}   \fi
\ifx \showURL      \undefined \def \showURL       {\relax}        \fi
\providecommand\bibfield[2]{#2}
\providecommand\bibinfo[2]{#2}
\providecommand\natexlab[1]{#1}
\providecommand\showeprint[2][]{arXiv:#2}

\bibitem[\protect\citeauthoryear{André, Kermarrec, and Scouarnec}{André
  et~al\mbox{.}}{2015}]%
        {Andre2015}
\bibfield{author}{\bibinfo{person}{Fabien André}, \bibinfo{person}{Anne-Marie
  Kermarrec}, {and} \bibinfo{person}{Nicolas~Le Scouarnec}.}
  \bibinfo{year}{2015}\natexlab{}.
\newblock \showarticletitle{Cache locality is not enough: High-Performance
  Nearest Neighbor Search with Product Quantization Fast Scan.}
\newblock \bibinfo{journal}{\emph{Proc. VLDB Endow.}} \bibinfo{volume}{9},
  \bibinfo{number}{4} (\bibinfo{year}{2015}), \bibinfo{pages}{288--299}.
\newblock


\bibitem[\protect\citeauthoryear{Arora, Sinha, Kumar, and Bhattacharya}{Arora
  et~al\mbox{.}}{2018}]%
        {Arora2018}
\bibfield{author}{\bibinfo{person}{Akhil Arora}, \bibinfo{person}{Sakshi
  Sinha}, \bibinfo{person}{Piyush Kumar}, {and} \bibinfo{person}{Arnab
  Bhattacharya}.} \bibinfo{year}{2018}\natexlab{}.
\newblock \showarticletitle{HD-Index: Pushing the Scalability-Accuracy Boundary
  for Approximate kNN Search in High-Dimensional Spaces.}
\newblock \bibinfo{journal}{\emph{Proc. VLDB Endow.}} \bibinfo{volume}{11},
  \bibinfo{number}{8} (\bibinfo{year}{2018}), \bibinfo{pages}{906--919}.
\newblock


\bibitem[\protect\citeauthoryear{Aumüller, Bernhardsson, and
  Faithfull}{Aumüller et~al\mbox{.}}{2020}]%
        {Aumuller2020}
\bibfield{author}{\bibinfo{person}{Martin Aumüller}, \bibinfo{person}{Erik
  Bernhardsson}, {and} \bibinfo{person}{Alexander~John Faithfull}.}
  \bibinfo{year}{2020}\natexlab{}.
\newblock \showarticletitle{ANN-Benchmarks: A benchmarking tool for approximate
  nearest neighbor algorithms.}
\newblock \bibinfo{journal}{\emph{Inf. Syst.}}  \bibinfo{volume}{87}
  (\bibinfo{year}{2020}).
\newblock


\bibitem[\protect\citeauthoryear{Barthel, Hezel, Schall, and Jung}{Barthel
  et~al\mbox{.}}{2019}]%
        {Barthel2019}
\bibfield{author}{\bibinfo{person}{Kai~Uwe Barthel}, \bibinfo{person}{Nico
  Hezel}, \bibinfo{person}{Konstantin Schall}, {and} \bibinfo{person}{Klaus
  Jung}.} \bibinfo{year}{2019}\natexlab{}.
\newblock \showarticletitle{Real-Time Visual Navigation in Huge Image Sets
  Using Similarity Graphs.}. In \bibinfo{booktitle}{\emph{ACM Multimedia}},
  \bibfield{editor}{\bibinfo{person}{Laurent Amsaleg}, \bibinfo{person}{Benoit
  Huet}, \bibinfo{person}{Martha~A. Larson}, \bibinfo{person}{Guillaume
  Gravier}, \bibinfo{person}{Hayley Hung}, \bibinfo{person}{Chong-Wah Ngo},
  {and} \bibinfo{person}{Wei~Tsang Ooi}} (Eds.). \bibinfo{publisher}{ACM},
  \bibinfo{address}{Nice, France}, \bibinfo{pages}{2202--2204}.
\newblock
\showISBNx{978-1-4503-6889-6}


\bibitem[\protect\citeauthoryear{Barthel, Hezel, Schall, and Jung}{Barthel
  et~al\mbox{.}}{2023}]%
        {Navigu2023}
\bibfield{author}{\bibinfo{person}{Kai~Uwe Barthel}, \bibinfo{person}{Nico
  Hezel}, \bibinfo{person}{Konstantin Schall}, {and} \bibinfo{person}{Klaus
  Jung}.} \bibinfo{year}{2023}\natexlab{}.
\newblock \showarticletitle{Navigu.Net: NAvigation in Visual Image Graphs Gets
  User-Friendly}. In \bibinfo{booktitle}{\emph{Proceedings of the 2023 ACM
  International Conference on Multimedia Retrieval}} (Thessaloniki, Greece)
  \emph{(\bibinfo{series}{ICMR '23})}. \bibinfo{publisher}{Association for
  Computing Machinery}, \bibinfo{address}{New York, NY, USA},
  \bibinfo{pages}{654–658}.
\newblock
\showISBNx{9798400701788}
\urldef\tempurl%
\url{https://doi.org/10.1145/3591106.3592248}
\showDOI{\tempurl}


\bibitem[\protect\citeauthoryear{Bentley}{Bentley}{1975}]%
        {Bentley1975}
\bibfield{author}{\bibinfo{person}{Jon~Louis Bentley}.}
  \bibinfo{year}{1975}\natexlab{}.
\newblock \showarticletitle{Multidimensional binary search trees used for
  associative searching}.
\newblock \bibinfo{journal}{\emph{Commun. ACM}}  \bibinfo{volume}{18}
  (\bibinfo{date}{September} \bibinfo{year}{1975}), \bibinfo{pages}{509--517}.
\newblock
Issue 9.
\showISSN{0001-0782}
\urldef\tempurl%
\url{https://doi.org/10.1145/361002.361007}
\showDOI{\tempurl}


\bibitem[\protect\citeauthoryear{Boutet, Kermarrec, Mittal, and Taïani}{Boutet
  et~al\mbox{.}}{2016}]%
        {Boutet2016}
\bibfield{author}{\bibinfo{person}{Antoine Boutet}, \bibinfo{person}{Anne-Marie
  Kermarrec}, \bibinfo{person}{Nupur Mittal}, {and} \bibinfo{person}{François
  Taïani}.} \bibinfo{year}{2016}\natexlab{}.
\newblock \showarticletitle{Being prepared in a sparse world: The case of KNN
  graph construction.}. In \bibinfo{booktitle}{\emph{ICDE}}.
  \bibinfo{publisher}{IEEE Computer Society}, \bibinfo{address}{Helsinky,
  Finland}, \bibinfo{pages}{241--252}.
\newblock
\showISBNx{978-1-5090-2020-1}


\bibitem[\protect\citeauthoryear{Chen, Zhao, Wang, Li, Liu, Li, Yang, and
  Wang}{Chen et~al\mbox{.}}{2021}]%
        {Chen2021}
\bibfield{author}{\bibinfo{person}{Qi Chen}, \bibinfo{person}{Bing Zhao},
  \bibinfo{person}{Haidong Wang}, \bibinfo{person}{Mingqin Li},
  \bibinfo{person}{Chuanjie Liu}, \bibinfo{person}{Zengzhong Li},
  \bibinfo{person}{Mao Yang}, {and} \bibinfo{person}{Jingdong Wang}.}
  \bibinfo{year}{2021}\natexlab{}.
\newblock \showarticletitle{SPANN: Highly-efficient Billion-scale Approximate
  Nearest Neighbor Search}. In \bibinfo{booktitle}{\emph{35th Conference on
  Neural Information Processing Systems (NeurIPS 2021)}}.
  \bibinfo{publisher}{Neural Information Processing Systems Foundation, Inc.},
  \bibinfo{address}{Online}.
\newblock


\bibitem[\protect\citeauthoryear{Costa, Girotra, and Hero}{Costa
  et~al\mbox{.}}{2005}]%
        {Costa2005}
\bibfield{author}{\bibinfo{person}{J.A. Costa}, \bibinfo{person}{A. Girotra},
  {and} \bibinfo{person}{A.O. Hero}.} \bibinfo{year}{2005}\natexlab{}.
\newblock \showarticletitle{Estimating Local Intrinsic Dimension with k-Nearest
  Neighbor Graphs}. In \bibinfo{booktitle}{\emph{IEEE/SP 13th Workshop on
  Statistical Signal Processing, 2005}}. \bibinfo{pages}{417--422}.
\newblock
\urldef\tempurl%
\url{https://doi.org/10.1109/SSP.2005.1628631}
\showDOI{\tempurl}


\bibitem[\protect\citeauthoryear{Delaunay}{Delaunay}{1933}]%
        {Delaunay1933}
\bibfield{author}{\bibinfo{person}{B. Delaunay}.}
  \bibinfo{year}{1933}\natexlab{}.
\newblock \showarticletitle{Neue Darstellung der geometrischen
  Kristallographie}.
\newblock \bibinfo{journal}{\emph{Zeitschrift für Kristallographie -
  Crystalline Materials}} \bibinfo{volume}{84}, \bibinfo{number}{1-6}
  (\bibinfo{year}{1933}), \bibinfo{pages}{109--149}.
\newblock
\urldef\tempurl%
\url{https://doi.org/doi:10.1524/zkri.1933.84.1.109}
\showDOI{\tempurl}


\bibitem[\protect\citeauthoryear{Deng, Yan, Kelvin, Jiang, and Cheng}{Deng
  et~al\mbox{.}}{2019}]%
        {Deng2019}
\bibfield{author}{\bibinfo{person}{Shiyuan Deng}, \bibinfo{person}{Xiao Yan},
  \bibinfo{person}{K.W.~Ng Kelvin}, \bibinfo{person}{Chenyu Jiang}, {and}
  \bibinfo{person}{James Cheng}.} \bibinfo{year}{2019}\natexlab{}.
\newblock \showarticletitle{Pyramid: A General Framework for Distributed
  Similarity Search on Large-scale Datasets}. In \bibinfo{booktitle}{\emph{2019
  IEEE International Conference on Big Data (Big Data)}}.
  \bibinfo{pages}{1066--1071}.
\newblock


\bibitem[\protect\citeauthoryear{Dong, Moses, and Li}{Dong
  et~al\mbox{.}}{2011}]%
        {Dong2011}
\bibfield{author}{\bibinfo{person}{Wei Dong}, \bibinfo{person}{Charikar Moses},
  {and} \bibinfo{person}{Kai Li}.} \bibinfo{year}{2011}\natexlab{}.
\newblock \showarticletitle{Efficient K-Nearest Neighbor Graph Construction for
  Generic Similarity Measures}. In \bibinfo{booktitle}{\emph{Proceedings of the
  20th International Conference on World Wide Web}} (Hyderabad, India)
  \emph{(\bibinfo{series}{WWW '11})}. \bibinfo{publisher}{ACM},
  \bibinfo{address}{New York, NY, USA}, \bibinfo{pages}{577–586}.
\newblock
\showISBNx{978-1-4503-0632-4}


\bibitem[\protect\citeauthoryear{Euler}{Euler}{1736}]%
        {Euler1736}
\bibfield{author}{\bibinfo{person}{Leonhard Euler}.}
  \bibinfo{year}{1736}\natexlab{}.
\newblock \showarticletitle{Solutio problematis ad geometriam situs
  pertinentis}.
\newblock \bibinfo{journal}{\emph{Commentarii Academiae Scientiarum Imperialis
  Petropolitanae}}  \bibinfo{volume}{8} (\bibinfo{year}{1736}),
  \bibinfo{pages}{128--140}.
\newblock


\bibitem[\protect\citeauthoryear{Flynn}{Flynn}{1966}]%
        {SIMD}
\bibfield{author}{\bibinfo{person}{M.J. Flynn}.}
  \bibinfo{year}{1966}\natexlab{}.
\newblock \showarticletitle{Very high-speed computing systems}.
\newblock \bibinfo{journal}{\emph{Proc. IEEE}} \bibinfo{volume}{54},
  \bibinfo{number}{12} (\bibinfo{year}{1966}), \bibinfo{pages}{1901--1909}.
\newblock
\urldef\tempurl%
\url{https://doi.org/10.1109/PROC.1966.5273}
\showDOI{\tempurl}


\bibitem[\protect\citeauthoryear{Fu, shuen Chan, Cheung, and Moon}{Fu
  et~al\mbox{.}}{2000}]%
        {Fu2000}
\bibfield{author}{\bibinfo{person}{Ada Wai-Chee Fu}, \bibinfo{person}{Polly~Mei
  shuen Chan}, \bibinfo{person}{Yin-Ling Cheung}, {and}
  \bibinfo{person}{Yiu~Sang Moon}.} \bibinfo{year}{2000}\natexlab{}.
\newblock \showarticletitle{Dynamic vp-Tree Indexing for n-Nearest Neighbor
  Search Given Pair-Wise Distances.}
\newblock \bibinfo{journal}{\emph{VLDB J.}} \bibinfo{volume}{9},
  \bibinfo{number}{2} (\bibinfo{year}{2000}), \bibinfo{pages}{154--173}.
\newblock


\bibitem[\protect\citeauthoryear{Fu and Cai}{Fu and Cai}{2016}]%
        {Cong2016}
\bibfield{author}{\bibinfo{person}{Cong Fu} {and} \bibinfo{person}{Deng Cai}.}
  \bibinfo{year}{2016}\natexlab{}.
\newblock \showarticletitle{EFANNA : An Extremely Fast Approximate Nearest
  Neighbor Search Algorithm Based on kNN Graph.}
\newblock \bibinfo{journal}{\emph{CoRR}}  \bibinfo{volume}{abs/1609.07228}
  (\bibinfo{year}{2016}).
\newblock


\bibitem[\protect\citeauthoryear{Fu, Wang, and Cai}{Fu et~al\mbox{.}}{2022}]%
        {Cong2021}
\bibfield{author}{\bibinfo{person}{Cong Fu}, \bibinfo{person}{Changxu Wang},
  {and} \bibinfo{person}{Deng Cai}.} \bibinfo{year}{2022}\natexlab{}.
\newblock \showarticletitle{High Dimensional Similarity Search With Satellite
  System Graph: Efficiency, Scalability, and Unindexed Query Compatibility}.
\newblock \bibinfo{journal}{\emph{IEEE Transactions on Pattern Analysis and
  Machine Intelligence}} \bibinfo{volume}{44}, \bibinfo{number}{8}
  (\bibinfo{year}{2022}), \bibinfo{pages}{4139--4150}.
\newblock
\urldef\tempurl%
\url{https://doi.org/10.1109/TPAMI.2021.3067706}
\showDOI{\tempurl}


\bibitem[\protect\citeauthoryear{Fu, Xiang, Wang, and Cai}{Fu
  et~al\mbox{.}}{2019}]%
        {Cong2019}
\bibfield{author}{\bibinfo{person}{Cong Fu}, \bibinfo{person}{Chao Xiang},
  \bibinfo{person}{Changxu Wang}, {and} \bibinfo{person}{Deng Cai}.}
  \bibinfo{year}{2019}\natexlab{}.
\newblock \showarticletitle{Fast Approximate Nearest Neighbor Search With The
  Navigating Spreading-out Graph.}
\newblock \bibinfo{journal}{\emph{Proc. VLDB Endow.}} \bibinfo{volume}{12},
  \bibinfo{number}{5} (\bibinfo{year}{2019}), \bibinfo{pages}{461--474}.
\newblock


\bibitem[\protect\citeauthoryear{Gong, Wang, Ogihara, and Xu}{Gong
  et~al\mbox{.}}{2020}]%
        {Gong2020}
\bibfield{author}{\bibinfo{person}{Long Gong}, \bibinfo{person}{Huayi Wang},
  \bibinfo{person}{Mitsunori Ogihara}, {and} \bibinfo{person}{Jun Xu}.}
  \bibinfo{year}{2020}\natexlab{}.
\newblock \showarticletitle{iDEC: Indexable Distance Estimating Codes for
  Approximate Nearest Neighbor Search.}
\newblock \bibinfo{journal}{\emph{Proc. VLDB Endow.}} \bibinfo{volume}{13},
  \bibinfo{number}{9} (\bibinfo{year}{2020}), \bibinfo{pages}{1483--1497}.
\newblock


\bibitem[\protect\citeauthoryear{Groh, Ruppert, Wieschollek, and Lensch}{Groh
  et~al\mbox{.}}{2019}]%
        {Groh2019}
\bibfield{author}{\bibinfo{person}{Fabian Groh}, \bibinfo{person}{Lukas
  Ruppert}, \bibinfo{person}{Patrick Wieschollek}, {and}
  \bibinfo{person}{Hendrik P.~A. Lensch}.} \bibinfo{year}{2019}\natexlab{}.
\newblock \showarticletitle{GGNN: Graph-based GPU Nearest Neighbor Search.}
\newblock \bibinfo{journal}{\emph{CoRR}}  \bibinfo{volume}{abs/1912.01059}
  (\bibinfo{year}{2019}).
\newblock


\bibitem[\protect\citeauthoryear{Hajebi, Abbasi-Yadkori, Shahbazi, and
  Zhang}{Hajebi et~al\mbox{.}}{2011}]%
        {Hajebi2011}
\bibfield{author}{\bibinfo{person}{Kiana Hajebi}, \bibinfo{person}{Yasin
  Abbasi-Yadkori}, \bibinfo{person}{Hossein Shahbazi}, {and}
  \bibinfo{person}{Hong Zhang}.} \bibinfo{year}{2011}\natexlab{}.
\newblock \showarticletitle{Fast Approximate Nearest-Neighbor Search with
  k-Nearest Neighbor Graph.}. In \bibinfo{booktitle}{\emph{IJCAI}},
  \bibfield{editor}{\bibinfo{person}{Toby Walsh}} (Ed.).
  \bibinfo{publisher}{IJCAI/AAAI}, \bibinfo{pages}{1312--1317}.
\newblock
\showISBNx{978-1-57735-516-8}


\bibitem[\protect\citeauthoryear{Harwood and Drummond}{Harwood and
  Drummond}{2016}]%
        {Harwood2016}
\bibfield{author}{\bibinfo{person}{Ben Harwood} {and} \bibinfo{person}{Tom
  Drummond}.} \bibinfo{year}{2016}\natexlab{}.
\newblock \showarticletitle{FANNG: Fast Approximate Nearest Neighbour Graphs.}.
  In \bibinfo{booktitle}{\emph{CVPR}}. \bibinfo{publisher}{IEEE Computer
  Society}, \bibinfo{pages}{5713--5722}.
\newblock
\showISBNx{978-1-4673-8851-1}


\bibitem[\protect\citeauthoryear{Huang, Feng, Fang, Ng, and Wang}{Huang
  et~al\mbox{.}}{2017}]%
        {Huang2017}
\bibfield{author}{\bibinfo{person}{Qiang Huang}, \bibinfo{person}{Jianlin
  Feng}, \bibinfo{person}{Qiong Fang}, \bibinfo{person}{Wilfred Ng}, {and}
  \bibinfo{person}{Wei Wang}.} \bibinfo{year}{2017}\natexlab{}.
\newblock \showarticletitle{Query-aware locality-sensitive hashing scheme for
  lp norm.}
\newblock \bibinfo{journal}{\emph{VLDB J.}} \bibinfo{volume}{26},
  \bibinfo{number}{5} (\bibinfo{year}{2017}), \bibinfo{pages}{683--708}.
\newblock


\bibitem[\protect\citeauthoryear{Indyk and Motwani}{Indyk and Motwani}{1998}]%
        {Indyk1998}
\bibfield{author}{\bibinfo{person}{Piotr Indyk} {and} \bibinfo{person}{Rajeev
  Motwani}.} \bibinfo{year}{1998}\natexlab{}.
\newblock \showarticletitle{Approximate Nearest Neighbors: Towards Removing the
  Curse of Dimensionality}. In \bibinfo{booktitle}{\emph{Proceedings of the
  Thirtieth Annual ACM Symposium on Theory of Computing}} (Dallas, Texas, USA)
  \emph{(\bibinfo{series}{STOC '98})}. \bibinfo{publisher}{Association for
  Computing Machinery}, \bibinfo{address}{New York, NY, USA},
  \bibinfo{pages}{604–613}.
\newblock
\showISBNx{0897919629}


\bibitem[\protect\citeauthoryear{Iwasaki}{Iwasaki}{2010}]%
        {Iwasaki2010}
\bibfield{author}{\bibinfo{person}{Masajiro Iwasaki}.}
  \bibinfo{year}{2010}\natexlab{}.
\newblock \showarticletitle{Proximity Search in Metric Spaces Using Approximate
  K Nearest Neighbor Graph}.
\newblock \bibinfo{journal}{\emph{IPSJ Trans. on Database}}
  \bibinfo{volume}{3} (\bibinfo{year}{2010}), \bibinfo{pages}{18--28}.
\newblock
\showISSN{1882-7772}


\bibitem[\protect\citeauthoryear{Iwasaki and Miyazaki}{Iwasaki and
  Miyazaki}{2018}]%
        {Iwasaki2018}
\bibfield{author}{\bibinfo{person}{Masajiro Iwasaki} {and}
  \bibinfo{person}{Daisuke Miyazaki}.} \bibinfo{year}{2018}\natexlab{}.
\newblock \showarticletitle{Optimization of Indexing Based on k-Nearest
  Neighbor Graph for Proximity Search in High-dimensional Data.}
\newblock \bibinfo{journal}{\emph{CoRR}}  \bibinfo{volume}{abs/1810.07355}
  (\bibinfo{year}{2018}).
\newblock


\bibitem[\protect\citeauthoryear{Jaromczyk and Toussaint}{Jaromczyk and
  Toussaint}{1992}]%
        {Jaromczyk1992}
\bibfield{author}{\bibinfo{person}{J.W. Jaromczyk} {and} \bibinfo{person}{G.T.
  Toussaint}.} \bibinfo{year}{1992}\natexlab{}.
\newblock \showarticletitle{Relative neighborhood graphs and their relatives}.
\newblock \bibinfo{journal}{\emph{Proc. IEEE}} \bibinfo{volume}{80},
  \bibinfo{number}{9} (\bibinfo{year}{1992}), \bibinfo{pages}{1502--1517}.
\newblock
\urldef\tempurl%
\url{https://doi.org/10.1109/5.163414}
\showDOI{\tempurl}


\bibitem[\protect\citeauthoryear{Jayaram~Subramanya, Devvrit, Simhadri,
  Krishnawamy, and Kadekodi}{Jayaram~Subramanya et~al\mbox{.}}{2019}]%
        {Subramanya2019}
\bibfield{author}{\bibinfo{person}{Suhas Jayaram~Subramanya},
  \bibinfo{person}{Fnu Devvrit}, \bibinfo{person}{Harsha~Vardhan Simhadri},
  \bibinfo{person}{Ravishankar Krishnawamy}, {and} \bibinfo{person}{Rohan
  Kadekodi}.} \bibinfo{year}{2019}\natexlab{}.
\newblock \showarticletitle{DiskANN: Fast Accurate Billion-point Nearest
  Neighbor Search on a Single Node}. In \bibinfo{booktitle}{\emph{Advances in
  Neural Information Processing Systems}},
  \bibfield{editor}{\bibinfo{person}{H.~Wallach},
  \bibinfo{person}{H.~Larochelle}, \bibinfo{person}{A.~Beygelzimer},
  \bibinfo{person}{F.~d\textquotesingle Alch\'{e}-Buc},
  \bibinfo{person}{E.~Fox}, {and} \bibinfo{person}{R.~Garnett}} (Eds.),
  Vol.~\bibinfo{volume}{32}. \bibinfo{publisher}{Curran Associates, Inc.}
\newblock


\bibitem[\protect\citeauthoryear{Johnson, Douze, and J{\'e}gou}{Johnson
  et~al\mbox{.}}{2019}]%
        {Johnson2019}
\bibfield{author}{\bibinfo{person}{Jeff Johnson}, \bibinfo{person}{Matthijs
  Douze}, {and} \bibinfo{person}{Herv{\'e} J{\'e}gou}.}
  \bibinfo{year}{2019}\natexlab{}.
\newblock \showarticletitle{Billion-scale similarity search with {GPUs}}.
\newblock \bibinfo{journal}{\emph{IEEE Transactions on Big Data}}
  \bibinfo{volume}{7}, \bibinfo{number}{3} (\bibinfo{year}{2019}),
  \bibinfo{pages}{535--547}.
\newblock


\bibitem[\protect\citeauthoryear{Jégou, Douze, and Schmid}{Jégou
  et~al\mbox{.}}{2011}]%
        {Jegou2011}
\bibfield{author}{\bibinfo{person}{Hervé Jégou}, \bibinfo{person}{Matthijs
  Douze}, {and} \bibinfo{person}{Cordelia Schmid}.}
  \bibinfo{year}{2011}\natexlab{}.
\newblock \showarticletitle{Product Quantization for Nearest Neighbor Search.}
\newblock \bibinfo{journal}{\emph{IEEE Trans. Pattern Anal. Mach. Intell.}}
  \bibinfo{volume}{33}, \bibinfo{number}{1} (\bibinfo{year}{2011}),
  \bibinfo{pages}{117--128}.
\newblock


\bibitem[\protect\citeauthoryear{Knuth}{Knuth}{1997}]%
        {Knuth1997}
\bibfield{author}{\bibinfo{person}{Donald~Ervin Knuth}.}
  \bibinfo{year}{1997}\natexlab{}.
\newblock \bibinfo{booktitle}{\emph{The Art of Computer Programming: Sorting
  and Searching [1973]} (\bibinfo{edition}{2nd} ed.)}.
  Vol.~\bibinfo{volume}{3}.
\newblock \bibinfo{publisher}{Addison Wesley}, \bibinfo{address}{Reading,
  Massachusetts}, Chapter 6.5. Retrieval on Secondary Keys.
\newblock
\showISBNx{0-201-89685-0}


\bibitem[\protect\citeauthoryear{Li, Zhang, Andersen, and He}{Li
  et~al\mbox{.}}{2020a}]%
        {Li2020}
\bibfield{author}{\bibinfo{person}{Conglong Li}, \bibinfo{person}{Minjia
  Zhang}, \bibinfo{person}{David~G. Andersen}, {and} \bibinfo{person}{Yuxiong
  He}.} \bibinfo{year}{2020}\natexlab{a}.
\newblock \showarticletitle{Improving Approximate Nearest Neighbor Search
  through Learned Adaptive Early Termination}. In
  \bibinfo{booktitle}{\emph{Proceedings of the 2020 ACM SIGMOD International
  Conference on Management of Data}} (Portland, OR, USA)
  \emph{(\bibinfo{series}{SIGMOD '20})}. \bibinfo{publisher}{Association for
  Computing Machinery}, \bibinfo{address}{New York, NY, USA},
  \bibinfo{pages}{2539–2554}.
\newblock
\showISBNx{9781450367356}


\bibitem[\protect\citeauthoryear{Li, Zhang, Sun, Wang, Li, Zhang, and Lin}{Li
  et~al\mbox{.}}{2020b}]%
        {DPG2020}
\bibfield{author}{\bibinfo{person}{Wen Li}, \bibinfo{person}{Ying Zhang},
  \bibinfo{person}{Yifang Sun}, \bibinfo{person}{Wei Wang},
  \bibinfo{person}{Mingjie Li}, \bibinfo{person}{Wenjie Zhang}, {and}
  \bibinfo{person}{Xuemin Lin}.} \bibinfo{year}{2020}\natexlab{b}.
\newblock \showarticletitle{Approximate Nearest Neighbor Search on High
  Dimensional Data - Experiments, Analyses, and Improvement.}
\newblock \bibinfo{journal}{\emph{IEEE Trans. Knowl. Data Eng.}}
  \bibinfo{volume}{32}, \bibinfo{number}{8} (\bibinfo{year}{2020}),
  \bibinfo{pages}{1475--1488}.
\newblock


\bibitem[\protect\citeauthoryear{Lin and Zhao}{Lin and Zhao}{2019}]%
        {Peng2019}
\bibfield{author}{\bibinfo{person}{Peng{-}Cheng Lin} {and}
  \bibinfo{person}{Wan{-}Lei Zhao}.} \bibinfo{year}{2019}\natexlab{}.
\newblock \showarticletitle{Graph based Nearest Neighbor Search: Promises and
  Failures}.
\newblock \bibinfo{journal}{\emph{CoRR}}  \bibinfo{volume}{abs/1904.02077}
  (\bibinfo{year}{2019}).
\newblock
\showeprint[arXiv]{1904.02077}


\bibitem[\protect\citeauthoryear{Lsyhprum}{Lsyhprum}{[n.d.]}]%
        {GithubWEAVESS}
\bibfield{author}{\bibinfo{person}{Lsyhprum}.}
  \bibinfo{year}{[n.d.]}\natexlab{}.
\newblock \bibinfo{title}{LSYHPRUM/WEAVESS: A comprehensive survey and
  experimental comparison of graph-based approximate nearest neighbor search}.
\newblock
\newblock
\urldef\tempurl%
\url{https://github.com/Lsyhprum/WEAVESS}
\showURL{%
\tempurl}
\newblock
\shownote{[Accessed 21-Mar-2023].}


\bibitem[\protect\citeauthoryear{Malkov, Ponomarenko, Logvinov, and
  Krylov}{Malkov et~al\mbox{.}}{2014}]%
        {Malkov2014}
\bibfield{author}{\bibinfo{person}{Yury Malkov}, \bibinfo{person}{Alexander
  Ponomarenko}, \bibinfo{person}{Andrey Logvinov}, {and}
  \bibinfo{person}{Vladimir Krylov}.} \bibinfo{year}{2014}\natexlab{}.
\newblock \showarticletitle{Approximate nearest neighbor algorithm based on
  navigable small world graphs.}
\newblock \bibinfo{journal}{\emph{Inf. Syst.}}  \bibinfo{volume}{45}
  (\bibinfo{year}{2014}), \bibinfo{pages}{61--68}.
\newblock


\bibitem[\protect\citeauthoryear{Malkov and Yashunin}{Malkov and
  Yashunin}{2020}]%
        {Malkov2020}
\bibfield{author}{\bibinfo{person}{Yury~A. Malkov} {and} \bibinfo{person}{D.~A.
  Yashunin}.} \bibinfo{year}{2020}\natexlab{}.
\newblock \showarticletitle{Efficient and Robust Approximate Nearest Neighbor
  Search Using Hierarchical Navigable Small World Graphs.}
\newblock \bibinfo{journal}{\emph{IEEE Trans. Pattern Anal. Mach. Intell.}}
  \bibinfo{volume}{42}, \bibinfo{number}{4} (\bibinfo{year}{2020}),
  \bibinfo{pages}{824--836}.
\newblock


\bibitem[\protect\citeauthoryear{Naidan, Boytsov, and Nyberg}{Naidan
  et~al\mbox{.}}{2015}]%
        {Naidan2015}
\bibfield{author}{\bibinfo{person}{Bilegsaikhan Naidan},
  \bibinfo{person}{Leonid Boytsov}, {and} \bibinfo{person}{Eric Nyberg}.}
  \bibinfo{year}{2015}\natexlab{}.
\newblock \showarticletitle{Permutation Search Methods are Efficient, Yet
  Faster Search is Possible.}
\newblock \bibinfo{journal}{\emph{Proc. VLDB Endow.}} \bibinfo{volume}{8},
  \bibinfo{number}{12} (\bibinfo{year}{2015}), \bibinfo{pages}{1618--1629}.
\newblock


\bibitem[\protect\citeauthoryear{Navarro}{Navarro}{2002}]%
        {Navarro2002}
\bibfield{author}{\bibinfo{person}{Gonzalo Navarro}.}
  \bibinfo{year}{2002}\natexlab{}.
\newblock \showarticletitle{Searching in metric spaces by spatial
  approximation.}
\newblock \bibinfo{journal}{\emph{VLDB J.}} \bibinfo{volume}{11},
  \bibinfo{number}{1} (\bibinfo{year}{2002}), \bibinfo{pages}{28--46}.
\newblock


\bibitem[\protect\citeauthoryear{O'Rourke}{O'Rourke}{1982}]%
        {ORourke1982}
\bibfield{author}{\bibinfo{person}{Joseph O'Rourke}.}
  \bibinfo{year}{1982}\natexlab{}.
\newblock \showarticletitle{Computing the relative neighborhood graph in the L1
  and Linfinity metrics}.
\newblock \bibinfo{journal}{\emph{Pattern Recognit.}}  \bibinfo{volume}{15}
  (\bibinfo{year}{1982}), \bibinfo{pages}{189--192}.
\newblock


\bibitem[\protect\citeauthoryear{Paredes and Chávez}{Paredes and
  Chávez}{2005}]%
        {Paredes2005}
\bibfield{author}{\bibinfo{person}{Rodrigo Paredes} {and}
  \bibinfo{person}{Edgar Chávez}.} \bibinfo{year}{2005}\natexlab{}.
\newblock \showarticletitle{Using the k-Nearest Neighbor Graph for Proximity
  Searching in Metric Spaces.}. In \bibinfo{booktitle}{\emph{SPIRE}}
  \emph{(\bibinfo{series}{Lecture Notes in Computer Science})},
  \bibfield{editor}{\bibinfo{person}{Mariano~P. Consens} {and}
  \bibinfo{person}{Gonzalo Navarro}} (Eds.), Vol.~\bibinfo{volume}{3772}.
  \bibinfo{publisher}{Springer}, \bibinfo{pages}{127--138}.
\newblock
\showISBNx{3-540-29740-5}


\bibitem[\protect\citeauthoryear{Park, Park, Jung, and goo Lee}{Park
  et~al\mbox{.}}{2015}]%
        {Park2015}
\bibfield{author}{\bibinfo{person}{Youngki Park}, \bibinfo{person}{Sungchan
  Park}, \bibinfo{person}{Woosung Jung}, {and} \bibinfo{person}{Sang goo Lee}.}
  \bibinfo{year}{2015}\natexlab{}.
\newblock \showarticletitle{Reversed CF: A fast collaborative filtering
  algorithm using a k-nearest neighbor graph.}
\newblock \bibinfo{journal}{\emph{Expert Syst. Appl.}} \bibinfo{volume}{42},
  \bibinfo{number}{8} (\bibinfo{year}{2015}), \bibinfo{pages}{4022--4028}.
\newblock


\bibitem[\protect\citeauthoryear{Pennington, Socher, and Manning}{Pennington
  et~al\mbox{.}}{2014}]%
        {Pennington2014}
\bibfield{author}{\bibinfo{person}{Jeffrey Pennington},
  \bibinfo{person}{Richard Socher}, {and} \bibinfo{person}{Christopher~D
  Manning}.} \bibinfo{year}{2014}\natexlab{}.
\newblock \showarticletitle{Glove: Global vectors for word representation}. In
  \bibinfo{booktitle}{\emph{Proceedings of the 2014 conference on empirical
  methods in natural language processing (EMNLP)}}.
  \bibinfo{pages}{1532--1543}.
\newblock


\bibitem[\protect\citeauthoryear{Ponomarenko, Avrelin, Naidan, and
  Boytsov}{Ponomarenko et~al\mbox{.}}{2014}]%
        {Ponomarenko2014}
\bibfield{author}{\bibinfo{person}{Alexander Ponomarenko},
  \bibinfo{person}{Nikita Avrelin}, \bibinfo{person}{Bilegsaikhan Naidan},
  {and} \bibinfo{person}{Leonid Boytsov}.} \bibinfo{year}{2014}\natexlab{}.
\newblock \showarticletitle{Comparative Analysis of Data Structures for
  Approximate Nearest Neighbor Search}.
\newblock


\bibitem[\protect\citeauthoryear{Schall, Hezel, Jung, and Barthel}{Schall
  et~al\mbox{.}}{2023}]%
        {Vibro2023}
\bibfield{author}{\bibinfo{person}{Konstantin Schall}, \bibinfo{person}{Nico
  Hezel}, \bibinfo{person}{Klaus Jung}, {and} \bibinfo{person}{Kai~Uwe
  Barthel}.} \bibinfo{year}{2023}\natexlab{}.
\newblock \showarticletitle{Vibro: Video Browsing With Semantic And Visual
  Image Embeddings}. In \bibinfo{booktitle}{\emph{MultiMedia Modeling: 29th
  International Conference, MMM 2023, Bergen, Norway, January 9–12, 2023,
  Proceedings, Part I}} (Bergen, Norway). \bibinfo{publisher}{Springer-Verlag},
  \bibinfo{address}{Berlin, Heidelberg}, \bibinfo{pages}{665–670}.
\newblock
\showISBNx{978-3-031-27076-5}
\urldef\tempurl%
\url{https://doi.org/10.1007/978-3-031-27077-2_56}
\showDOI{\tempurl}


\bibitem[\protect\citeauthoryear{Shimomura, Oyamada, Vieira, and
  Kaster}{Shimomura et~al\mbox{.}}{2021}]%
        {Shimomura2021}
\bibfield{author}{\bibinfo{person}{Larissa~Capobianco Shimomura},
  \bibinfo{person}{Rafael~Seidi Oyamada}, \bibinfo{person}{Marcos~R. Vieira},
  {and} \bibinfo{person}{Daniel~S. Kaster}.} \bibinfo{year}{2021}\natexlab{}.
\newblock \showarticletitle{A survey on graph-based methods for similarity
  searches in metric spaces.}
\newblock \bibinfo{journal}{\emph{Inf. Syst.}}  \bibinfo{volume}{95}
  (\bibinfo{year}{2021}), \bibinfo{pages}{101507}.
\newblock


\bibitem[\protect\citeauthoryear{Simonyan and Zisserman}{Simonyan and
  Zisserman}{2015}]%
        {Simonyan2015}
\bibfield{author}{\bibinfo{person}{Karen Simonyan} {and}
  \bibinfo{person}{Andrew Zisserman}.} \bibinfo{year}{2015}\natexlab{}.
\newblock \showarticletitle{Very Deep Convolutional Networks for Large-Scale
  Image Recognition.}. In \bibinfo{booktitle}{\emph{ICLR}},
  \bibfield{editor}{\bibinfo{person}{Yoshua Bengio} {and} \bibinfo{person}{Yann
  LeCun}} (Eds.).
\newblock


\bibitem[\protect\citeauthoryear{Sugawara, Kobayashi, and Iwasaki}{Sugawara
  et~al\mbox{.}}{2016}]%
        {Sugawara2016}
\bibfield{author}{\bibinfo{person}{Kohei Sugawara}, \bibinfo{person}{Hayato
  Kobayashi}, {and} \bibinfo{person}{Masajiro Iwasaki}.}
  \bibinfo{year}{2016}\natexlab{}.
\newblock \showarticletitle{On Approximately Searching for Similar Word
  Embeddings.}. In \bibinfo{booktitle}{\emph{ACL (1)}}. \bibinfo{publisher}{The
  Association for Computer Linguistics}, \bibinfo{address}{Berlin, Germany}.
\newblock
\showISBNx{978-1-945626-00-5}


\bibitem[\protect\citeauthoryear{Sun, Wang, Qin, Zhang, and Lin}{Sun
  et~al\mbox{.}}{2014}]%
        {Enron}
\bibfield{author}{\bibinfo{person}{Yifang Sun}, \bibinfo{person}{Wei Wang},
  \bibinfo{person}{Jianbin Qin}, \bibinfo{person}{Ying Zhang}, {and}
  \bibinfo{person}{Xuemin Lin}.} \bibinfo{year}{2014}\natexlab{}.
\newblock \showarticletitle{SRS: Solving c-Approximate Nearest Neighbor Queries
  in High Dimensional Euclidean Space with a Tiny Index.}
\newblock \bibinfo{journal}{\emph{Proc. VLDB Endow.}} \bibinfo{volume}{8},
  \bibinfo{number}{1} (\bibinfo{year}{2014}), \bibinfo{pages}{1--12}.
\newblock


\bibitem[\protect\citeauthoryear{Toussaint}{Toussaint}{1980}]%
        {Toussaint1980}
\bibfield{author}{\bibinfo{person}{Godfried~T. Toussaint}.}
  \bibinfo{year}{1980}\natexlab{}.
\newblock \showarticletitle{The relative neighbourhood graph of a finite planar
  set}.
\newblock \bibinfo{journal}{\emph{Pattern Recognition}} \bibinfo{volume}{12},
  \bibinfo{number}{4} (\bibinfo{year}{1980}), \bibinfo{pages}{261--268}.
\newblock
\showISSN{0031-3203}
\urldef\tempurl%
\url{https://doi.org/10.1016/0031-3203(80)90066-7}
\showDOI{\tempurl}


\bibitem[\protect\citeauthoryear{Wang, Yi, Guo, Jin, Xu, Li, Wang, Guo, Li, Xu,
  Yu, Yuan, Zou, Long, Cai, Li, Zhang, Mo, Gu, Jiang, Wei, and Xie}{Wang
  et~al\mbox{.}}{2021b}]%
        {Wang2021}
\bibfield{author}{\bibinfo{person}{Jianguo Wang}, \bibinfo{person}{Xiaomeng
  Yi}, \bibinfo{person}{Rentong Guo}, \bibinfo{person}{Hai Jin},
  \bibinfo{person}{Peng Xu}, \bibinfo{person}{Shengjun Li},
  \bibinfo{person}{Xiangyu Wang}, \bibinfo{person}{Xiangzhou Guo},
  \bibinfo{person}{Chengming Li}, \bibinfo{person}{Xiaohai Xu},
  \bibinfo{person}{Kun Yu}, \bibinfo{person}{Yuxing Yuan},
  \bibinfo{person}{Yinghao Zou}, \bibinfo{person}{Jiquan Long},
  \bibinfo{person}{Yudong Cai}, \bibinfo{person}{Zhenxiang Li},
  \bibinfo{person}{Zhifeng Zhang}, \bibinfo{person}{Yihua Mo},
  \bibinfo{person}{Jun Gu}, \bibinfo{person}{Ruiyi Jiang}, \bibinfo{person}{Yi
  Wei}, {and} \bibinfo{person}{Charles Xie}.} \bibinfo{year}{2021}\natexlab{b}.
\newblock \showarticletitle{Milvus: A Purpose-Built Vector Data Management
  System}. In \bibinfo{booktitle}{\emph{Proceedings of the 2021 International
  Conference on Management of Data}} (Virtual Event, China)
  \emph{(\bibinfo{series}{SIGMOD/PODS '21})}. \bibinfo{publisher}{Association
  for Computing Machinery}, \bibinfo{address}{New York, NY, USA},
  \bibinfo{pages}{2614–2627}.
\newblock
\showISBNx{9781450383431}


\bibitem[\protect\citeauthoryear{Wang, Xu, Yue, and Wang}{Wang
  et~al\mbox{.}}{2021a}]%
        {Wang2021Survey}
\bibfield{author}{\bibinfo{person}{Mengzhao Wang}, \bibinfo{person}{Xiaoliang
  Xu}, \bibinfo{person}{Qiang Yue}, {and} \bibinfo{person}{Yuxiang Wang}.}
  \bibinfo{year}{2021}\natexlab{a}.
\newblock \showarticletitle{A Comprehensive Survey and Experimental Comparison
  of Graph-Based Approximate Nearest Neighbor Search}.
\newblock \bibinfo{journal}{\emph{Proc. VLDB Endow.}} \bibinfo{volume}{14},
  \bibinfo{number}{11} (\bibinfo{date}{jul} \bibinfo{year}{2021}),
  \bibinfo{pages}{1964–1978}.
\newblock
\showISSN{2150-8097}
\urldef\tempurl%
\url{https://doi.org/10.14778/3476249.3476255}
\showDOI{\tempurl}


\bibitem[\protect\citeauthoryear{Wang, Dong, Josephson, Lv, Charikar, and
  Li}{Wang et~al\mbox{.}}{2007}]%
        {Audio}
\bibfield{author}{\bibinfo{person}{Zhe Wang}, \bibinfo{person}{Wei Dong},
  \bibinfo{person}{William Josephson}, \bibinfo{person}{Qin Lv},
  \bibinfo{person}{Moses Charikar}, {and} \bibinfo{person}{Kai Li}.}
  \bibinfo{year}{2007}\natexlab{}.
\newblock \showarticletitle{Sizing sketches: a rank-based analysis for
  similarity search.}. In \bibinfo{booktitle}{\emph{SIGMETRICS}},
  \bibfield{editor}{\bibinfo{person}{Leana Golubchik},
  \bibinfo{person}{Mostafa~H. Ammar}, {and} \bibinfo{person}{Mor
  Harchol-Balter}} (Eds.). \bibinfo{publisher}{ACM}, \bibinfo{pages}{157--168}.
\newblock
\showISBNx{978-1-59593-639-4}


\bibitem[\protect\citeauthoryear{Weber, Schek, and Blott}{Weber
  et~al\mbox{.}}{1998}]%
        {Weber1998}
\bibfield{author}{\bibinfo{person}{Roger Weber}, \bibinfo{person}{Hans-J\"{o}rg
  Schek}, {and} \bibinfo{person}{Stephen Blott}.}
  \bibinfo{year}{1998}\natexlab{}.
\newblock \showarticletitle{A Quantitative Analysis and Performance Study for
  Similarity-Search Methods in High-Dimensional Spaces}. In
  \bibinfo{booktitle}{\emph{VLDB '98: Proceedings of the 24rd International
  Conference on Very Large Data Bases}}. \bibinfo{publisher}{Morgan Kaufmann
  Publishers Inc.}, \bibinfo{address}{San Francisco, CA, USA},
  \bibinfo{pages}{194--205}.
\newblock
\showISBNx{1-55860-566-5}


\bibitem[\protect\citeauthoryear{Wei, Wu, Wang, Lou, Zhan, Li, and Cai}{Wei
  et~al\mbox{.}}{2020}]%
        {Wei2020}
\bibfield{author}{\bibinfo{person}{Chuangxian Wei}, \bibinfo{person}{Bin Wu},
  \bibinfo{person}{Sheng Wang}, \bibinfo{person}{Renjie Lou},
  \bibinfo{person}{Chaoqun Zhan}, \bibinfo{person}{Feifei Li}, {and}
  \bibinfo{person}{Yuanzhe Cai}.} \bibinfo{year}{2020}\natexlab{}.
\newblock \showarticletitle{AnalyticDB-V: A Hybrid Analytical Engine Towards
  Query Fusion for Structured and Unstructured Data.}
\newblock \bibinfo{journal}{\emph{Proc. VLDB Endow.}} \bibinfo{volume}{13},
  \bibinfo{number}{12} (\bibinfo{year}{2020}), \bibinfo{pages}{3152--3165}.
\newblock


\bibitem[\protect\citeauthoryear{Zhao, Pan, Zheng, Zhang, Wang, Zhang, Xu, and
  Jin}{Zhao et~al\mbox{.}}{2019}]%
        {Zhao2019}
\bibfield{author}{\bibinfo{person}{Kang Zhao}, \bibinfo{person}{Pan Pan},
  \bibinfo{person}{Yun Zheng}, \bibinfo{person}{Yanhao Zhang},
  \bibinfo{person}{Changxu Wang}, \bibinfo{person}{Yingya Zhang},
  \bibinfo{person}{Yinghui Xu}, {and} \bibinfo{person}{Rong Jin}.}
  \bibinfo{year}{2019}\natexlab{}.
\newblock \showarticletitle{Large-Scale Visual Search with Binary Distributed
  Graph at Alibaba.}. In \bibinfo{booktitle}{\emph{CIKM}},
  \bibfield{editor}{\bibinfo{person}{Wenwu Zhu}, \bibinfo{person}{Dacheng Tao},
  \bibinfo{person}{Xueqi Cheng}, \bibinfo{person}{Peng Cui},
  \bibinfo{person}{Elke~A. Rundensteiner}, \bibinfo{person}{David Carmel},
  \bibinfo{person}{Qi~He}, {and} \bibinfo{person}{Jeffrey~Xu Yu}} (Eds.).
  \bibinfo{publisher}{ACM}, \bibinfo{address}{Beijing, China},
  \bibinfo{pages}{2567--2575}.
\newblock
\showISBNx{978-1-4503-6976-3}


\bibitem[\protect\citeauthoryear{Zhao, Tan, and Li}{Zhao et~al\mbox{.}}{2020}]%
        {Zhao2020}
\bibfield{author}{\bibinfo{person}{Weijie Zhao}, \bibinfo{person}{Shulong Tan},
  {and} \bibinfo{person}{Ping Li}.} \bibinfo{year}{2020}\natexlab{}.
\newblock \showarticletitle{SONG: Approximate Nearest Neighbor Search on GPU.}.
  In \bibinfo{booktitle}{\emph{ICDE}}. \bibinfo{publisher}{IEEE},
  \bibinfo{pages}{1033--1044}.
\newblock
\showISBNx{978-1-7281-2903-7}


\bibitem[\protect\citeauthoryear{ZJULearning}{ZJULearning}{[n.d.]a}]%
        {GithubEFANNA}
\bibfield{author}{\bibinfo{person}{ZJULearning}.}
  \bibinfo{year}{[n.d.]}\natexlab{a}.
\newblock \bibinfo{title}{ZJULearning/efanna\_graph: An extremely fast
  approximate nearest neighbor graph construction algorithm framework}.
\newblock
\newblock
\urldef\tempurl%
\url{https://github.com/ZJULearning/efanna\_graph}
\showURL{%
\tempurl}
\newblock
\shownote{[Accessed 21-Mar-2023].}


\bibitem[\protect\citeauthoryear{ZJULearning}{ZJULearning}{[n.d.]b}]%
        {GithubNSG}
\bibfield{author}{\bibinfo{person}{ZJULearning}.}
  \bibinfo{year}{[n.d.]}\natexlab{b}.
\newblock \bibinfo{title}{ZJULearning/NSG: Navigating spreading-out graph for
  approximate nearest neighbor search}.
\newblock
\newblock
\urldef\tempurl%
\url{https://github.com/ZJULearning/nsg}
\showURL{%
\tempurl}
\newblock
\shownote{[Accessed 21-Mar-2023].}


\bibitem[\protect\citeauthoryear{ZJULearning}{ZJULearning}{[n.d.]c}]%
        {GithubSSG}
\bibfield{author}{\bibinfo{person}{ZJULearning}.}
  \bibinfo{year}{[n.d.]}\natexlab{c}.
\newblock \bibinfo{title}{ZJULearning/SSG: Code for Satellite System graphs}.
\newblock
\newblock
\urldef\tempurl%
\url{https://github.com/ZJULearning/SSG}
\showURL{%
\tempurl}
\newblock
\shownote{[Accessed 21-Mar-2023].}


\end{thebibliography}

\clearpage 
\section*{APPENDIX}
\appendix

\section{Evaluation of graph properties}
\label{appendix:evaluatingGraphProperties}

The properties of the graphs listed in Table \ref{tab:graph_props} have been assessed using the algorithms outlined in their respective papers. While there is no theoretical support for some of these properties, they are likely to be true and will be discussed in the following paragraphs.

None of the compared graph implementations offers the capability to delete a vertex. Although HNSW allows for marking vertices as deleted, they remain in the index and thus affect the search speed. The DEG is the only fully dynamic graph.

The connectivity of a graph may not have an impact on search efficiency. However, exploring a disconnected graph can result in a loss of accuracy because not all vertices can reach each other. In our experiments kGraph, Efanna and HNSW frequently consist of multiple strongly connected components, making them less suitable for exploring large neighborhoods.

Furthermore, these three graphs contain \textit{source vertices} without incoming edges, which makes them unreachable (see Table \ref{tab:graph_props}). As a result, an approximate nearest neighbor search starting from a seed vertex cannot reach all the vertices in the graph. 
To address this, DPG converts the directed edges of kGraph into undirected edges, effectively solving the issue in most cases. However, in certain scenarios, this approach may not be sufficient, resulting in the graph having multiple components \cite{Wang2021Survey}.

In Table \ref{tab:graph_props}, the search reachability of NSSG is marked as true, although it is not guaranteed by its implementation. Despite the fact that the authors proposed a solution in their paper and implemented it in NSG, they opted for a different approach for NSSG. 
While this alternative implementation does not guarantee search reachability, it did not cause any problems in our experiments.
\vspace{3mm} 
\\
The \textbf{kGraph} is neither an fully dynamic nor an incremental graph. In theory, the incremental property could be achieved by adding random edges to new vertices and applying the \textit{NN-descent} algorithm only to the involved vertices. After multiple iterations, the new vertices would be well connected to their nearest neighbors, but this approach may lead to a deterioration in the quality of the neighborhood for other vertices. 
Since \textit{kGraph} does not pay special attention to routing during its construction, there is no guarantee of connectivity or search reachability for a randomly selected seed vertex. In our experiments with various datasets, \textit{kGraph} fragmented into hundreds to thousands of strongly connected components.
\vspace{3mm} 
\\
By transforming all the edges of an initial \textit{kGraph} into undirected edges, a \textbf{Diversified Proximity Graph} (DPG) reduces the likelihood of the graph to disconnect. However, if the strongly connected components within the original \textit{kGraph} lack inter-component edges, DPG cannot reconnect them. Fortunately, such situations rarely occur in practice. Despite this improvement, the subsequent edge selection and pruning process still focuses solely on the local neighborhood, rather than considering routing through the graph. 
\vspace{3mm} 
\\
The construction process of an \textbf{Optimized Nearest Neighbors Graph} (ONNG) involves multiple steps and is based on an existing ANNG, which makes incremental updates quite complex. Specifically, if new vertices were added regularly, the conversion from undirected to bi-directed edges and the verification of alternative paths would require careful consideration. Furthermore, ONNG does not support vertex removal. While ANNG guarantees both \textit{connectivity} and \textit{search reachability}, ONNG loses these properties due to the pruning of bi-directional edges. \textit{Search reachability} is restored by a \textit{VP-tree}.
\vspace{3mm} 
\\
The \textbf{Extremely Fast Approximate Nearest Neighbor Search Algorithm} (EFANNA) not only utilizes \textit{KD-trees} for graph construction but also employs them during the search process to identify seed vertices in proximity to the query. However, these KD-trees do not offer any advantages in exploration tasks. Similar to \textit{kGraph}, EFANNA suffers from the same limitations due to its construction process, namely being a non-incremental graph without guarantees of connectivity.
\vspace{3mm} 
\\
While a \textbf{Navigating Spreading-out Graph} (NSG), employs an incremental construction strategy, its dependency on EFANNA as the initial graph, limits its ability to add new vertices at any time. 
An option would be to replace the underlying graph with ANNG or NSW. Nevertheless the vertices already incorporated into NSG would not be updated by subsequent changes, leading to a deterioration of the search accuracy. In the final stage of the construction process, a seed vertex is chosen for future search tasks, and a depth-first-search is conducted to identify all reachable vertices. Any none reachable vertex receive an additional edge to the closest reachable vertex, thereby ensuring overall \textit{search reachability}. During this process, the upper bound of the out-degree per vertex may be exceeded.
\vspace{3mm} 
\\
The \textbf{Navigating Satellite System Graph} (NSSG) shares similarities with NSG as it relies on EFANNA and inherits many of its characteristics. One major difference is the higher number of seed vertices. After the construction phase, a depth-first search (DFS) is performed for each seed vertex to identify all connected vertices. For the remaining unreachable vertices, edges are established with randomly selected reachable vertices that have not exceeded their edge limit. 
In the event no vertex meets these criteria, the graph may become disconnected. However, based on our experiments, the probability of such disconnection is very low and has not been observed. 
This problem is not explicitly mentioned in the original paper and can be fixed with the NSG version, where the upper bound on edges per vertex can be exceeded.
\vspace{3mm} 
\\
In the \textbf{Hierarchical Navigable Small World} (HNSW) algorithm, new vertices are incorporated into the structure by identifying the most similar vertices in the hierarchical layers. Directed edges are then added to some of these vertices, while simultaneously updating the neighborhoods of the visited vertices during the search. This incremental approach improves the reachability of the search, as all results are obtained from the same seed vertex. 
The graphs of the lower layers of the hierarchy may be disconnected and their individual components can only be reached from the layers above. 
When exploring the lowest layer from a specific seed vertex, there is an increased likelihood to miss some vertices. The entire hierarchy structure requires additional memory to store.
\vspace{3mm} 
\\
The \textbf{Dynamic Exploration Graph} (DEG) proposed in this paper is an undirected and even-regular graph. The first few vertices are fully connected and any subsequent vertex searches for suitable neighbors in the current graph index. Some of the neighbors edges are replaced with two new edges, integrating the new vertex in the graph by adding detours of the previously adjacent vertices. After enough edges have been replaced, the graph becomes even-regular and connected again. The carefully designed edge optimization algorithm of DEG also preserves these properties. Although the procedure for removing vertices from the graph is not addressed in this paper, its inclusion makes the graph fully dynamic.

\section{Connectivity Guarantees}
\label{appendix:connectivityGuarantees}
In graph theory a vertex-induced subgraph $G[S]$ contains only edges where both endpoints are in the vertex set $S \subset V$. Two vertex-induced subgraphs of $\DEG_d$ based on the disjoint subsets $S_1 = S$ and $S_2 = V \setminus S$ can define a cut $C = (S_1,S_2)$ and its cut-set $B$ containing the edges $\{(u,v) \in E | u \in S_1, v \in S_2\}$. Since every edge in the cut-set has one endpoint in one of the two vertex-induced subgraphs, these subgraphs can not be d-regular and have less than $(|S| \cdot d)/2$ edges. Removing all edges of the cut-set, spanning the two subgraphs $S_1$ and $S_2$, would disconnect $\DEG$. The following sections provide more details on how to calculate the size of the cut-set and its lower bound with respect to the size of the subgraphs.
\\
\\
\textbf{Cut-set size of DEG.}
Let $\DEG[S_1]$ and $\DEG[S_2]$ be two vertex-induced subgraphs of $\DEG$ and $S_1 = S, S_2 = V \setminus S$ their sets of vertices. Further denote $L = \{(u,v) \in E | u \in S, v \in S\}$ as the set of edges of $\DEG[S]$. Then the cut-set size $|B|$ is the difference of the number of edges in $\DEG$ and the combined number of edges in both subgraphs (see the first part of the equation \ref{eq:cutset_size}). 

Since each crossing edge of the cut is undirected, both subgraphs are missing the same number of edges to be d-regular again. The cut-set size can therefore also be computed by comparing the number of edges $|L|$ in $\DEG[S]$ with the number of edges in a $d$-regular undirected graph containing the vertices $S$. The resulting difference must be doubled to account for the two subsets (each missing the same edge).

\begin{equation} \label{eq:cutset_size}
|B| = |E| - (|L_1| + |L_2|) = 2\left(\frac{|S| \cdot d}{2} - |L|\right)
\end{equation}
\\
\textbf{Lower bound of the cut-set size of DEG.} The cut-set size of $\DEG_d$ depends on the size of the two vertex sets $S_1$ and $S_2$. Because of the 2-edge connectivity of $\DEG$, a cut-set size of two can exists. However this lower bound is only true if the resulting components have at least $(d+1)$ vertices.
Any cut of $\DEG$ where one of the subsets $S$ has less than $(d+1)$ vertices would require a cut-set size of $d$ or more edges. The precise lower bound for $|S| < (d+1)$ can be determined based on the observation that each vertex in $\DEG_d$ has more connections than the number of available vertices in $S$. 

Let $S \subset V$ be the smaller vertex set of a cut with $|S| \leq d$ and $G[S]$ its vertex induced subgraph. Furthermore let all vertices of the set be adjacent to each other in order to create a complete subgraph while also reducing the number of edges in the cut-set $B = E \setminus \{L_1 \cup L_2\}$. The number of undirected edges in the complete subgraph $G[S]$ is $|L| = \frac{|S|}{2}(|S|-1)$. Since each vertex in $S$ has $d$ edges in $\DEG_d$ but can only be adjacent to $|S|-1$ vertices in the subgraph, the remaining edges must cross the cut. A lower bound for the cut-set size of such a subgraph is given by:
\begin{equation} 
\begin{split}
\label{eq:cutsetSize}
|S| < (d+1) : |B| &\geq 2\left(\frac{|S| \cdot d}{2} - \frac{|S|}{2}(|S|-1)\right) \\
&= |S| \cdot (d+1) - |S|^2 \geq d
\end{split}
\end{equation}

\begin{figure}
\centering
\includegraphics[trim=1cm 10cm 10cm 1cm,clip=true,width=\linewidth]{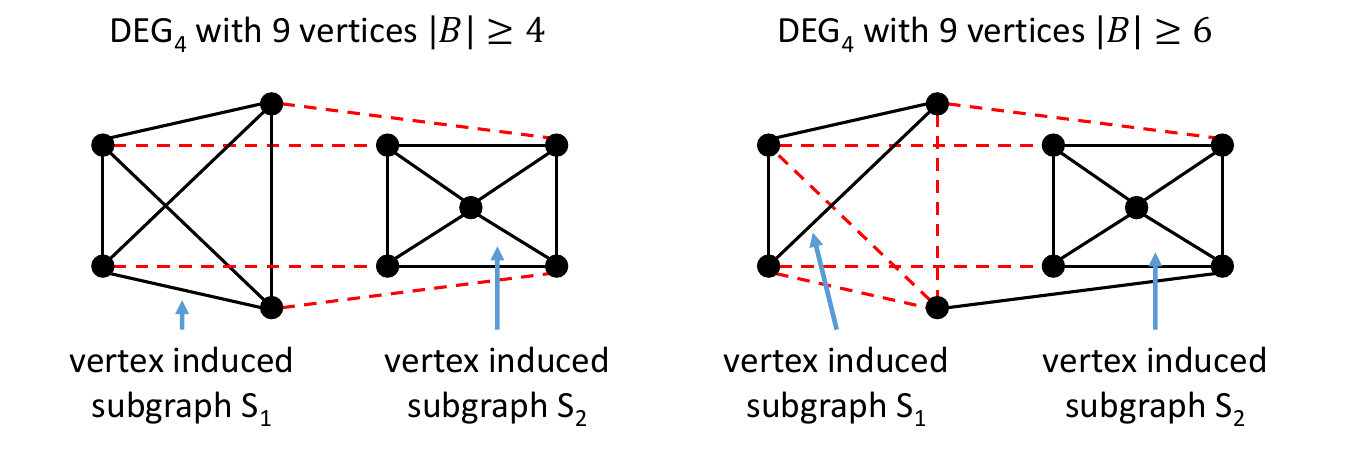}
\captionsetup{font=small}
\caption{Example of a $DEG_4$. Left: A vertex induced subgraph $S_1$ with four vertices. The remaining vertices form the vertex induced subgraph $S_2$ and the four crossing edges are the cut-set $B$. Right: the same example with a smaller $S_1$ and a larger cut-set size of six edges.} \label{fig:deg_cutset}
\end{figure}

Any vertex-induced subgraph of $\DEG_d$ with less than $(d+1)$ vertices has a cut-set size of $d$ or more edges and the resulting graph components of the cut will no longer be d-regular (see Figure \ref{fig:deg_cutset}). 
\\
\\
This inherent characteristic of $\DEG$ makes it quite resilient against disconnections of small components. 
Cutting off a component with 15 vertices of a $\DEG_{30}$ requires at least 240 specific edge removals, which is very unlikely. Nevertheless, our algorithms for adding vertices and updating edges thoroughly consider this minor possibility and perform corresponding checks. When the graph disconnects, the checks involved become more computationally intensive. Hence, a low probability of disconnections remains advantageous.

\section{DEG is approximating a DG and RNG}
\label{appendix:approximatingDGandRNG}

In a comprehensive study conducted in 2021 \cite{Wang2021Survey}, most state-of-the-art graph algorithms are based on an approximation of the Delaunay Graph (DG) or Relative Neighborhood Graph (RNG). 
The following section describes the similarities in the construction strategies of DEG and HNSW, as well as its predecessor, NSW. 
During the construction process, additional constraints are imposed on the edges to approximate a Monotonic Relative Neighborhood Graph, which increases the search efficiency.
\\
\\
\textbf{Construction:} 
Generally, DEG employs a similar approach to NSW \cite{Malkov2014} for constructing a graph. 
Both incrementally add new vertices to an undirected graph while ensuring global connectivity. 
When connecting a new vertex to the best possible vertex in the graph, the resulting edge is guaranteed to exist in a Delaunay graph. 
Since the greedy and range search algorithms only provide an approximation of the best search results, NSW as well as DEG just approximate a Delaunay graph.
While NSW does limit the amount of edges, its successor HNSW \cite{Malkov2020} imposes an upper bound on the number of incoming directed edges per vertex and introduces a hierarchy to store edges of different lengths at different levels. 
DEG on the other hand does not use hierarchies and integrates new vertices by replacing existing undirected edges with two new undirected edges.
In conjunction with its regularity, the resulting edge distribution is very similar to the edge distribution of HNSW, containing short edges for high search accuracy and some longer edges for fast navigation to other graph regions. 
This approximation of RNG \cite{Wang2021Survey} provides a similar search complexity as HNSW, as demonstrated in Section \ref{sec:scalabilityAndComplexity}.

\textbf{Edge constrains:}
DG and RNG restrict the number of valid edges \cite{Jaromczyk1992}, by enforcing an empty space between two adjacent vertices.
In the case of DG, two vertices share an edge if and only if the smallest possible open ball enclosing the vertices contains no other vertex \cite{ORourke1982}.
RNG, on the other hand, utilizes a lune comprised of the intersection of two open balls to increase the restricted area and further reduce the number of allowed edges. 
The emptiness of the lune between vertex $v_1$ and $v_2$ can be verified by Equation \ref{eq:rng}.

\begin{equation}
\label{eq:rng}
\delta(v_1, v_2) \stackrel{!}{<} \max_{u \in V} \left( \max(\delta(v_1, u), \delta(v_2, u)) \right)
\end{equation}

The naive construction of RNG requires repeating the test in Equation \ref{eq:rng} for all combinations of $v_1$ and $v_2$ in $V$, resulting in a complexity of $O(|V|^3)$. 
As the quality of all graph manipulating operations presented in this paper is evaluated using the average neighbor distance ($\AND$), which only takes into account the neighbors of a vertex, we approximate RNG by only checking the neighborhoods of the involved vertices. 
In addition, DEG stores the weights of the edges during the construction phase, eliminating the need to recalculate most distances of Equation \ref{eq:rng}. The approximation of RNG is expressed by only connecting $v_1$ and $v_2$ if
Equation \ref{eq:rng_approximiation} is true.

\begin{equation}
\label{eq:rng_approximiation}
\delta(v_1, v_2) \stackrel{!}{<} \max_{u \in (N(G, v_1) \cap N(G, v_2)) }({\max(w(v_1, u), w(v_2, u))})
\end{equation}

With this optimization, the complexity is reduced to $O(|V|^2 \cdot d)$, where $d$ is the number of edges per vertex. 
Checking all combinations of $v_1$ and $v_2$ when constructing DEG is not necessary. Since the main objective is to minimize the average neighbor distance within a given neighbor limit per vertex, the algorithm can prioritize similar vertex pairs identified through an ANN search.
Consequently, Equation \ref{eq:rng_approximiation} (implemented in Algorithm \ref{alg:checkMRNG}) is primarily a tool for verifying the RNG compliance of potential neighbors, with a complexity of $O(|V| \cdot k_{\extend} \cdot d)$, where $k_{\extend}$ corresponds to the $k$ parameter of the ANNS.

Applying the checks only to the existing edges of the graph also makes this an approximation of a Monotonic Relative Neighborhood Graph (MRNG). 
A proof is provided in appendix \ref{appendix:approximatingMRNG}. Although these tests are not essential for constructing DEG, they can improve the search quality as shown in Appendix \ref{appendix:neighborChoices}. 

\section{DEG's approximation of RNG is an approximation of MRNG}
\label{appendix:approximatingMRNG}

The paper \cite{Jaromczyk1992} explores the geometric aspects of the edges in a \textit{Relative Neighborhood Graph} (RNG). To define these edges, several concepts are introduced. First, let $B(u, r) = \{x \in \mathbb{R}^m : \delta(u, x) < r\}$ represents an open sphere centered at point $u$ with a radius of $r$. Additionally, let $\lune_{uv}$ be the region where two spheres intersect, given by $\lune_{uv} = B(u, \delta(u, v)) \cap B(v, \delta(v, u))$. 
Based on these definitions, the edges of RNG can be described as follows: $(u,v) \in E$ if and only if $\lune_{uv} \cap V = \emptyset$.

For DEG, this rule was modified in Section \ref{appendix:approximatingDGandRNG} and does no longer consider all vertices in $V$ to exclude some edges, but only those that are neighbors of $u$ or $v$. The set of edges in DEG can therefore be defined as $(u,v) \in E$ if and only if $\lune_{uv} \cap N(G,u) \cap N(G,v) = \emptyset$ with $N(G, x)$ being the set of adjacent vertices to $x$. As the addition of new edges alters the neighborhoods of both $u$ and $v$, the order of operations matters and makes the resulting graph just an approximation of RNG. For DEG the edge checks are exclusively performed when a new vertex $v$ is added using Algorithm \ref{alg:extendGraph} and the neighbors candidates $u \in U$ are acquired with the help of an approximate nearest neighbor search. An evaluation sequence is established by sorting the candidates in descending order of similarity to $v$.

The authors of NSG \cite{Cong2019} define a \textit{Monotonic Relative Neighborhood Graph} as follows: Let $G = \MRNG(V, E)$ be a directed graph and its edge set $(u,v) \in E$ if and only if $\lune_{uv} \cap V = \emptyset$ or $\forall q \in (\lune_{uv} \cap V), (u,q) \notin E$. They also specify an order in which edges are checked, which is consistent to the DEG approach.
We argue that if the vertices of MRNG are sequentially added and are not initially part of set $V$, the second part of the condition can be reformulate as $\forall q \in (\lune_{uv} \cap V), q \notin N(G,u)$.
Conversely, this equation can be inverted by considering all neighbors of $u$, not just those within $(\lune_{uv} \cap V)$. In this case, the formulation becomes $\forall q \in N(G,u), q \notin (\lune_{uv} \cap V)$.
In conjunction with the first part of the initial condition, the whole check can be simplified to: $(u,v) \in E$ if and only if $lune_{uv} \cap N(G,u) = \emptyset$.
This representation is very similar to DEG's approximation of RNG. The key difference lies in the direction of the edges. MRNG only needs to examine its edges in one direction, whereas DEG, with its undirected edges, must consider the neighborhoods of both vertices of an edge. DEG's approximation of RNG can be viewed as a specialized approximation of MRNG designed specifically for undirected graphs.

\section{Parameters of compared methods}
\label{appendix:parameters}

The graphs in the experiments of Section (\ref{sec:searchExperiments}) are created using the indexing parameters provided by the original authors. If those parameters were not available, they were obtained from the Github repository \cite{GithubWEAVESS} of the latest survey \cite{Wang2021Survey} on graph-based approximated nearest neighbor search. In this survey, optimal parameters were determined through a grid search. The Tables (\ref{tab:kgraph_parameters}, \ref{tab:dpg_parameters}, \ref{tab:efanna_parameters}, \ref{tab:nsg_parameters}, \ref{tab:ssg_parameters}, \ref{tab:hnsw_parameters}, \ref{tab:onng_parameters}) are included for the sake of completeness and reproducibility. For detailed explanations of the parameter symbols, please refer to the respective sources.

\begin{table}
    \centering
    \begin{tabularx}{\linewidth} { 
   >{\centering\arraybackslash}l  
   >{\centering\arraybackslash}X  
   >{\centering\arraybackslash}X   
   >{\centering\arraybackslash}X  
   >{\centering\arraybackslash}X  
   >{\centering\arraybackslash}X  
   >{\centering\arraybackslash}r   }
        \hline
        \textbf{dataset} & 
        \textbf{source} & 
        \textbf{K} & 
        \textbf{L} & 
        \textbf{iter} & 
        \textbf{S} & 
        \textbf{R}  \\ 
        \hline
        Audio & \cite{GithubWEAVESS} & 40 & 60 & 5 & 20 & 100\\
        Enron & \cite{GithubWEAVESS} & 50 & 80 & 7 & 15 & 100\\ 
        SIFT1M & \cite{GithubWEAVESS} & 90 & 130 & 12 & 20 & 50\\
        GloVe & \cite{GithubWEAVESS} & 100 & 150 & 12 & 35 & 150\\
        \hline
    \end{tabularx}
    \caption{Parameters of kgraph for various datasets.}
    \label{tab:kgraph_parameters}
\end{table}

\begin{table}
    \centering
    \setlength\tabcolsep{6pt} 
    \begin{tabularx}{\linewidth} { 
   >{\centering\arraybackslash}l  
   >{\centering\arraybackslash}X  |
   >{\centering\arraybackslash}c   
   >{\centering\arraybackslash}c  
   >{\centering\arraybackslash}c  
   >{\centering\arraybackslash}c  
   >{\centering\arraybackslash}c  |
   >{\centering\arraybackslash}c   }
        \hline
        \multicolumn{2}{c}{} & 
        \multicolumn{5}{c}{\textbf{kGraph}} & 
        \multicolumn{1}{c}{\textbf{DPG}} \\
        \hline
        \textbf{dataset} & 
        \textbf{source} & 
        \textbf{K} & 
        \textbf{L} & 
        \textbf{iter} & 
        \textbf{S} & 
        \textbf{R} & 
        \textbf{L} \\ 
        \hline
        Audio & \cite{GithubWEAVESS} & 100 & 130 & 5 & 25 & 100 & 25*\\
        Enron & \cite{GithubWEAVESS} & 100 & 120 & 7 & 15 & 200 & 60\\ 
        SIFT1M & \cite{GithubWEAVESS} & 100 & 100 & 12 & 25 & 300 & 30*\\
        GloVe & \cite{GithubWEAVESS} & 100 & 130 & 12 & 20 & 100 & 50\\
        \hline
    \end{tabularx}
    \caption{Parameters of DPG for various datasets. DPG's L parameters marked with * are lowered to get reading across the 95-100\% accuracy range, its default value is L=K/2.}
    \label{tab:dpg_parameters}
\end{table}

\begin{table}
    \centering
    \setlength\tabcolsep{3pt} 
    \begin{tabularx}{\linewidth} { 
   >{\centering\arraybackslash}l  
   >{\centering\arraybackslash}X  |
   >{\centering\arraybackslash}c   
   >{\centering\arraybackslash}c  |  
   >{\centering\arraybackslash}c  
   >{\centering\arraybackslash}c  
   >{\centering\arraybackslash}c  
   >{\centering\arraybackslash}c  
   >{\centering\arraybackslash}c   }
        \hline
        \multicolumn{2}{c}{} & 
        \multicolumn{2}{c}{\textbf{kdtree}} & 
        \multicolumn{5}{c}{\textbf{EFANNA}} \\
        \hline
        \textbf{dataset} & 
        \textbf{source} & 
        \textbf{nTrees} & 
        \textbf{mLevel} & 
        \textbf{K} & 
        \textbf{L} & 
        \textbf{iter} & 
        \textbf{S} & 
        \textbf{R} \\ 
        \hline
        Audio & \cite{GithubWEAVESS} & 16 & 8 & 40 & 40 & 10 & 30 & 100\\
        Enron & \cite{GithubWEAVESS} & 4 & 8 & 40 & 140 & 5 & 35 & 150\\ 
        SIFT1M & \cite{GithubEFANNA} & - & - & 50 & 70 & 10 & 10 & 50 \\
        GloVe & \cite{GithubSSG} & - & - & 400 & 420 & 12 & 15 & 200\\
        \hline
    \end{tabularx}
    \caption{Parameters of EFANNA for various datasets. \cite{GithubWEAVESS} uses kdtrees to initialize kGraph. }
    \label{tab:efanna_parameters}
\end{table}

\begin{table}
    \centering
    \setlength\tabcolsep{4pt} 
    \begin{tabularx}{\linewidth} { 
   >{\centering\arraybackslash}l  
   >{\centering\arraybackslash}X  |
   >{\centering\arraybackslash}c   
   >{\centering\arraybackslash}c    
   >{\centering\arraybackslash}c  
   >{\centering\arraybackslash}c  
   >{\centering\arraybackslash}c  |
   >{\centering\arraybackslash}c      
   >{\centering\arraybackslash}c  
   >{\centering\arraybackslash}c   }
        \hline
        \multicolumn{2}{c}{} &         
        \multicolumn{5}{c}{\textbf{EFANNA}} &
        \multicolumn{3}{c}{\textbf{NSG}} \\
        \hline
        \textbf{dataset} & 
        \textbf{source} & 
        \textbf{K} & 
        \textbf{L} & 
        \textbf{iter} & 
        \textbf{S} & 
        \textbf{R} &
        \textbf{L} & 
        \textbf{R} & 
        \textbf{C} \\ 
        \hline
        Audio & \cite{GithubWEAVESS} & 200 & 230 & 5 & 10 & 100 & 200 & 30 & 600\\
        Enron & \cite{GithubWEAVESS} & 200 & 200 & 7 & 25 & 200 & 150 & 60 & 600\\ 
        SIFT1M & \cite{GithubNSG} & 200 & 200 & 10 & 10 & 100 & 40 & 50 & 500 \\
        GloVe & \cite{Cong2021} & 400 & 420 & 12 & 15 & 200 & 50 & 70 & 500\\
        \hline
    \end{tabularx}
    \caption{Parameters of NSG for various datasets. For GloVe the EFANNA parameters of NSSG \cite{GithubSSG} are used.}
    \label{tab:nsg_parameters}
\end{table}

\begin{table}
    \centering
    \setlength\tabcolsep{4pt} 
    \begin{tabularx}{\linewidth} { 
   >{\centering\arraybackslash}l  
   >{\centering\arraybackslash}X  |
   >{\centering\arraybackslash}c   
   >{\centering\arraybackslash}c    
   >{\centering\arraybackslash}c  
   >{\centering\arraybackslash}c  
   >{\centering\arraybackslash}c  |
   >{\centering\arraybackslash}c      
   >{\centering\arraybackslash}c  
   >{\centering\arraybackslash}c   }
        \hline
        \multicolumn{2}{c}{} &         
        \multicolumn{5}{c}{\textbf{EFANNA}} &
        \multicolumn{3}{c}{\textbf{NSSG}} \\
        \hline
        \textbf{dataset} & 
        \textbf{source} & 
        \textbf{K} & 
        \textbf{L} & 
        \textbf{iter} & 
        \textbf{S} & 
        \textbf{R} &
        \textbf{L} & 
        \textbf{R} & 
        \textbf{C} \\ 
        \hline
        Audio & \cite{GithubWEAVESS} & 400 & 400 & 5 & 25 & 200 & 50 & 20 & 60\\
        Enron & \cite{GithubWEAVESS} & 100 & 110 & 7 & 20 & 300 & 300 & 30 & 60\\ 
        SIFT1M & \cite{GithubSSG} & 200 & 200 & 12 & 10 & 100 & 100 & 50 & 60 \\
        GloVe & \cite{GithubSSG} & 400 & 420 & 12 & 15 & 200 & 500 & 50 & 60\\
        \hline
    \end{tabularx}
    \caption{Parameters of NSSG for various datasets.}
    \label{tab:ssg_parameters}
\end{table}

\begin{table}
    \centering
    \begin{tabularx}{\linewidth} { 
   >{\centering\arraybackslash}l  
   >{\centering\arraybackslash}X  
   >{\centering\arraybackslash}X  
   >{\centering\arraybackslash}X  
   >{\centering\arraybackslash}r   }
        \hline
        \textbf{Dataset} & 
        \textbf{source} & 
        \textbf{max\_m} & 
        \textbf{max\_m0} & 
        \textbf{ef\_const}  \\ 
        \hline
        Audio & \cite{GithubWEAVESS} & 10 & 50 & 700\\
        Enron & \cite{GithubWEAVESS} & 50 & 80 & 900\\ 
        SIFT1M & \cite{GithubWEAVESS} & 40 & 50 & 800\\
        GloVe & \cite{GithubWEAVESS} & 50 & 60 & 700\\
        \hline
    \end{tabularx}
    \caption{Parameters of HNSW for various datasets.}
    \label{tab:hnsw_parameters}
\end{table}

\begin{table}
    \centering
    \setlength\tabcolsep{4pt} 
    \begin{tabularx}{\linewidth} { 
   >{\centering\arraybackslash}l  
   >{\centering\arraybackslash}X  |
   >{\centering\arraybackslash}X   
   >{\centering\arraybackslash}X   | 
   >{\centering\arraybackslash}X  
   >{\centering\arraybackslash}X   }
        \hline
        \multicolumn{2}{c}{} &         
        \multicolumn{2}{c}{\textbf{ANNG}} &
        \multicolumn{2}{c}{\textbf{ONNG}} \\
        \hline
        \textbf{dataset} & 
        \textbf{source} & 
        $\bm{K_c}$ & 
        $L$ & 
        $\bm{e_o}$ & 
        $\bm{e_i}$ \\ 
        \hline
        Audio & \cite{GithubWEAVESS} & 200 & 200 & 40 & 100 \\
        Enron & \cite{GithubWEAVESS} & 200 & 200 & 20 & 100 \\ 
        SIFT1M & \cite{Iwasaki2018} & 200 & 200 & 30 & 110 \\
        GloVe & \cite{Iwasaki2018} & 200 & 200 & 15 & 155 \\
        \hline
    \end{tabularx}
    \caption{Parameters of ONNG for various datasets.}
    \label{tab:onng_parameters}
\end{table}

\section{Graph Statistics}
\label{appendix:graphStatistics}

In order to investigate why some graphs outperform others in certain tests, several statistical graph properties are measured and documented in Table \ref{tab:exploration_properties}. All graphs are created using the parameters specified in Appendix \ref{appendix:parameters} and Section \ref{sec:parameters} for the four datasets described in Section \ref{sec:datasets}.

Among the calculated metrics is the \textit{graph quality} \cite{Boutet2016}, which assists in detecting k-NN graphs and assessing the tendency of other graphs to primarily connect their most similar vertices. A high \textit{graph quality} implies that the majority of edges in the graph are as short as possible, making it challenging to navigate to different regions of the graph. In combination with a high average degree the performance deteriorates even further.

While vertices with a large number of outgoing edges \textit{max out} can facilitate exploration, they can also slow down navigation. Conversely, vertices with a low number of outgoing edges \textit{min out} can be quickly examined but offer limited new information. 
An optimal graph strikes a balance by incorporating an appropriate number of short edges connecting nearest neighbors for efficient local exploration, as well as longer edges for effective navigation across the graph.

Based on the \textit{min out} value of 1 for NSG and NSSG in Table \ref{tab:exploration_properties}, it can be concluded that these graphs had more than one strongly connected component during their construction and were subsequently reconnected using their VP tree.

Reaching a vertex becomes difficult when it has only a small number of incoming edges (\textit{min in}). In contrast, having a large number of incoming edges (\textit{max in}) can create hub vertices, which may not be relevant for every search task but are frequently visited due to multiple paths leading to them. If these vertices also possess numerous outgoing edges, they can unnecessarily slow down the search process.

\textit{Source vertices} with no incoming edges are even more problematic as they affect the search- and exploration reachability. The former measures the percentage of vertices reachable during a search, while the latter calculates the average number of vertices accessible from each vertex. The number of not reachable \textit{source vertices} is represented by the "source count" value in Table \ref{tab:exploration_properties}.

In order to measure the search reachability in HNSW, we conduct an approximate nearest neighbor (ANN)  search on the upper layers to identify a suitable entry point into the base layer. Then a breadth-first search was used to reach a certain vertex within the base layer. This process is repeated for every vertex in the graph and the resulting percentage of reachable vertices is recorded as "search reach" in Table \ref{tab:exploration_properties}. To calculate the "explore reach", the breadth-first search begins directly on the base layer and counts the average number of reachable vertices.

\begin{table*}
    \centering
    \begin{tabularx}{\textwidth} { 
   >{\centering\arraybackslash}c  |
   >{\centering\arraybackslash}X   
   >{\raggedleft\arraybackslash}X  
   >{\raggedleft\arraybackslash}X   
   >{\raggedleft\arraybackslash}X  
   >{\raggedleft\arraybackslash}X   
   >{\raggedleft\arraybackslash}X   
   >{\raggedleft\arraybackslash}X  
   >{\raggedleft\arraybackslash}X   
   >{\raggedleft\arraybackslash}X  
   >{\raggedleft\arraybackslash}X  }
        \hline
        \textbf{Dataset} & 
        \textbf{Algorithm} & 
        \textbf{graph quality} & 
        \textbf{avg degree} & 
        \textbf{min out} & 
        \textbf{max out} & 
        \textbf{min in} & 
        \textbf{max in} & 
        \textbf{source count} & 
        \textbf{search reach \%} & 
        \textbf{explore reach \%} \\ 
        \hline
        \multirow{7}{*}{Audio} & DEG & 0.48 & 20.0 & 20 & 20 & 20 & 20 & 0 & 100.00 & 100.00 \\
                               & HNSW & 0.40 & 16.9 & 1 & 50 & 1 & 50 & 0 & 100.00 & 100.00 \\ 
                               & kGraph & 1.00 & 32.7 & 20 & 47 & 0 & 332 & 754 & 98.34 & 98.33 \\
                               & DPG & 0.68 & 30.6 & 20 & 158 & 20 & 158 & 0 & 100.00 & 100.00 \\
                               & EFANNA & 1.00 & 40.0 & 40 & 40 & 0 & 421 & 440 & 99.04 & 99.04 \\
                               & NSG & 0.41 & 17.7 & 1 & 31 & 1 & 51 & 0 & 100.00 & 100.00 \\
                               & NSSG & 0.37 & 18.6 & 1 & 20 & 1 & 54 & 0 & 100.00 & 100.00 \\
                               & ONNG & 0.37 & 40.3 & 10 & 326 & 18 & 246 & 0 & 100.00 & 100.00 \\
        \hline
        \multirow{7}{*}{Enron} & DEG & 0.36 & 30.0 & 30 & 30 & 30 & 30 & 0 & 100.00 & 100.00 \\
                               & HNSW & 0.33 & 21.9 & 1 & 80 & 0 & 243 & 34 & 99.96 & 99.96 \\ 
                               & kGraph & 0.99 & 39.3 & 15 & 78 & 0 & 154 & 3351 & 90.92 & 87.95 \\
                               & DPG & 0.62 & 99.8 & 60 & 1366 & 60 & 1366 & 0 & 100.00 & 100.00 \\
                               & EFANNA & 1.00 & 40.0 & 40 & 40 & 0 & 1312 & 2829 & 91.97 & 89.01 \\
                               & NSG & 0.36 & 15.3 & 1 & 62 & 1 & 127 & 0 & 100.00 & 100.00 \\
                               & NSSG & 0.37 & 18.7 & 1 & 30 & 1 & 154 & 0 & 100.00 & 99.96 \\
                               & ONNG & 0.31 & 45.4 & 5 & 787 & 12 & 511 & 0 & 100.00 & 100.00 \\
        \hline
        \multirow{7}{*}{SIFT1M} & DEG & 0.49 & 30.0 & 30 & 30 & 30 & 30 & 0 & 100.00 & 100.00 \\
                                & HNSW & 0.37 & 32.8 & 1 & 50 & 1 & 151 & 0 & 100.00 & 100.00 \\ 
                                & kGraph & 1.00 & 77.2 & 20 & 150 & 0 & 1159 & 437 & 99.95 & 99.95 \\
                                & DPG & 0.65 & 45.8 & 30 & 248 & 30 & 248 & 0 & 100.00 & 100.00 \\
                                & EFANNA & 0.97 & 50.0 & 50 & 50 & 0 & 694 & 1688 & 99.79 & 99.79 \\
                                & NSG & 0.39 & 29.2 & 1 & 50 & 1 & 142 & 0 & 100.00 & 100.00 \\
                                & NSSG & 0.42 & 39.2 & 3 & 50 & 3 & 179 & 0 & 100.00 & 100.00 \\
                                & ONNG & 0.40 & 50.4 & 10 & 516 & 17 & 248 & 0 & 100.00 & 100.00 \\
        \hline
        \multirow{7}{*}{GloVe} & DEG & 0.21 & 30.0 & 30 & 30 & 30 & 30 & 0 & 100.00 & 100.00 \\
                               & HNSW & 0.29 & 14.8 & 1 & 60 & 0 & 2161 & 41294 & 96.46 & 96.46 \\ 
                               & kGraph & 0.92 & 124.6 & 35 & 150 & 0 & 149251 & 138317 & 77.61 & 77.61 \\
                               & DPG & 0.49 & 92.7 & 50 & 63943 & 50 & 63943 & 0 & 100.00 & 100.00 \\
                               & EFANNA & 0.82 & 400.0 & 400 & 400 & 0 & 418292 & 90588 & 84.83 & 84.83 \\
                               & NSG & 0.33 & 13.0 & 1 & 150 & 1 & 1008 & 0 & 100.00 & 100.00 \\
                               & NSSG & 0.29 & 21.2 & 1 & 50 & 1 & 1134 & 0 & 100.00 & 100.00 \\
                               & ONNG & 0.16 & 81.9 & 4 & 24541 & 23 & 16017 & 0 & 100.00 & 100.00 \\
        \hline
    \end{tabularx}
    \vspace{0.2cm}
    \caption{The table presents details on different attributes of the graphs used in the search and exploration tasks. It includes the \textit{graph quality}, the average degree, as well as the minimum and maximum out- and in-degrees. Additionally, the number of \textit{source vertices} is provided, which refers to all vertices with no incoming edges. The reachability values are expressed as a ratio to the total number of vertices.}
    \label{tab:exploration_properties}
\end{table*}

\section{Neighbor choice} \label{appendix:neighborChoices}
\begin{figure}
    \begin{subfigure}{0.49\columnwidth}
        \includegraphics[trim=2.0cm 5cm 2.0cm 7cm,clip=true,width=\columnwidth]{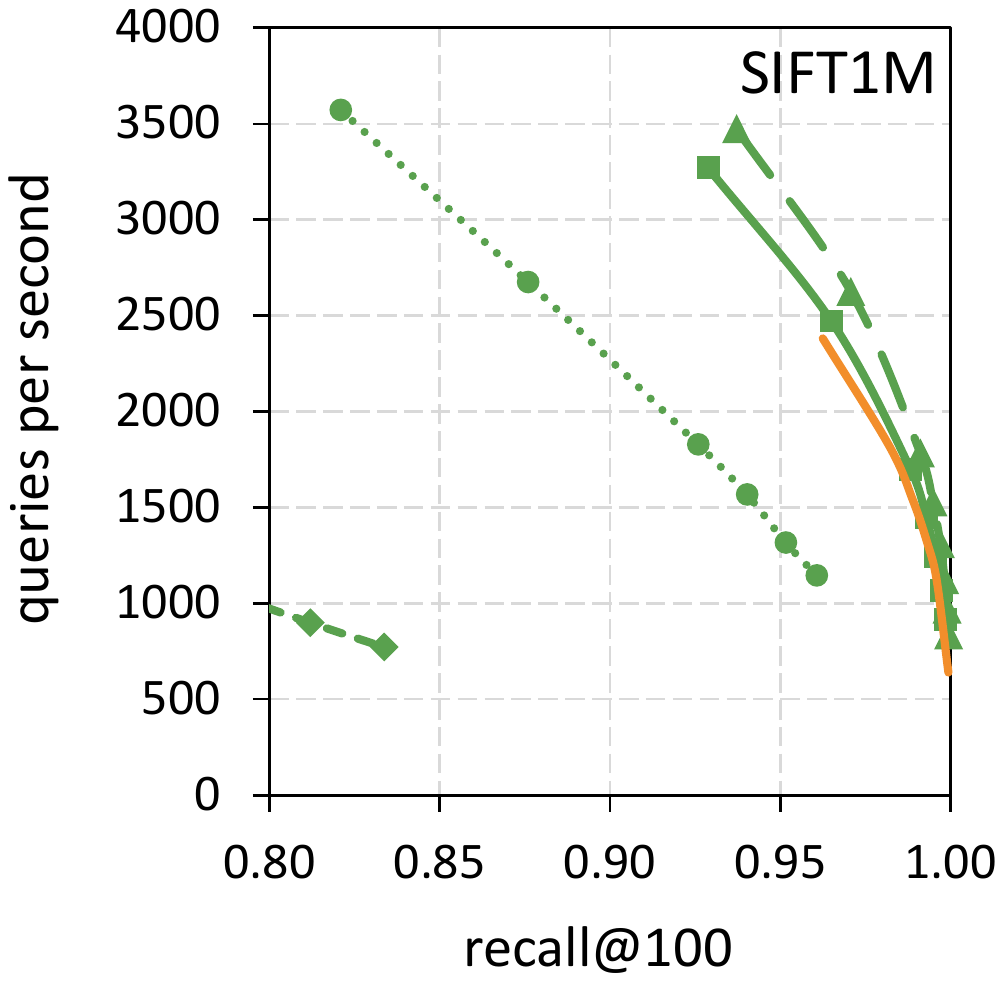}
    \end{subfigure}
    \hfill
    \begin{subfigure}{0.49\columnwidth}
        \includegraphics[trim=2.0cm 5cm 2.0cm 7cm,clip=true,width=\columnwidth]{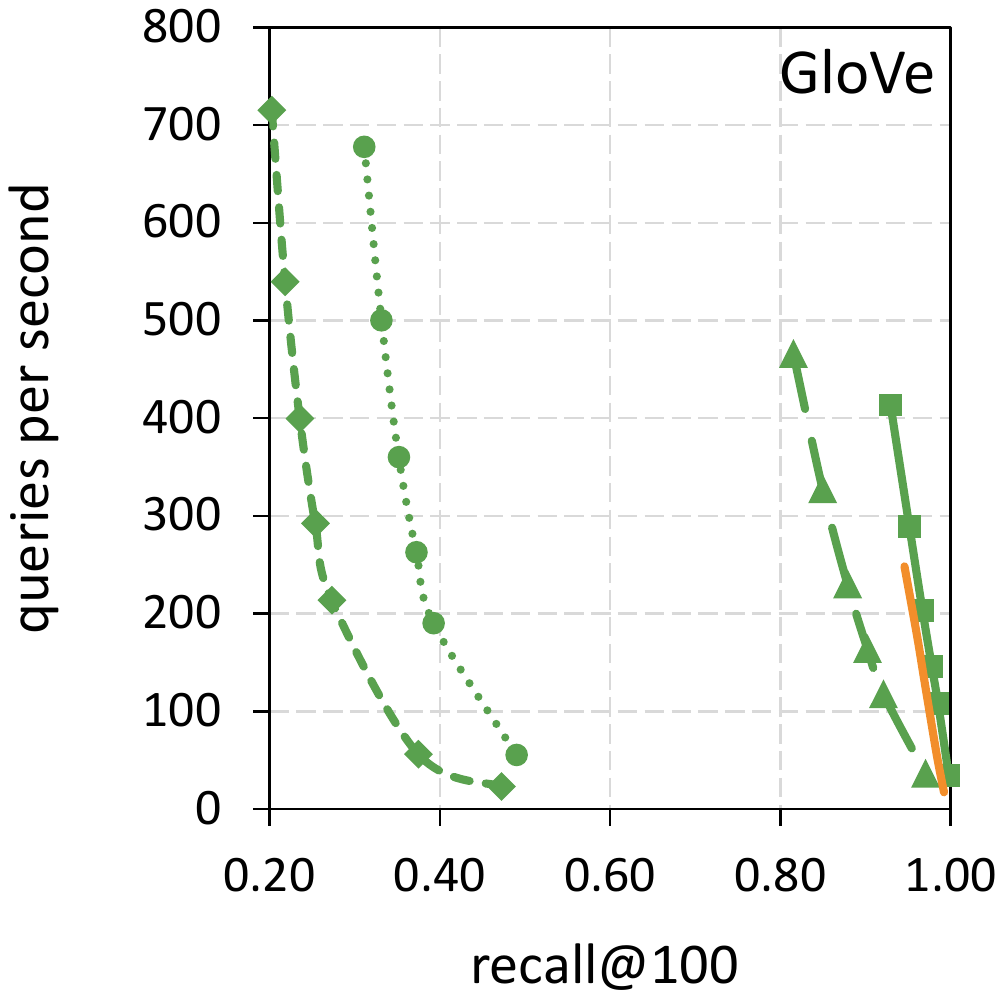}
    \end{subfigure}
  
    \begin{subfigure}{1\columnwidth}
        \includegraphics[trim=1.5cm 13.85cm 1.5cm 15.3cm,clip=true,width=\columnwidth]{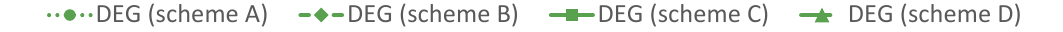}
    \end{subfigure}
    \caption{Constructing DEG without any edge optimization, yields competitive search performance and fast indexing speed. HNSW (in orange) was added for reference.}
    \label{fig:build_only_performance}
\end{figure}

\begin{figure}[ht!]
    \begin{subfigure}{0.49\columnwidth}
        \includegraphics[trim=2.0cm 5cm 2.0cm 7.1cm,clip=true,width=\columnwidth]{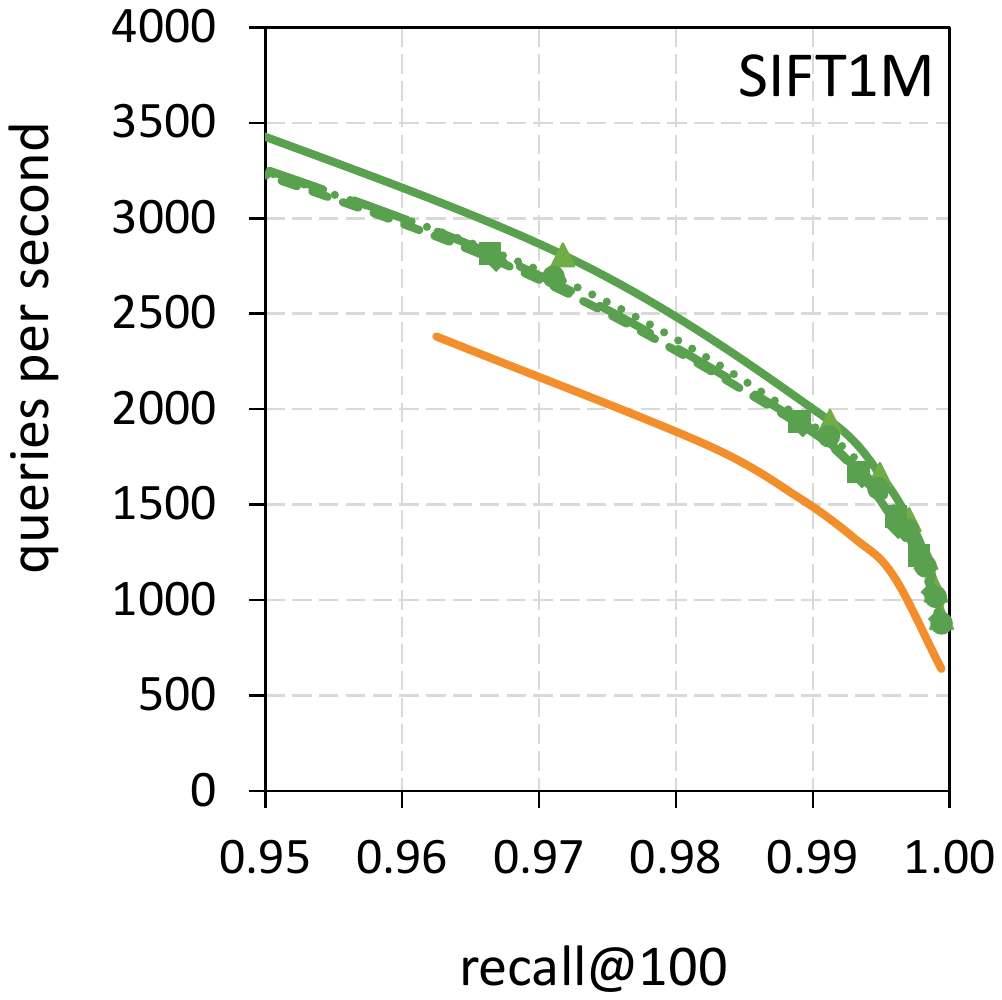}
    \end{subfigure}
    \hfill
    \begin{subfigure}{0.49\columnwidth}
        \includegraphics[trim=2.0cm 5cm 2.0cm 7.1cm,clip=true,width=\columnwidth]{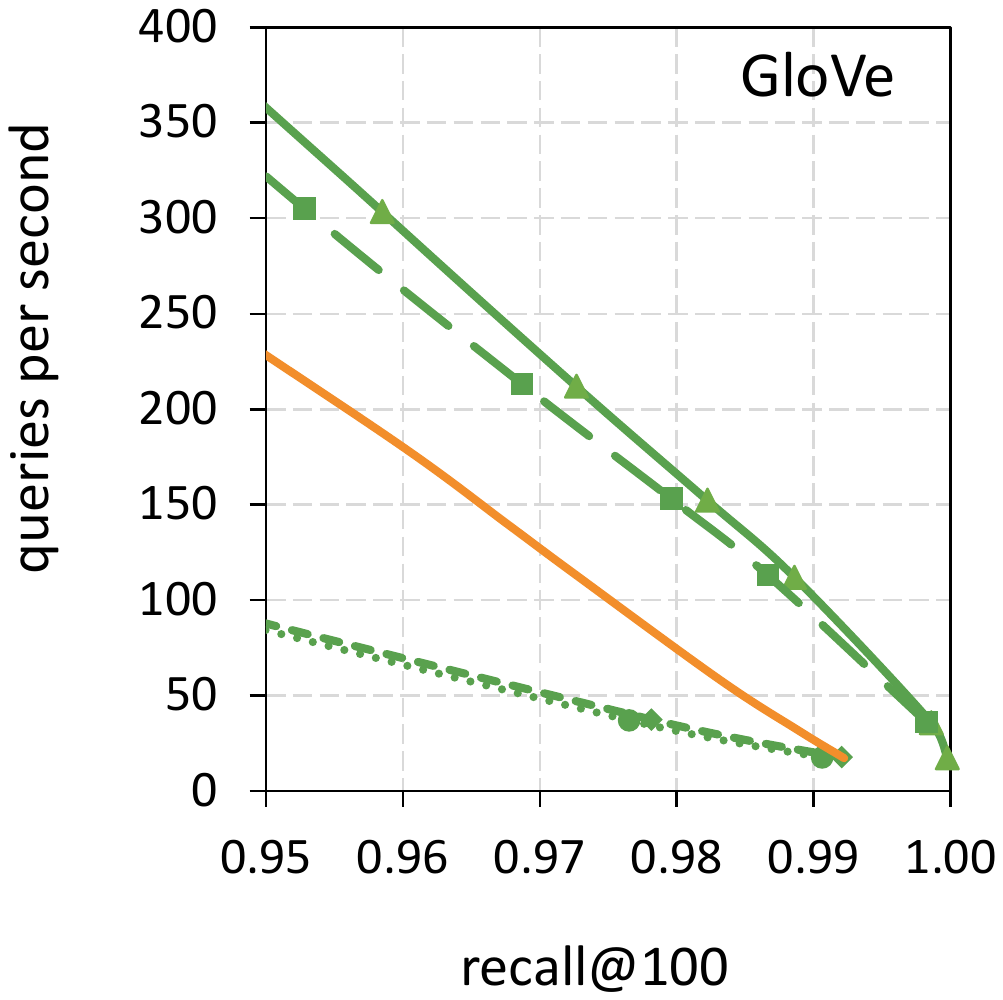}
    \end{subfigure}
  
    \begin{subfigure}{1\columnwidth}
        \includegraphics[trim=2cm 13.85cm 2cm 15.3cm,clip=true,width=\columnwidth]{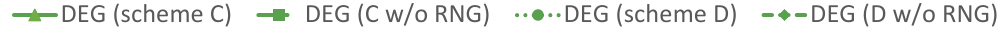}
    \end{subfigure}
    \caption{Extending DEG with scheme C or D from Section \ref{sec:incrementalConstruction} and performing edge optimization with scheme D has proven to be effective across all datasets. Adding RNG-checks (Algorithm \ref{alg:checkMRNG}) consistently yields slightly better results. HNSW (in orange) was added for reference.}
    \label{fig:rng_check_performance}
\end{figure}
The graph extension in Algorithm \ref{alg:extendGraph} of DEG substitutes existing edges with two new ones to integrate new vertices into the graph. Numerous options are available for selecting these edges. Figure \ref{fig:add_vertex} shows four variations, all focused on decreasing the average neighbor distance. The subsequent investigation aims to identify the best scheme in order to achieve the fastest search times. To assess the effects in varying data distributions, all experiments are conducted for both the SIFT1M and GloVe dataset. 
Four DEG graphs are constructed for each dataset, following the steps of Algorithm \ref{alg:extendGraph}. 
The edge selection is determined by scheme A, B, C or D as described in Section \ref{sec:incrementalConstruction}, and no edge optimization is applied to avoid distorting the results. Consequently, the last line in Algorithm \ref{alg:extendGraph} is ignored. 
The parameters used are consistent with those of the other experiments listed in Table \ref{tab:deg_parameters}. Following the graph construction, the same queries as in Section \ref{sec:searchExperiments} were used to evaluate search time and quality. 

The time required to build the graph indices varies only slightly between the schemes. The average times are 12 min for SIFT and 30 min for the GloVe dataset.
This is much faster than HNSW (35 min for SIFT1M, 54 min for Glove) which is included in the plots for reference. The resulting search speed highly depends on the chosen build scheme, as illustrated by Figure \ref{fig:build_only_performance}. It is apparent that datasets with low local intrinsic dimensions (LID) achieve better results using scheme D, while those with high LID benefit more from scheme C. The performance curves for the Audio dataset are similar to those for SIFT1M, while those for Enron are comparable to GloVe. Scheme C creates more short edges than D, but D has a lower average neighbor distance. This indicates that in high LID domains, having a few very short edges is more important than reducing the length of all edges.

However, this behavior changes if the edges of a new vertex get further optimized using the edge-swapping Algorithm \ref{alg:optimizeEdge} (reactivating the last line in Algorithm \ref{alg:extendGraph}). The idea of different decision-making schemes during graph extension can also be applied to the edge optimization. In a series of experiments, which were omitted for space reason, \textit{optimizing edges} according to scheme D produced the best results. In order to evaluate the performance of combining the graph extension using either scheme C or D together with the edge optimization, tests were conducted and presented in Figure \ref{fig:rng_check_performance}. Although scheme C was inferior to scheme D during the construction-only tests of SIFT1M, it outperformed scheme D in the construction and optimization tests. Additional tests were carried out to evaluate the impact of RNG checks, which, as discussed in Section \ref{sec:incrementalConstruction}, are not mandatory but can improve search performance in large datasets, as shown in Figure \ref{fig:rng_check_performance}. Following these finding, DEG's default setting used in all other experiments is: scheme C with RNG checks for extending the graph and scheme D for optimizing the edges.

\end{document}